\documentclass[prb,aps,twocolumn,amsfonts,showpacs,eqsecnum,superscriptaddress,floatfix]{revtex4}

\usepackage{epsfig}
\usepackage{psfrag}
\usepackage{color}
\usepackage{graphicx}
\usepackage{subfigure}

\begin{document}

\title{Fano Resonances in Quantum Dots: A Non-Perturbative Role for Potential-Like Scattering}

\author{Robert M. Konik}
\affiliation{Physics Department, Brookhaven National Laboratory, Upton NY 11973}
\affiliation{Physics Department, University of Virginia, Charlottesville VA 22904}
\date{\today}

\begin{abstract}
We consider the physics of transport through quantum dots in the
presence of two tunneling paths.  The first path sees electrons hopping
on and off the dot while the second path is modeled through
a potential scattering-like term.  To study
the effects of potential scattering, we employ a modified version of the
Anderson model.  Such a model can be exactly solved through
the Bethe ansatz, thus allowing a comprehensive and exact
analysis of the zero temperature linear response
conductance.  We find transport properties to be extremely sensitive to
the introduction of a potential scattering term.  Indeed the presence
of such a scattering term, {\it inter alia}, induces a series of first order quantum phase transitions.
Focusing on the Kondo regime of the quantum dot, the non-perturbative effect of
potential-like scattering can be directly tied to both the breaking of
particle-hole symmetry and the interlinking of charge and spin degrees
of freedom in the Anderson model.
The sensitivity to potential scattering
is also reflected in a set of complementary
exact diagonalization computations.
The consequences of this analysis extends to
observations in general of Fano resonances in quantum dots as well as
to the physics of transport through
quantum dots embedded in Aharonov-Bohm rings.
\end{abstract}
\maketitle

%
%
%
\newcommand{\lb}{\langle}
\newcommand{\rb}{\rangle}
%
%
\newcommand{\zb}{\bar{z} }
\newcommand{\half}{{1 \over 2}}
%
%
\newcommand{\ga}{\gamma}         \newcommand{\Ga}{\Gamma}
\newcommand{\be}{\beta}
\newcommand{\al}{\alpha}
\newcommand{\ep}{\varepsilon}
\newcommand{\la}{\lambda}        \newcommand{\La}{\Lambda}
\newcommand{\de}{\delta}         \newcommand{\De}{\Delta}
\newcommand{\om}{\omega}         \newcommand{\Om}{\Omega}
\newcommand{\sig}{\sigma}        \newcommand{\Sig}{\Sigma}
\newcommand{\vphi}{\varphi}
%
%
\newcommand{\CA}{{\cal A}}       
\newcommand{\CB}{{\cal B}}       
\newcommand{\CC}{{\cal C}}
\newcommand{\CD}{{\cal D}}       
\newcommand{\CE}{{\cal E}}       
\newcommand{\CF}{{\cal F}}
\newcommand{\CG}{{\cal G}}       
\newcommand{\CH}{{\cal H}}       
\newcommand{\CI}{{\cal J}}
\newcommand{\CJ}{{\cal J}}       
\newcommand{\CK}{{\cal K}}       
\newcommand{\CL}{{\cal L}}
\newcommand{\CM}{{\cal M}}       
\newcommand{\CN}{{\cal N}}       
\newcommand{\CO}{{\cal O}}
\newcommand{\CP}{{\cal P}}       
\newcommand{\CQ}{{\cal Q}}       
\newcommand{\CR}{{\cal R}}
\newcommand{\CS}{{\cal S}}       
\newcommand{\CT}{{\cal T}}       
\newcommand{\CU}{{\cal U}}
\newcommand{\CV}{{\cal V}}       
\newcommand{\CW}{{\cal W}}       
\newcommand{\CX}{{\cal X}}
\newcommand{\CY}{{\cal Y}}       
\newcommand{\CZ}{{\cal Z}}

\newcommand{\rvac}{\hbox{$\vert 0\rangle$}}
\newcommand{\lvac}{\hbox{$\langle 0 \vert $}}

%
%
%
\font\numbers=cmss12
\font\upright=cmu10 scaled\magstep1
\newcommand{\stroke}{\vrule height8pt width0.4pt depth-0.1pt}
\newcommand{\topfleck}{\vrule height8pt width0.5pt depth-5.9pt}
\newcommand{\botfleck}{\vrule height2pt width0.5pt depth0.1pt}
\newcommand{\Z}{\ifmmode\Zmath\else$\Zmath$\fi}
\newcommand{\Q}{\ifmmode\Qmath\else$\Qmath$\fi}
\newcommand{\N}{\ifmmode\Nmath\else$\Nmath$\fi}
\newcommand{\C}{\ifmmode\Cmath\else$\Cmath$\fi}
\newcommand{\R}{\ifmmode\Rmath\else$\Rmath$\fi}


\newcommand{\del}{\partial}
\newcommand{\js}{J_s}
\newcommand{\rs}{R_s}
\newcommand{\bjs}{\bar{J}_s}
\newcommand{\brs}{\bar{R}_s}
\newcommand{\jnm}{J^{n,m}_s}
\newcommand{\rnm}{R^{n,m}_s}

\newcommand{\prnm}{R^{n',m'}_s}
\newcommand{\bjnm}{\bar{J}^{n,m}_s}
\newcommand{\brnm}{\bar{R}^{n,m}_s}
\newcommand{\pbrnm}{\bar{R}^{n',m'}_s}
\newcommand{\bt}{\beta}
\newcommand{\bs}{{\hat{\beta}^2}}
\newcommand{\tcor}{\langle T(G^{12}_1(x,\tau)G^{12}_1(0,0))\rangle}

\newcommand{\nup}{n_{\rm imp \uparrow}}
\newcommand{\ndown}{n_{\rm imp \downarrow}}
\newcommand{\nboth}{n_{\rm imp \uparrow/\downarrow}}
\newcommand{\vc}{V^{\rm crit}_p}
\newcommand{\vp}{V_p}
\newcommand{\ec}{\epsilon_d^{\rm crit}}
\newcommand{\nt}{n_{\rm imp}}

\newcommand{\ted}{\tilde\epsilon_d}
\newcommand{\ed}{\epsilon_d}
\newcommand{\G}{\Gamma}
\newcommand{\bG}{\bar\Gamma}

\section{Introduction}

Fano resonances are a commonplace phenomena arising from a competition
between two competing scattering paths, one path sensitive to the energy of the scattering
state, one path not.  These resonances were first studied
by the eponymous U. Fano\cite{fano} in the context of
photoionization experiments in atomic systems where 
a continuum of states interacting with a single resonant level serve as the active ionization pathways.  
This work\cite{fano} predicted that interference
between two scattering paths will in general lead to asymmetrically shaped resonances as a function of
scattering energy.  Research on Fano resonances in atomic systems remains a topic of active
interest.\cite{eich,ph_det,ares}  Resonances of similar origin have long been known to characterize scattering off 
nuclei.\cite{bw}  More recently they have been observed in the context of interference in quantum
well systems,\cite{kw} polymer films,\cite{pf} tunneling
into and between multi-wall carbon nanotubes,\cite{tubes} and STM measurements of
adsorbed magnetic atoms on metallic substrates,\cite{madhavan,li,mano} particularly interesting from
a mesoscopic viewpoint in that Kondo physics comes into play.

In this article we will be interested in yet another manifestation of Fano resonances, that involving single
electron transistors (SETs).  Interference between two tunneling paths, one path involving an SET embedded
in an arm of an Aharonov-Bohm ring, was first
studied in Ref. \onlinecite{yacoby}.  
However asymmetric resonances
typical of the phenomena were only first reported in the conductance profiles of SETs several years thereafter.\cite{gores,zach}
In this work, the exact
geometry of the interfering tunneling paths through the SET was left unspecified, 
a situation again seen in the experiments of Ref. \onlinecite{haug}.
More recently, Kobayashi et al.\cite{kobayashia,kobayashib}
have reported Fano resonances in SETs embedded in Aharonov-Bohm rings where necessarily the two tunneling paths are clearly
delineated.

An SET (or more colloquially, a quantum dot)
is fabricated from a gated semiconductor heterostructure such as GaAs/AlGaAs
and is formed from a segregated nanometer scale sized region of a two dimensional electron gas (2DEG).
Such quantum dots are highly tunable.  Both the hopping between the dot and the bulk of the 2DEG
as well as the dot's energy levels relative to the 2DEG's Fermi surface can be adjusted through metallic
gates placed on the heterostructure.
At sufficiently small temperatures (much smaller than the dot level spacing and 
tunneling widths) only the electron levels on the dot nearest in 
energy to the Fermi energy of the bulk 2DEG contribute to transport, turning
the system into a quantum impurity problem.
As is the case in such problems, the
effective dimensionality of the bulk 2DEG is $d_{\rm eff} = 1$ in regards to questions
concerning the impurity, a result of focusing
upon low energy impurity S-wave scattering.\cite{oned}  
With $d_{\rm eff} =1$, the bulk electrons can be treated as living in leads.

The remarkable ability to control the various parameters of a quantum dot has allowed
Kondo physics to be observed in such a setting.\cite{gold,kondo,otherdot}  Tuning the dot such that one electron sits on the dot
level nearest the Fermi surface gives rise to the much of the same physics as an isolated
magnetic impurity.\cite{glazman}.  With the introduction of a second tunneling path, Kondo
physics does not disappear,\cite{gores} but as we will argue, takes on a considerably
different guise.

\begin{figure}
\begin{center}
\noindent
\epsfysize=0.35\textwidth
\epsfbox{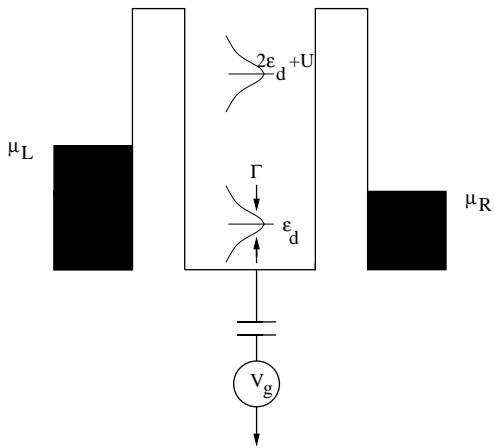}
\end{center}
\caption{Sketched are the relevant energy scales of the single active dot level.}
\end{figure}

In Figure 1 we sketch out the relevant energy scales of the dot with only its energy level nearest the bulk Fermi surface
drawn.  The energy levels of the two `leads'
are marked as forming a continuum filled to levels marked by two chemical potentials, $\mu_L$ and $\mu_R$.
The active dot level is characterized by two energy scales, one marking the energy of single occupancy,
$\epsilon_d$, the other marking the energy of double occupancy, $U+2\epsilon_d$.  $U$ here represents roughly the charging
energy incurred from the addition of the second electron into the level.  Separating the active level from
the two leads are two tunnel barriers.  As drawn, the hybridization of the electrons living on
the level with those in the leads gives rise to the broadening of the level to a width, $\Gamma$.  Represented
in the capacitively coupled gate voltage, $V_g$, is the ability to shift $\epsilon_d$ relative to $\mu_L$ and
$\mu_R$.

The properties of a single level SET are believed to be well captured by the Anderson model.\cite{gold,kondo,otherdot}
Theoretical computations of the finite temperature linear response conductance in the Kondo regime, a non-trivial quantity, 
both using numerical renormalization group techniques\cite{costi} and 
the Anderson's model underlying integrability\cite{long} well match the observations.\cite{gold}  Moreover
qualitative
predictions resulting from the out-of-equilibrium Anderson model,\cite{wingreen} namely
the splitting of the zero bias anomaly into two peaks, have been observed\cite{gold,kondo} (although
the predicted zero temperature features of the peaks \cite{long,moore} have not yet been seen
\cite{gg}).

A continuum Anderson model governing the quantum dot is given by 
\begin{eqnarray}\label{eIi}
{\cal H}_{\rm And.} &=& \sum_{\sigma=\uparrow/\downarrow \atop \alpha=L,R} \bigg\{-i\int dx 
\big(c_{\alpha\sigma}^\dagger(x)\partial_xc_{\alpha\sigma}(x)\big) \cr\cr
&& \hskip -1.in + V(c^\dagger_{\alpha\sigma}d_\sigma + d^\dagger_\sigma c_{\alpha\sigma})|_{x=0}\bigg\}
+ \epsilon_d \sum_{\sigma = \uparrow,\downarrow}n_\sigma + U n_\uparrow n_\downarrow.
\end{eqnarray}
Here $c_{\sigma\alpha}/d$ are electrons living in the leads/dot.
Hopping between the two is governed by $V$.  As indicated in Figure 1, $\ed$ marks the dot single level energy while $U$ describes 
the additional cost of double occupancy.

This description of an SET yields a single, resonant tunneling path for electrons traveling from the right to the left lead.
A second non-resonant tunneling path in the vicinity of the dot can arise in a variety of manners.
This path can occur through the dot itself,\cite{gores,haug} a product perhaps of the detailed
geometry of the confining region of the 2DEG.\cite{clerk}  Alternatively, the second path can be spatially
separated as occurs when a quantum dot is embedded 
in an arm of an Aharonov-Bohm ring,\cite{yacoby,kobayashia,kobayashib} as pictured in Figure 2.  
This is in some sense preferable as the two tunneling paths
are then marked concretely by the two arms of the ring.

We model the second tunneling path in the most minimal way possible by adding to the Anderson model 
above (\ref{eIi}) the term,
\begin{eqnarray}\label{eIii}
{\cal \delta H} &=& V_{LR} \big(c^\dagger_{L\sigma}c_{R\sigma} + {\rm h.c.}\big)\bigg|_{x=0}.
\end{eqnarray}
Such a term has been considered previously in a number of places.\cite{bulka,hofstetter,kor,imry}
It allows the electrons to bypass the dot through a direct lead-lead tunneling term.
In adding $\cal\delta H$, we allow the possibility of multiple windings as electrons transit
successively from one lead to the other via different paths.  This modified Hamiltonian is thus appropriate 
to describing an Aharonov-Bohm ring in a two terminal geometry,\cite{yacoby,kobayashia,kobayashib} not the four terminal
setup of Ref. \onlinecite{schuster}.
The second tunneling path is potential-like in structure.  Indeed with 
the introduction of even and odd combinations of electrons,
$$
c_{e/o} = {1\over \sqrt{2}}(c_L\pm c_R),
$$
the addition is precisely a potential scattering term.

At the centre of this work lies the discovery that this
Hamiltonian, ${\cal H_{\rm And.}}+{\cal \delta H}$,
is integrable.  While it has long been known that the Anderson model alone is integrable,\cite{kao,wie}
only recently has it been realized that the addition, ${\cal \delta H}$, to the Hamiltonian does not spoil
the integrability.\cite{konik}  As will be shown here, the perturbation, ${\cal \delta H}$, 
modifies the integrable structure of the Anderson model in a straightforward way.
While this modification is minor in a technical sense, it will have profound effects on
the transport properties of the system.

\begin{figure}
\begin{center}
\noindent
\epsfysize=0.35\textwidth
\epsfbox{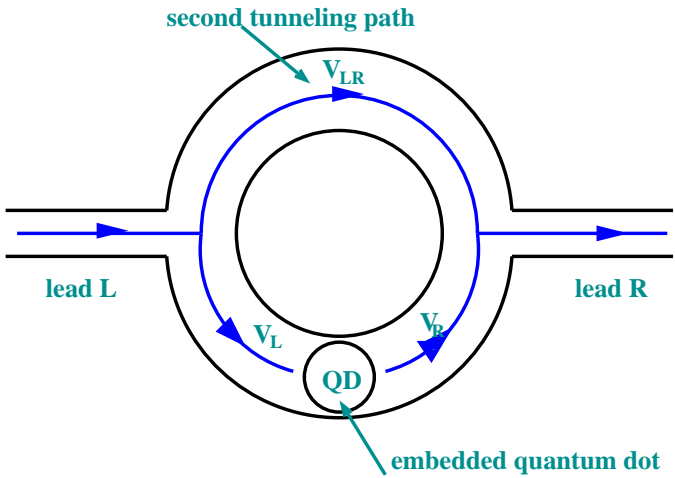}
\end{center}
\caption{Drawn is a ring connected to two leads.  In the lower arm of the 
ring a quantum dot is embedded.
Electrons transit from lead $L$ to lead $R$ by either hopping on and off the dot 
embedded in the lower arm
of the ring or by traveling through the upper arm of the ring.  Electrons may also move from lead $L$
to lead $R$ by executing multiple windings about the ring.}
\end{figure}

\begin{figure*}
\begin{center}
\noindent
\epsfysize=0.55\textwidth
\epsfbox{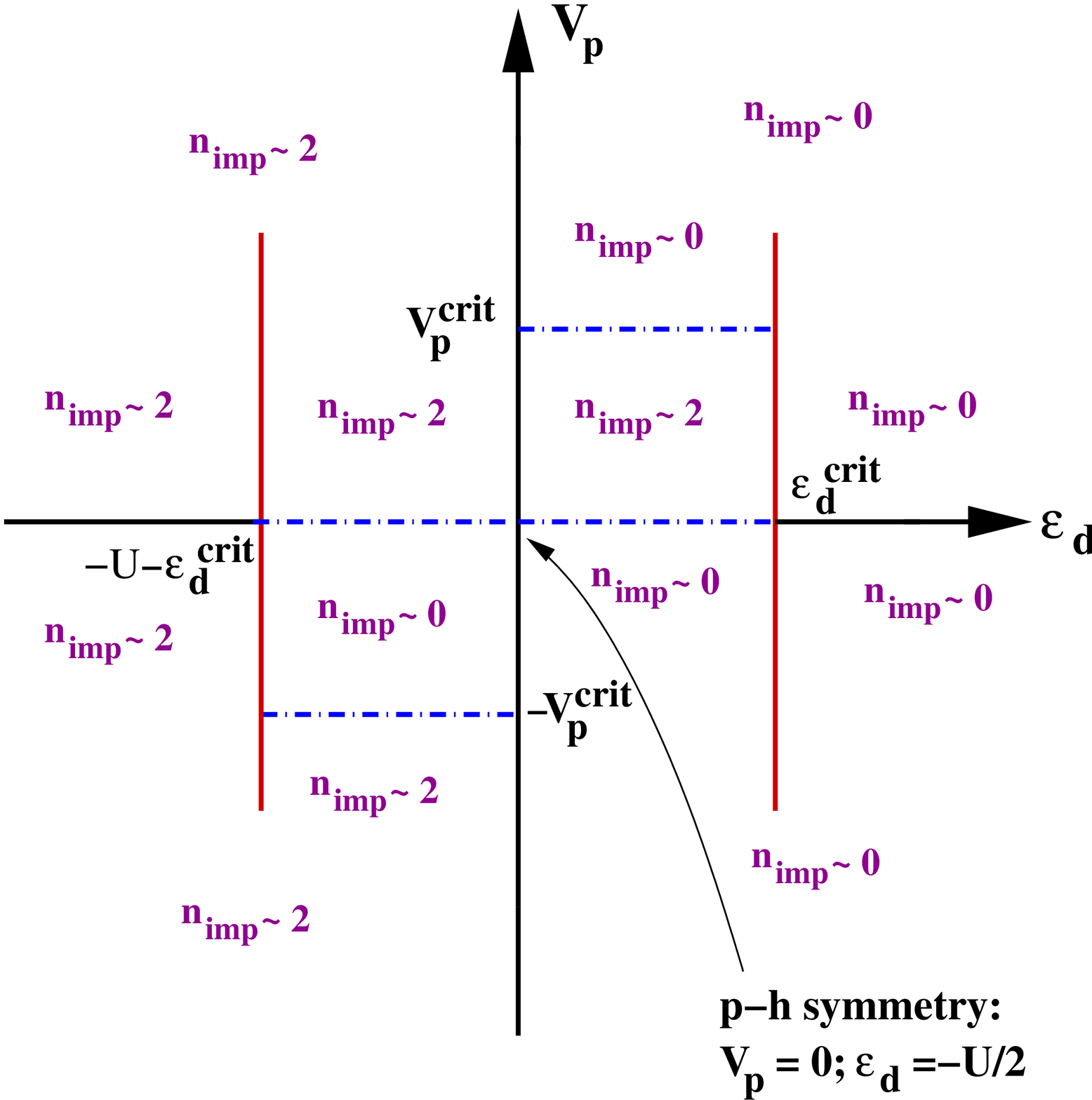}
\end{center}
\caption{A phase diagram of the Anderson model as a function of $\ed$ and $\vp$.  The horizontal dashed
(blue) lines mark first order quantum phase transitions while the vertical solid (red) lines mark
both the termini of the phase transitions as a function of $\ed$ and regions of rapid variation in $n_{\rm imp}$.}
\end{figure*}

At the heart of the model's integrability lies the knowledge of the model's exact eigenfunctions.
From the knowledge of the exact eigenfunctions, we are able to extract the behaviour of excitations
above an interacting Fermi sea.  In particular we can compute the {\it dressed} scattering matrices
of the excitations both with themselves and off the impurity.
While the integrability of the Anderson model has been long established, it was only recently
understood\cite{long} how to wed integrable data
with a Landauer-B\"uttiker framework to access transport properties.  In particular,
in Ref. \onlinecite{long} it was shown how the knowledge of the excitations above the Fermi sea
and their associated scattering phases allow the computation of both the in-equilibrium and out-of-equilibrium
conductances.

Much of the content of Ref. \onlinecite{long} was devoted to understanding how to treat
both finite temperature, $T$, and voltage, $V$, transport in the context of integrability.  At finite $T$ and
$V$ various uncertainties plagued the discussion.  Away from the Fermi surface, electronic excitations
could be constructed only in an ambiguous fashion.
The association of a finite energy electron with a particular
integrable excitation was then left unfixed.  This ambiguity was overcome only in a number of cases with some educated
guesswork.  In this way Ref. \onlinecite{long} was  
able to reproduce the finite temperature linear
response conductance determined from numerical renormalization group calculations.\cite{costi}
Here, in contrast, we will focus upon zero temperature, linear response transport.  There is no
ambiguity in identifying the necessary electronic excitations and their scattering phases at the zero-temperature 
Fermi surface.  As such the results contained herein should be considered exact and without approximation.

\newcommand{\dd}{\dagger}
\newcommand{\ua}{\uparrow}
\newcommand{\da}{\downarrow}
\newcommand{\cu}{c^\dagger_\uparrow}
\newcommand{\cd}{c^\dagger_\downarrow}

At zero temperature, the linear response conductance in the $V_{LR}=0$ Anderson model, $\cal H_{\rm And.}$,
is described by the Friedel sum rule.  The Friedel sum rule relates the scattering phase, $\delta_{e\uparrow/\downarrow}$,
at the Fermi surface
to the number of electrons sitting on the dot,
\begin{equation}\label{eIiii}
\delta_{e\uparrow /\downarrow} = 2\pi n_{d\uparrow /\downarrow}.
\end{equation}
With the second tunneling path activated, we find that a variant of the Friedel sum rule still holds.
But now the scattering at the Fermi surface is related to the total number of electrons displaced by
the impurity (here the impurity should be thought of as both the dot and $\delta {\cal H}$):
\begin{equation}\label{eIiv}
\delta_{e\uparrow /\downarrow} = 2\pi n_{{\rm imp}\uparrow /\downarrow} - 2\tan^{-1}(\vp ).
\end{equation}
where $\vp = V_{LR}/2$.  While the additive constant, $2\tan^{-1} (\vp )$, is important,
the crucial difference between the two expressions for the scattering phase lies in that $n_{\rm imp}$
is distinct from $n_d$.  At least for $U>0$, exact diagonalization computations that complement
the integrable analysis indicate that $n_{{\rm imp}\uparrow/\downarrow}$ involves both
contributions from the dot degrees of freedom as well as bulk electrons in the leads:
\begin{eqnarray}\label{eIv}
n_{\rm imp \ua} &=& n_{d\ua} + \int dx \bigg[ \langle c^\dd_{e\ua}(x) c_{e\ua}(x)\rangle - \rho_{\rm bulk}\bigg];\cr\cr
n_{\rm imp \da} &=& n_{d\da} + \int dx \bigg[ \langle c^\dd_{e\da}(x) c_{e\da}(x)\rangle - \rho_{\rm bulk}\bigg].
\cr &&
\end{eqnarray}
Here $\rho_{\rm bulk}$ equals the unperturbed (absent the impurity) density of states of the lead electrons.

It is hardly surprising that deviations of the electron density in the leads should affect scattering at
the Fermi surface.  This was specifically forewarned as a possibility in the proof of the Friedel sum rule for
the Anderson model by Langreth.\cite{langreth}  And we know such deviations in the lead electron density are important for
determining the finite field scattering phase (and so the magnetoconductance) 
in the Kondo model.\cite{andrei,hewson}
Instead what turns out here to be surprising is that $n_{\rm imp} \equiv n_{\rm imp \ua}+ n_{\rm imp \da}$ 
is not a continuous function of $\vp$.

As a function of $\vp$, $n_{\rm imp}$ sees two discontinuities.  The first discontinuity occurs 
at $\vp$ non-zero, i.e. at $\vp = 0^\pm$.  The origin of this discontinuity is perhaps clearest
at $\ed = -U/2$, at what would be the particle-hole symmetric point of the Anderson model if $\vp = 0$.  $\vp$ finite
then breaks particle-hole symmetry.  In this breaking of the symmetry, a two-fold degeneracy
of the ground state is broken leading to a first order quantum
phase transition.  The two putative ground states carry different electron occupancy.  Their breaking leads 
then to a discontinuous change in $n_{\rm imp}$
by one electron.  At least at the point $\ed = -U/2$, this conclusion is supported by an
exact diagonalization study.

A second first order transition as a function of $\vp$ occurs approximately when the two tunneling paths become
equal in strength.  The sign of $\vp$ at which the transition occurs is a function of $\ed$.  If $\ed > -U/2$,
transport is naturally expressed in terms of particles (as opposed to holes) and the transition
occurs at $\vp = \vc > 0$ where $\vc$ is given by (for $\Gamma \ll U$)
\begin{equation}\label{eIvi}
\vp^{\rm crit} \simeq {2 \Gamma \over U}.
\end{equation}
This positive value of $\vp$ marks the energy necessary to ionize an
electron from the vicinity of the impurity.  If $\ed < -U/2$, transport is most naturally given
in terms of holes and the second transition occurs at $\vp = -\vc < 0$.  $\vp = -\vc$ then marks the 
energy necessary to ionize a hole from the impurity site.

These two transitions in $\vp$ do not persist to all $\ed$.  Rather they are bounded (roughly)
to fall in the range $-U-\ed^{\rm crit} < \ed < \ed^{\rm crit}$ where $\ed^{\rm crit}$ equals
\begin{equation}\label{eIvii}
\ed^{\rm crit} = -U/2 + \sqrt{U\over 8\Gamma}{U\over 2} - {\sqrt{2U\Gamma} \over 2\pi}\log (2\pi e {U\over 8\Gamma}).
\end{equation}
$\ed^{\rm crit}$ marks the point where the disturbance in the impurity density of states due to $\vp$ occurs
above the Fermi surface and so is excluded from contributing to the scattering phase.

These transitions are encoded in Figure 3, a phase diagram of the Anderson model as a function of $\ed$ and
$\vp$.  Each area of the phase diagram is marked by its characteristic value of $n_{\rm imp}$.
The horizontal dashed (blue) line segments mark the two sets of first order phase transition.
On the phase transition lines themselves, $n_{\rm imp} \sim 1$ (not marked in Figure 3).  The termini of these phase
transitions are delineated by solid vertical (red) lines.  Although the phase transitions at
$\vp =0$ and $\vp = \vc$ are drawn so as to end both simultaneously at $\ed^{\rm crit}$ and
$-U-\ed^{\rm crit}$, the termini differ by a value of order $\Gamma$.  The vertical lines
also mark regions of rapid (although not discontinuous) variation in $n_{\rm imp}$.

The conductance, $G$, is simply expressible in terms of the scattering phase 
(at $H=0$ where $\delta_{e\uparrow}=\delta_{e\downarrow}\equiv\delta_e$),
\begin{eqnarray}\label{eIviii}
G = 2{e^2\over h} \sin^2({1\over 2}(\delta_e-2\tan^{-1}(\vp )).
\end{eqnarray}
By making the identifications,
\begin{eqnarray}\label{eIix}
\tilde e &=& \cot\big({1\over 2}(\delta_e + 2\tan^{-1}(\vp ))\big);\cr\cr
\tilde q &=& -\cot (2\tan^{-1}(\vp )),
\end{eqnarray}
$G$ can be put into a Fano-like form,
\begin{equation}\label{eIx}
G = 2{e^2 \over h} {4\vp^2 \over 1 + \vp^2} {(\tilde e + \tilde q)^2 \over \tilde e^2 +1}.
\end{equation}
However this form is somewhat misleading.  $\tilde e$ is not be identified with $\ed/\Gamma$ as is true
in the $U=0$ case.  In fact unlike the non-interacting case, we find the many-body correlations introduced by finite $U$ can
lead to the non-vanishing of $G$ as $\ed$ is varied with $q$ still real.  It would be interesting
to reconsider then the analysis of the Fano resonances found in Ref. \onlinecite{kobayashib} (although 
the large magnetic fields at which the resonances were observed might largely eliminate many-body effects).

The origin of the phase transitions and their attendant consequences
can be understood not only as a consequence of many-body interactions
but of mere quantum mechanics as well.  The exact solution of the $\vp \neq 0$ Anderson model
dictates that the ground state consists of a sea of two-electron repulsively interacting bound states.  
(This feature of the ground state is independent of $\vp$ and holds true for the $\vp =0$ Anderson
model.)
While this feature results
from the exact treatment of the many body physics in the problem, the origin of the discontinuous
behaviour in $n_{\rm imp}$ can be seen in the bare bound states themselves.  If one computes the finite $\vp$ scattering
phase of a two-electron bound state off the impurity (a quantum mechanical problem), 
one sees that it is characterized as a function of energy by a number of non-analyticities.
The close relationship between the undressed bound state scattering phase, the impurity occupancy, $n_{\rm imp}$, 
and so ultimately the dressed scattering phase at the Fermi surface results in transport properties sharing these
same non-analyticities.  The explicit construction of the bound state wave function and a display
of the non-analyticities are carried out in Section II.B.3.

The appearance of non-analyticities in the bound state scattering phase and so transport properties is in part a consequence of 
the treatment of the two tunneling paths,
\begin{eqnarray}\label{eIxi}
{\cal H}_{\rm tun} &=& V\sum_{{l=L,R}\atop{\sigma=\uparrow ,\downarrow}}
\big( c^\dagger_{l\sigma}d_\sigma + d^\dagger_\sigma c_{l\sigma}\big)|_{x=0}\cr\cr
&& \hskip -.4in + 2V_{p}\sum_{\sigma=\uparrow ,\downarrow}
\big( c^\dagger_{L\sigma}c_{R\sigma} + c^\dagger_{R\sigma} c_{L\sigma}\big) |_{x=0}.
\end{eqnarray}
In ${\cal H}_{\rm tun}$, the two tunneling terms are presented as both infinitely narrow with delta-function strength.
However in any real experimental system
the two tunneling terms will have finite, differing widths.
In general unequal widths can lead to changes in the physics.
If for example the width of the lead-lead term 
is much greater than the width of the dot-lead coupling, i.e. $w_{\rm lead-lead} \gg w_{\rm dot-lead}$,
the effect of the second tunneling path on the physics
becomes relatively trivial.  The results presented in this article are presumed valid only if
$w_{\rm lead-lead} < w_{\rm dot-lead}$.
More is said on this in Section II.B.1.

${\cal H}_{\rm tun}$ also treats the two tunnelings
as occurring at the same
point, $x=0$.  If in the effective theory we were to displace the scattering terms by some arbitrarily small distance
the non-analyticities disappear.  The distinction can be made clear by considering the single electron
$U=0$ scattering phase.  The scattering phase of $c_e = (c_L + c_R)/\sqrt{2}$ at wavevector q is given by
\begin{equation}\label{eIxii}
\delta_e(q) = -2\tan^{-1}({\Gamma \over q-\ed} + \vp).
\end{equation}
Notice that $\vp$ appears within the argument of the arctan.
In the case of a quantum dot embedded in an Aharonov-Bohm ring, this form of the scattering phase results 
from topology, from the possibility of multiple windings around the ring.  While we are ultimately interested
in the scattering phase of bound states, this form of the scattering phase is crucial if non-analyticities 
in $\vp$ are to arise.
Now if instead we shift
(say) the potential scattering to $x=\epsilon$, i.e.
\begin{eqnarray}\label{eIxiii}
{\cal H}_{\rm tun} &=& V\sum_{{l=L,R}\atop{\sigma=\uparrow ,\downarrow}}
\big( c^\dagger_{l\sigma}d_\sigma + d^\dagger_\sigma c_{l\sigma}\big)|_{x=0}\cr\cr
&& \hskip -.4in + 2V_{p}\sum_{\sigma=\uparrow ,\downarrow}
\big( c^\dagger_{L\sigma}c_{R\sigma} + c^\dagger_{R\sigma} c_{L\sigma}\big)|_{x=\epsilon},
\end{eqnarray}
the scattering phase reads instead
\begin{equation}\label{eIxiv}
\delta_e(q) = -2\tan^{-1}({\Gamma \over q-\ed}) - 2\tan^{-1}(\vp ).
\end{equation}
We now have excluded the possibility of multiple windings and
the limit of the bound state scattering phase as a function of $\vp$ will be entirely well behaved.

Generically the tunneling terms will not exactly coincide and 
we will have the situation as given in (\ref{eIxiv}).
This is reassuring.  If potential scattering was always to have drastic consequences we would need to
explain why its presence is not always felt in experiments in magnetic impurities in bulk metals.
The effective Anderson Hamiltonian of a magnetic impurity in a bulk metal will generally not possess particle-hole
symmetry.  Thus it will include, however small, a potential scattering term. 
However the coincidence of dot and potential scattering centers as given in (\ref{eIxi}) requires some enforcement mechanism
(such as perhaps the topological possibility of multiple windings arising for a quantum dot embedded in a ring).  
This enforcement mechanism will not
be present in the general case and potential scattering will be innocuous.

What then is the likelihood that the scattering phase appearing in (\ref{eIxii}) is relevant to understanding
Fano resonances in quantum dots?
Certainly this phase is generic theoretically.\cite{bulka,hofstetter}
The differences between our work and other analyses detailed in this article arise not because
of the form of $\delta_e(q)$, but because of the treatment of interactions on the quantum dot.  But what is the situation
experimentally?  While Fano resonances have been observed, their observation does not necessarily preclude the
form of the scattering phase in Eqn. (\ref{eIxiv}).  One possible means of distinguishing
between the two forms of the scattering phase lies in the effect of finite interactions.  As we present in Section VI, 
various features of the observations found in Ref. \onlinecite{gores} suggest a role for finite interactions
only consistent with (\ref{eIxii}).  However this cannot be considered definitive while the role interactions
play in other reports of Fano resonances\cite{haug,kobayashia,kobayashib} is undetermined.

The dramatic effect potential scattering presents is not only a consequence of the coincidence of scattering
centers.  It reflects the entanglement of charge and spin degrees of freedom that occurs in the Anderson model.
In the Kondo model, one may add a potential scattering term with strength, $\vp$,
and maintain the model's integrability.  
If the scattering phase at the Fermi surface is then computed one finds
\begin{equation}\label{eIxv}
\delta_{e\ua /\da} = \pi + f(\vp ),
\end{equation}
where $\pi$ is the original contribution of the screened spin to the scattering phase while 
$f$ is a smooth function of the strength, $\vp$.  This scattering phase then behaves in the same fashion
as the scattering phase for an Anderson model with potential scattering but without coincident scattering centers.
The trivial effect of potential scattering is a result of the complete
uncoupling of spin and charge degrees of freedom in the Kondo model.  Potential scattering acts directly only in the charge sector
while all the non-perturbative physics occurs in the spin sector.  Thus potential
scattering is unable to effect the Kondo physics in an interesting fashion.  This limitation is removed
however for the Anderson model where the two sectors can interact.

A consequence of this behaviour is that the Anderson model with potential scattering cannot be treated
via a Schrieffer-Wolff (SW) transformation.\cite{sw}  To implement the standard 
SW transformation on ${\cal H}_{\rm And.}+\delta {\cal H}$,
one should first transform away the potential scattering term via
\begin{eqnarray}\label{eIxvi}
c_e (x\neq 0) &=& e^{-i\delta/2}\theta(-x)\tilde c_e (x) + e^{i\delta /2}\theta (x) \tilde c_e (x);\cr\cr
c_e (0) &=& \cos (\delta /2) \tilde c(0) + V\vp \cos^2(\delta /2) d;\cr\cr
\delta &=& -2\tan^{-1}(\vp ).
\end{eqnarray}
This gauge transformation transforms $\CH_e$ into the ordinary Anderson model with renormalized
parameters, $\tilde\ed$ and $\tilde V$:
\begin{eqnarray}\label{eIxvii}
\tilde\ed &=& \ed - V^2 \vp\cos^2(\delta /2);\cr\cr
\tilde V &=& V\cos (\delta /2).
\end{eqnarray}
The SW transformation taking the Anderson model
to a Kondo model goes forward as usual by freezing out charge excitations.  The scattering
of the transformed electrons at the Fermi surface
is the Kondo scattering phase
\begin{eqnarray}\label{eIxviii}
\tilde\delta = \delta_{\rm Kondo} = \pi ,
\end{eqnarray}
while the scattering phase of the original electrons will
have the form
\begin{eqnarray}\label{eIxix}
\delta = \delta_{\rm Kondo} - 2\tan^{-1}(V_p) ,
\end{eqnarray}
that is, one restores the scattering phase due to potential
scattering that was originally transformed away.  Comparing this result with (\ref{eIxiv}) or (\ref{eIxv}),
suggests that an SW transformation performed in such a fashion
implicitly
assumes the potential scattering term is not exactly at $x=0$.  
Alternatively, the gauge transformation in (\ref{eIxvi}) 
together with the SW transformation seem to mistreat the coupling between the spin and charge sectors of the
theory.  We will sharpen our critique of the gauge transformation in (\ref{eIxvi}) in what follows (see in particular
Section V).

While the SW transformation fails to describe the physics of the $\vp \neq 0$ Anderson model, this failure does not imply that
Kondo physics is absent.  One hallmark of Kondo physics is the presence of the Abrikosov-Suhl resonance.  This is
an enhancement in the low energy 
spectral weight of the impurity degrees of freedom.  This ensures quantities 
such as the zero temperature magnetoconductance vary rapidly on the order of the Kondo
temperature, $T_k$.  At $\vp \neq 0$,
low energy spectral weight is in general still present in the Kondo regime ($\ed \sim -U/2$) of the Anderson
model but in modified form.  With $\vp$ finite,
spectral weight present near the Fermi surface is shifted.  The nature of this shift can be read off from the value
of $n_{\rm imp}$ in Figure 3.  For $\vp =0$ we have $n_{\rm imp} \sim 1$ for $\ed \sim -U/2$.  Taking $\vp$ finite,
$n_{\rm imp}$ shifts to $\sim 0$ or $2$.  A shift of $n_{\rm imp}$ to 2 coincides with low lying spectral weight
just above the Fermi surface being transferred to just below.  
The corresponding absence of spectral weight at low but positive energies leads to the magnetoconductance being
insensitive to small changes in $H$.
Analogously, a shift of $n_{\rm imp}$ from $1$ to $0$
implies that low energy weight has been transferred from just below the Fermi surface to just above.  
Here in contrast the magnetoconductance becomes all the more sensitive to small changes in $H$.
The magnetoconductance is discussed in detail in Section II.B.

With these comments in hand, we now outline the course of the paper.  A preliminary, considerably more
terse, treatment of this subject has already appeared.\cite{konik}  In Section II we outline the
exact solution of the finite $\vp \neq 0$ Anderson model.  We begin by introducing the model followed by the development of the 
Bethe ansatz solution in several steps.  By constructing the single and two-particle eigenstates of the model
we are able to extract the scattering matrices/phases.  The structure of these scattering matrices
demonstrate that the model is integrable.  Beyond the scattering in the system, the two
particle eigenstates allow us to exhibit the nature of the bound states.  As we have already indicated,
much of the physics is encoded in the scattering of the bound states off the impurity.  By further sketching
the development of multi-particle eigenstates, we are able to argue that in all details but the impurity scattering
phase, the equations describing the quantization of momentum in the system, the Bethe ansatz equations,
are identical to the $\vp =0$ case.  This allows us readily to develop the method by which dressed scattering
phases of electrons at the Fermi surface are computed.  In the first part of Section II we limit ourselves
to the case $\ed \geq -U/2$.  We so end Section II by extending the analysis to $\ed \leq -U/2$.  This extension
will prove essential when we come to compute the conductance exactly at $\ed = -U/2$.

In Section III we develop the resulting properties of the linear response conductance.  We begin by computing the zero
field conductance.  We carefully associate discontinuous changes in $n_{\rm imp}$ with similar changes in the conductance.
The appearance of the discontinuous changes finds support in exact diagonalization computations which show similar
discontinuous changes in the ground state occupancy.  Key features of $G$ as both a function of $\vp$ and
$\ed$ are described here.  In particular the ingredients going into the phase diagram of Figure 3 are established.
Having described $G(H=0)$, we then turn to understanding the magnetoconductance.  Here we are able 
to characterize analytically the behaviour of the $H\neq 0$ conductance at $\ed = -U/2$.  Away from
$\ed = -U/2$ we rely upon numerical solutions of the equations describing the dressed scattering phases.

The role two-electron bound states play in the exact solution of the $\vp\neq 0$ Anderson model is ultimately
a consequence of a finite $U$.  With $U=0$, the behaviour of the conductance, $G$, as a function of $\vp$ is
relatively bland.  This points to a non-perturbative role for $U$.  To reinforce this, in Section IV we consider the solution
of the $\vp \neq 0$ Anderson model for $U<0$.  With $U<0$, all non-analyticities in $\vp$ disappear suggesting
that various impurity quantities do not possess well-behaved power series expansions in $U/\Gamma$ as with
the $\vp =0$ model.

In the penultimate Section V, we analyze an alternative approach to this problem, an approach predicated on
the use of Dyson equations and developed in Ref. \onlinecite{bulka} and Ref. \onlinecite{hofstetter}.  
This approach leads to transport quantities being analytic in both $U$
and $\vp$ and so to considerably different results than presented in this article.  
We will argue that
the differences arise because the Dyson equations treat the impurity occupancy, $n_{\rm imp}$,
as equivalent to $n_d$, the number of electrons on the dot itself.
We will equate the Dyson equations to the results obtained through applying the gauge
transformation in (\ref{eIxvi}).  This will enable us to argue that even at the quantum mechanical
level, the Dyson equations seem to incorrectly treat finite interactions.
We believe that the origin of the discrepancy between the two approaches lies in that the Dyson equations
presume implicitly both $\vp$ and $U$ are perturbative quantities 
and so end with quantities well behaved in $\vp$ and $U$.  

In Section VI, the final section, we consider the observations reported in G\"ores et al.\cite{gores} as
these served as the initial motivation for this work.  While hardly determinative, 
some generic features present in these experiments are well explained using the formalism
developed in this work.

\section{Exact Solution}

\begin{figure}
\begin{center}
\noindent
\epsfysize=0.1\textwidth
\epsfbox{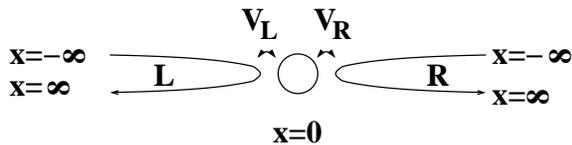}
\end{center}
\caption{The unfolding of the left (L)/right (R) leads into chiral fermions extending
from $x=-\infty$ to $x=\infty $.}
\end{figure}

\subsection{Description of System}

We use a more general Hamiltonian than 
in (\ref{eIi}) and (\ref{eIii}) to describe the lead-impurity system,
\begin{eqnarray}\label{eIIi}
{\cal H} &=& H_{\rm leads} + H^1_{\rm tun} + H^2_{\rm tun} + H_{\rm dot};\cr\cr
H_{\rm leads} &=& \sum_{{l=L,R}\atop{\sigma=\uparrow ,\downarrow}}\int^{\infty}_{-\infty}dx
\big(-ic^\dagger_{l\sigma}(x)\partial_xc_{l\sigma}(x)\big);\cr\cr
H^1_{\rm tun} &=& \sum_{{l=L,R}\atop{\sigma=\uparrow ,\downarrow}}V_l
\big(c^\dagger_{l\sigma}d_\sigma + d^\dagger_\sigma c_{l\sigma}\big)|_{x=0};\cr\cr
H^2_{\rm tun} &=& V_{LR}\sum_{\sigma=\uparrow ,\downarrow}
\big(c^\dagger_{L\sigma}c_{R\sigma} + c^\dagger_{R\sigma} c_{L\sigma}\big)|_{x=0};\cr\cr
H_{\rm ref} &=& \sum_{\sigma=\uparrow ,\downarrow}
\big(V_{LL}c^\dagger_{L\sigma}c_{L\sigma} + V_{RR}c^\dagger_{R\sigma}c_{R\sigma}\big)|_{x=0};\cr\cr
H_{\rm dot} &=& \epsilon_d \sum_{\sigma = \uparrow,\downarrow}n_\sigma + U n_\uparrow n_\downarrow.
\end{eqnarray}
Here again $c^\dagger_{l\sigma} / c_{l\sigma}$ 
are electron operators in the leads while $d^\dagger_\sigma / d_\sigma$
are the corresponding operators on the dot
with $n_\sigma = d^\dagger_\sigma d_\sigma$.  The leads in Figure 4
are semi-infinite and have electrons which move both to the left and the right.  By ``unfolding'' the leads
(in the continuum limit), each lead can be thought to run from $x=-\infty$ to $x=\infty$ but with chiral
electrons which are solely right moving.

The path from $L$ to $R$ taking the electron through the dot is modeled by $H^1_{\rm tun}$.  We have allowed
for the possibility that tunneling between the dot and the two leads is asymmetric, i.e. $V_L \neq V_R$.  The second
path from $L$ to $R$ (for example, the upper arm of the ring in Figure 2) 
is modeled by $H^2_{\rm tun}$, a potential
scattering-like term.  We have also included a term in the Hamiltonian, $H_{\rm ref}$, by which electrons again
experience a single-body potential but one which does not scatter the electrons into the other lead.  Although
likely to be generically present, we here consider
such a term for technical reasons alone, as will be seen shortly.

$H_{\rm dot}$ describes the energies associated with the dot.  The dot single particle Hilbert space is
two fold: one for occupancy by a spin $\uparrow$-electron and one for occupancy by a spin-$\downarrow$ electron.
The chemical potential of these levels is controlled by $\epsilon_d\sum_\sigma n_\sigma$ while
Coulomb interactions on the dot are represented by the term $Un_\uparrow n_\downarrow$.  This is 
in fact the only interacting term in the problem -- the leads are treated as Fermi liquids.  This
is reasonable as the one-dimensional leads ultimately arise out of a dimensional reduction of electrons
living in two or three dimensions.  
Although $U$ is the only non-trivial interaction in the problem, it presents the same set
of complications seen in a fully interacting system.

To treat this Hamiltonian we want first to map it to a model involving a single electron specie.  This cannot
be done for the general values of $V_{LL}$, $V_{RR}$, $V_{LR}$, and $V_L/V_R$.  However if we restrict
ourselves to the parameter space,
\begin{equation}\label{eIIii}
V_{LL} = -V_{RR} = {V_{LR}\over 2V_L V_R}(V_L^2-V_R^2),
\end{equation}
we can proceed.
Note that if the dot-lead coupling is symmetric (or if $V_{LR}=0$) then $V_{LL}$ and $V_{RR}$ must vanish.
With this constraint on the parameter space we introduce even and odd electrons, $c_{e/o}$,
via
\begin{equation}\label{eIIiii}
c_{e/o} = (V_{L/R}c_L \pm V_{R/L}c_R)/\sqrt{V_L^2+V_R^2}.
\end{equation}
The Hamiltonian in this basis then reads
\begin{equation}\label{eIIiv}
{\cal H} = H_e + H_o,
\end{equation}
where
\begin{eqnarray}\label{eIIv}
H_e &=& \sum_\sigma \bigg\{\int dx -i c_{e\sigma}^\dagger(x)\partial_xc_{e\sigma}(x) \cr\cr
&& + (2\Gamma)^{1/2}(c^\dagger_{e\sigma}d_\sigma + d^\dagger_\sigma c_{e\sigma})|_{x=0}
+2V_p c_{e\sigma}^\dagger c_{e\sigma}|_{x=0} \bigg\}\cr\cr
&& + \epsilon_d \sum_{\sigma = \uparrow,\downarrow}n_\sigma + 
U n_\uparrow n_\downarrow;\cr\cr
H_o &=& \sum_\sigma \int dx -ic_{o\sigma}^\dagger(x)\partial_xc_{o\sigma}(x) - 2V_p c_{o\sigma}^\dagger 
c_{o\sigma}|_{x=0}.\cr&&
\end{eqnarray}
The parameters $\Gamma$ and $V_p$ are given by
\begin{eqnarray}\label{eIIvi}
\Gamma &=& {V_L^2 + V_R^2 \over 2};\cr\cr
V_p &=& {\Gamma V_{LR} \over 2V_LV_R}.
\end{eqnarray}
$\Gamma$ represents the total broadening of the electron dot level due to coupling to the leads
while $V_p$ represent the effective potential scattering seen by the even and odd electrons.
The odd sector of the theory, uncoupled to the dot, is trivial.  The even sector,
we will argue, is solvable via Bethe ansatz.

Having undertaken this recasting of the Hamiltonian we must describe how to maintain contact with the
original problem: the transport of electrons from the left lead to the right lead and vice versa.
We can do this as follows.  Let ${\cal T}_{ll'}$ be the transmission amplitude of an electron from
lead $l$ to lead $l'$ and ${\cal R}_{l}$ the reflection amplitude of an electron in lead $l$ back
into the same lead.
In the even and odd basis it will turn out we can readily compute the scattering phases, $\delta_e/\delta_o$,
of an even/odd electron off the impurity (here the scattering must be a pure phase as the even/odd
sectors are uncoupled and both chiral). ${\cal T/R}$ can be readily expressed in terms 
of $\delta_e/\delta_o$:\cite{long}
\begin{eqnarray}\label{eIIvii}
V_R e^{i\delta_e} &=& V_R{\cal R}_R+V_L{\cal T}_{LR};\cr\cr
V_L e^{i\delta_o} &=& V_L{\cal R}_R-V_R{\cal T}_{LR}.
\end{eqnarray}
Similar relations hold with ${\cal T}_{RL}$ and ${\cal R}_L$.
Thus 
\begin{eqnarray}\label{eIIviii}
{\cal T}_{LR} &=& {V_LV_R\over V_L^2 + V_R^2}(e^{i\delta_e}-e^{i\delta_o});\cr\cr
{\cal R}_R &=& {V_R^2 e^{i\delta_e}+ V_L^2 e^{i\delta_o} \over V_L^2+V_R^2}.
\end{eqnarray}
By parity, we must have $|{\cal T}_{LR}|=|{\cal T}_{RL}|$ and $|{\cal R}_R |= |{\cal R}_L|$.
The zero temperature linear response conductance is then given by 
\begin{equation}\label{eIIix}
G=2{e^2 \over h}|{\cal T}_{LR}|^2=2{e^2\over h} {V_L^2V_R^2 \over \Gamma^2}\sin^2({\delta_e-\delta_o \over 2})
\end{equation}
For most of the paper we will assume $V_L=V_R$.  Only when comparing theory with the experimental observations
of Ref. \onlinecite{gores} will we relax this constraint.

The prescription at finite energy/finite temperature 
for relating $\delta_e/\delta_o$ to $\cal T$ and $\cal R$ is in general fraught with ambiguities.
This is discussed in scholastic detail in Ref. \onlinecite{long}.  In this paper we will focus upon $T=0$ linear
response quantities.  Taking this limit lifts these ambiguities making the above transformation
precise.

\subsection{Construction of Multiparticle Eigenstates for ${\cal H}_e$}

\subsubsection{Single Particle Eigenstates}

We first examine the single particle eigenstates of ${\cal H}_e$.  This will give the bare scattering phase off
the impurity.  We want to look for states of the form
\begin{eqnarray}\label{eIIx}
|\psi_\sigma\rangle = \bigg[ \int^\infty_{-\infty}dx \{ g_\sigma (x) c^\dagger_\sigma(x) \} 
+ e_\sigma d^\dagger_\sigma\bigg]|0\rangle .
\end{eqnarray}
Solving Schr\"odinger's equation, ${\cal H}_e |\psi_\sigma\rangle = q |\psi_\sigma\rangle$, yields the
following constraints upon $g_\sigma (x)$ and $e_\sigma$:
\begin{eqnarray}\label{eIIxi}
g_\sigma (0) (2\Gamma)^{1/2} + \epsilon_d e_\sigma &=& q e_\sigma ;\cr\cr
-i\partial_x g_\sigma (x) + \delta (x)((2\Gamma )^{1/2}e_\sigma + 2V_p g_\sigma (0)) &=& q g_\sigma (x). \cr &&
\end{eqnarray}
Taking $g_\sigma (x)$ to be of the form
\begin{eqnarray}\label{eIIxii}
g_\sigma (x) &=& \theta (x) e^{iqx + i\delta/2} + \theta (-x) e^{iqx - i\delta/2};\cr\cr
g_\sigma (0) &=& \cos (\delta(q)/2) ,
\end{eqnarray}
we can readily solve (\ref{eIIxi}) finding expressions for $e_\sigma$ and $\delta (q)$,
\begin{eqnarray}\label{eIIxiii}
e_\sigma &=& (2\Gamma)^{1/2}{\cos(\delta(q)/2) \over q -\ed}\cr\cr
\delta (q) &=& -2\tan^{-1}({\Gamma \over q - \epsilon_d} + V_p).
\end{eqnarray}
Note that the potential scattering strength appears inside the argument of the $\tan^{-1}$.  This, ultimately,
is a reflection of the coupling between the spin and charge sectors of the Anderson model 
(and if appropriate, the possibility of multiple windings about an Aharonov-Bohm ring).  It thus
sets the scene for the non-trivial physics of the potential scattering term.

$\delta (q)$ contains information on the non-interacting $(U=0)$ states of the retarded dot Greens function:
\begin{eqnarray}\label{eIIxiv}
-{\rm Im}G^{\rm ret}_d(\omega ) &=& {1\over 2\pi}\partial_q \delta (q) |_{q=\omega}\cr\cr
&=& {\bG \over (q-\ted)^2 + \bG^2}
\end{eqnarray}
where $\bG$ and $\ted$ are the renormalized values of the dot level broadening and dot chemical potential
in the presence of a finite $V_p$
\begin{eqnarray}\label{eIIxv}
\bG &=& {\G \over 1+ V_p^2};\cr\cr
\ted &=& \ed - V_p\bG.
\end{eqnarray}
We see the presence of the potential scattering term serves to narrow the dot resonance as well as shifting
the position of the energy level.  In terms of correlators involving dot degrees of freedom alone,
these renormalizations, we believe, will be the sole response to the potential scattering.  However the correlators
relevant to transport go beyond dot correlations.  In
these correlators the effects of potential scattering will be non-trivial.

The resulting single particle phase in (\ref{eIIxiii}) results from treating the lead-lead coupling, $V_p$,
as well as the dot-lead coupling, $\Gamma$, as delta-function scatterers.  However as pointed out in the
introduction, in any experimental system the two couplings will have some finite widths, $w_{\rm dot-lead}$ and
$w_{\rm lead-lead}$.  It is relatively
easy to show that in order to see the physics encoded in (\ref{eIIxiii}) 
we require $w_{\rm dot-lead} \geq w_{\rm lead-lead}$.
To see an indication of this, suppose we broaden out the lead-lead scattering term, replacing $V_p\delta (x)$ in
(\ref{eIIxi}) with $V_p w(x)$ where $w(x)$ is some even
smooth function with unit weight.  The resulting solution
to the modified (\ref{eIIxi}) is
\begin{eqnarray*}
g_\sigma (x) &=& \exp\big( -i\int_{-\infty}^x dx' 2 V_p w(x')\big)\cr\cr
&& \times \bigg(\theta (x) e^{iqx + i\delta/2} 
+ \theta (-x) e^{iqx - i\delta/2}\bigg);\cr\cr
e_\sigma &=& e^{-iV_p}(2\Gamma)^{1/2}{\cos(\delta(q)/2) \over q -\ed}\cr\cr
\delta (q) &=& -2\tan^{-1}({\Gamma \over q - \epsilon_d}).
\end{eqnarray*}
Thus the effective scattering phase of an electron as it traverses from $x=-\infty$ to $x=\infty$
is
\begin{eqnarray*}
\delta^{\rm eff}(q) = -2\tan^{-1}({\Gamma \over q - \epsilon_d}) - 2V_p
\end{eqnarray*}
Unlike with the electron scattering phase in (\ref{eIIxiii}), 
we would expect all finite $U$ transport properties to be smooth functions of $V_p$.
Then in order to see the effects predicated upon (\ref{eIIxiii}), we (at least) require
$w_{\rm dot-lead} \geq w_{\rm lead-lead}$.

\subsubsection{Two-Particle Eigenstates}

The examination of the two-particle eigenstates will allow us to construct the S-matrix for the model.
In principle we will then be ready to compute the multi-particle eigenstates.  The only non-trivial
two-particle states are those with total $S_z=0$.  We thus focus on these.  Such states have the form
\begin{eqnarray}\label{eIIxvi}
|\psi\rangle &=& \bigg[ \int^{\infty}_{-\infty} dx_1dx_2 g(x_1,x_2)c^\dagger_\uparrow (x_1) 
c^\dagger_\downarrow (x_2)\cr\cr
&& \hskip -.75in + \int^\infty_{-\infty} dx e(x)
(c^\dagger_\uparrow (x) d^\dagger_\downarrow - c^\dagger_\downarrow (x)d^\dagger_\uparrow)
+ fd^\dagger_\uparrow d^\dagger_\downarrow\bigg]|0\rangle .
\end{eqnarray}
Solving the Schr\"odinger equation, ${\cal H}_e|\psi\rangle = E |\psi\rangle$, leaves us with the
following set of consistency equations,
\begin{eqnarray}\label{eIIxvii}
Eg(x_1,x_2) &=& -i(\partial_{x_1}+\partial_{x_2})g(x_1,x_2)\cr\cr 
&+& \sqrt{2\Gamma}(\delta (x_2) e(x_1)+ \delta (x_1) e(x_2))\cr\cr
&+& V_p g(x_1,x_2)(\delta (x_1)+\delta (x_2));\cr\cr
Ee(x) &=& -i\partial_x e(x) + \sqrt{2\Gamma}g(x,0) + \sqrt{2\Gamma}f\delta(x) \cr\cr 
&+& \ed e(x) + \delta (x) V_p e(0);\cr\cr
Ef &=& 2\sqrt{2\Gamma}e(0) +2\ed f +Uf,
\end{eqnarray}
together with assuming $g(x_1,x_2)$ is symmetric.
To solve these relations, we employ the following ansatz
\begin{eqnarray}\label{eIIxviii}
g(x_1,x_2) &=& g_q (x_1)g_p(x_2)\phi(x_1-x_2)\cr\cr 
&&\hskip .3in + g_p(x_1)g_q(x_2)\phi(x_2-x_1);\cr\cr
e(x) &=& e_p\phi(x)g_q(x) + e_q\phi(-x)g_p (x);\cr\cr
E &=& q+p.
\end{eqnarray}
Here $g_{q/p}(x)$ and $e_{q/p}$ are the coefficients appearing in the one particle wavefunction corresponding to
energies $q/p$ (i.e. (\ref{eIIxii}) and (\ref{eIIxiii})).  
$\phi(x)$ is an expression of the phase change incurred when two electrons are interchanged,
$x_1\leftrightarrow x_2$.  We assume it takes the form
\begin{equation}\label{eIIxix}
\phi (x) = 1+i\alpha (p,q) {\rm sign} (x) .
\end{equation}
Under this ansatz, we find
\begin{eqnarray}\label{eIIxx}
f &=& -2\sqrt{2\Gamma}{e_p cos (\delta (q)/2) + e_q \cos (\delta (p)/2) \over 2\ed + U - (q+p)};\cr\cr
\alpha (p,q) &=& {1\over q-p}{2U\Gamma \over 2\ed +U - q -p}.
\end{eqnarray}
From $\alpha (q,p)$ we can determine the scattering matrix for two electrons.

This S-matrix, $S^{\sigma_1'\sigma_2'}_{\sigma_1\sigma_2}(p,q)$, where $\sigma_1$ and $\sigma_2$ are the spins of 
ingoing electrons while $\sigma_1'$ and $\sigma_2'$ are the spins of the outgoing electrons.  This
S-matrix, on the grounds of $SU(2)$ invariance, must have the form
\begin{eqnarray}\label{eIIxxi}
S^{\sigma_1'\sigma_2'}_{\sigma_1\sigma_2} = b(p,q) I^{\sigma_1'\sigma_2'}_{\sigma_1\sigma_2}
+ c(p,q) P^{\sigma_1'\sigma_2'}_{\sigma_1\sigma_2},
\end{eqnarray}
where $I^{\sigma_1'\sigma_2'}_{\sigma_1\sigma_2} = \delta^{\sigma_1'}_{\sigma_1}\delta^{\sigma_2'}_{\sigma_2}$
is the identity matrix while 
$P^{\sigma_1'\sigma_2'}_{\sigma_1\sigma_2} = \delta^{\sigma_2'}_{\sigma_1}\delta^{\sigma_1'}_{\sigma_2}$ is
the permutation matrix.  The coefficients, $b(p,q)$ and $c(p,q)$, are related to $\alpha(p,q)$ via
\begin{eqnarray}\label{eIIxxii}
b(p,q)-c(p,q) &=& {\phi(x_1-x_2)\over \phi(x_2-x_1)} = {1+i\alpha (p,q) \over 1-i\alpha(p,q)}\cr\cr
&=& {g(p)-g(q)+i \over g(p)-g(q)-i},
\end{eqnarray}
where $g(p)=(p-\ed-U/2)^2/(2U\G )$.  With this relation in hand, $b$ and $c$ can then be 
fixed by considering the scattering of two
electrons with equal spin.  As two such electrons do not interact, 
necessarily $S^{\sigma\sigma}_{\sigma\sigma} =1 $.
Hence 
\begin{equation}\label{eIIxxiii}
b(p,q)+c(p,q) = 1.
\end{equation}
So then 
\begin{eqnarray}\label{eIIxxiv}
b(p,q) &=& {g(p)-g(q) \over g(p)-g(q) -i};\cr\cr  
c(p,q) &=& {-i \over g(p)-g(q) -i}.
\end{eqnarray}
What is to be noticed is that this is exactly the same S-matrix as for the Anderson model
with $V_p=0$.  The sole change $V_p$ finite introduces to the analysis is a change in the impurity
scattering phase.

\begin{figure}
\vskip .28in
\begin{center}
\noindent
\epsfysize=0.35\textwidth
\psfrag{y}{{\hskip -1in \rm bd. state scattering phase}, $\delta_{\rm bd. state} (E)$}
\epsfbox{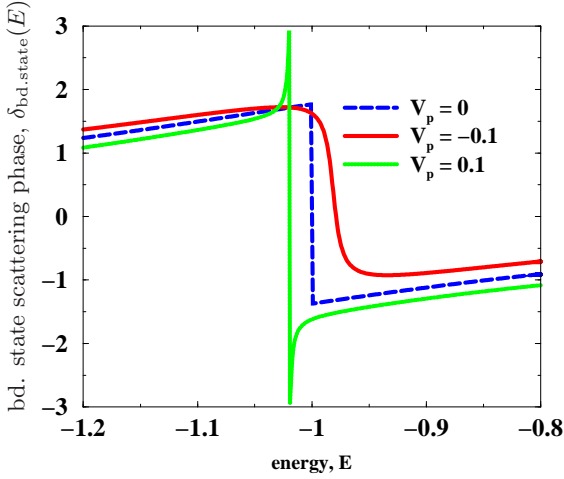}
\end{center}
\caption{Plots of the bound state impurity scattering phase as a function
of energy of the bound state.  We take
$U+2\epsilon_d = 0$, $U=1$ and $\Gamma = .1$.}
\end{figure}

\subsubsection{Two-Particle Boundstates}

In constructing the two-particle eigenstates we determined the S-matrix of scattering between opposite spins
of momentum $p$ and $q$.  In particular the state in (\ref{eIIxvi}) is a singlet with an associated
scattering phase,
\begin{equation}\label{eIIxxv}
S_{\rm singlet} = b(p,q)-c(p,q) = {g(p) - g(q) + i \over g(p) - q(q) - i}.
\end{equation}
This S-matrix has poles in it corresponding to the choice
$$
g(p) = \lambda + i/2; ~~ g(q) = \lambda - i/2,
$$
where $\lambda$ is for now an arbitrary real parameter.  At these values, the pole in $S_{\rm singlet}$ has
a positive imaginary residue indicative of a stable bound state.

\begin{figure}
\vskip .28in
\begin{center}
\noindent
\epsfysize=0.33\textwidth
\psfrag{y}{$\partial_E \delta_{\rm bd.~state} (E)$}
\epsfbox{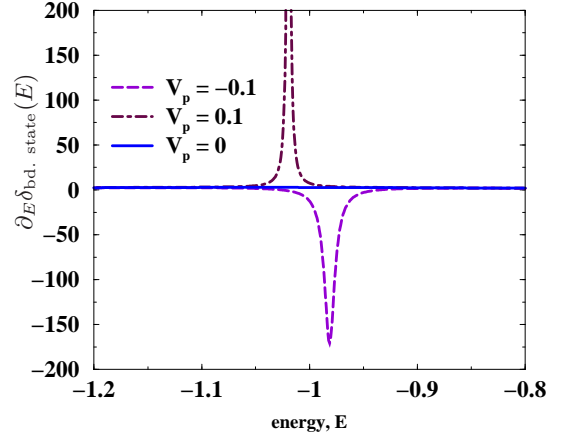}
\end{center}
\caption{Plots of the derivative with respect to energy of the
bound state impurity scattering phase.  We take
$U+2\epsilon_d = 0$, $U=1$ and $\Gamma = .1$.}
\end{figure}

For $U>0$, the ground state is constructed out of such bound states.  Much of the physics is thus contained
in the behaviour of the bound state scattering from the impurity.  As such it is worthwhile to construct
the bound state wave function explicitly.  We assume the bound state has the same form as (\ref{eIIxvi}).
The various components of the eigenstate, $g(x_1,x_2)$, $e(x)$, and $f$, still satisfy
(\ref{eIIxvii}).  However instead of (\ref{eIIxviii}), we take as an ansatz for these functions the
following
\begin{eqnarray}\label{eIIxxvi}
g(x_1,x_2) &=& \theta(x_1-x_2)h(x_1,x_2)f_+(x_1)f_-(x_1) \cr\cr 
&& \hskip .1in +~\theta(x_2-x_1)h(x_2,x_1)f_+(x_2)f_-(x_1);\cr\cr
e(x) &=& e_-\theta (x) h(x,0) f_+(x) \cr\cr
&& \hskip .5in +~e_+\theta (-x)h(0,x)f_-(x);
\end{eqnarray}
where
\begin{eqnarray}\label{eIIxxvii}
f_\pm (x) &=& \theta (x) e^{i\delta_\pm/2} + \theta(-x)e^{-i\delta_\pm/2};\cr\cr
h(x_1,x_2) &=& e^{ix\cdot(x_1+x_2)-y(x_1,x_2)};\cr\cr
\delta_\pm &=& - 2\tan^{-1}({\Gamma \over x \pm i y -\epsilon_d}+V_p);\cr\cr
e_\pm &=& {\sqrt{2\Gamma}\cos(\delta_\pm/2) \over x \pm i y -\epsilon_d};\cr\cr
E &=& 2x.
\end{eqnarray}
Using $e(x=0) = (e_-f_+(0)+e_+f_-(0))/2$, one can readily show this ansatz provides a solution to
(\ref{eIIxvii}) provided $x$ and $y$, the real and complex parts of the momentum of the
two particles forming the bound state, satisfy,  $\Gamma U + 2y(x-U/2-\ed ) = 0$.

The scattering phase of the bound state off the impurity is given by
\begin{eqnarray}\label{eIIxxviii}
\delta_{\rm bd.~state} &=& \delta_+ + \delta_-\cr\cr
&=& {1\over i}\log\bigg({(x-\epsilon_d-iy)(1-iV_p)-i\Gamma \over (x-\epsilon_d-iy)(1+iV_p)+i\Gamma}\cr\cr
&& \times {(x-\epsilon_d+iy)(1-iV_p)-i\Gamma \over (x-\epsilon_d+iy)(1+iV_p)+i\Gamma}\bigg).
\end{eqnarray}
We plot the bound state scattering phase as function of the energy $E=2x$.  At $V_p=0$ the scattering phase
undergoes a discontinuous jump at $E \approx -U$.  This jump is of magnitude $\pi$.  It however does not
in the end produce any discontinuous behaviour as the derivative of $\delta_{\rm bd. state}$ 
with respect to the energy, $E$,
is continuous.
This in turn will imply that the bare impurity density of states associated with the bound state is smooth.
However small changes in $V_p$ from 0 introduce radical changes.  If $V_p>0$, the size of the discontinuity
jumps to $2\pi$ and $\partial_E \delta_{\rm bd.~state}$ sees a large positive variation in the vicinity of
$E\approx -U$.  As $V_p\rightarrow 0^+$, this variation becomes a delta function, $\delta(E+U)$.  If
$V_p<0$, the $V_p=0$ discontinuity in $\delta_{\rm bd.~state}$ is smoothed out as evidenced in Figure 5.
However this smoothing also leads to a large but negative variation in $\delta_{\rm bd. state}$ (Figure 6).
As $V_p\rightarrow 0^-$, this variation becomes approximately $-\delta (E+U)$.

These large variations in the scattering phase of the bound states are determinative of the physics,
having profound effects on the linear response conductance.  It is worthwhile
to stress then that the origin of any novel behaviour in the linear response conductance is found at the level
of quantum mechanics.  When we construct the ground state out of bound states, this behaviour is only
modified quantitatively not qualitatively.  The dressing of the scattering phase due to interactions between
the bound states does not fundamentally modify the discontinuity structure of the bound state scattering phase.

\subsubsection{Multiparticle Eigenstates}

Multiparticle eigenstates see the same phenomena: the role $V_p$ plays is solely encoded in the single
particle scattering phase.  To see this, we write down an N-particle eigenstate with M $\downarrow$ electrons.
We will suppose $N>M>2$.  Similar arguments work for the other cases.  This wavefunction has the general
form
\begin{widetext}
\begin{eqnarray}\label{eIIxxix}
|\psi_{N,M}\rangle &=& \int dx_1 \cdots dx_N g(x_1,\ldots,x_N)\cd(x_1)\cdots\cd(x_M)\cu(x_{M+1})\cdots\cu(x_N)\cr\cr
&+& \int dx_1 \cdots dx_{M-1}dx_{M+1}\cdots dx_N 
e_\da(x_1,\ldots,x_{M-1},x_{M+1},\ldots,x_N)\cr
&&\hskip 2.7in \times\cd(x_1)\cdots\cd(x_{M-1})\cu(x_{M+1})\cdots\cu(x_N)
d^\dd_\da\cr\cr
&+& \int dx_1 \cdots dx_M dx_{M+1}\cdots dx_{N-1}\
e_\ua(x_1,\ldots,x_{M},x_{M+1},\ldots,x_{N-1})\cr
&&\hskip 2.7in \times\cd(x_1)\cdots\cd(x_M)\cu(x_{M+1})\cdots\cu(x_{N-1})
d^\dd_\ua\cr\cr
&+& \int dx_1 \cdots dx_{M-1} dx_{M+1}\cdots dx_{N-1}\
f(x_1,\ldots,x_{M-1},x_{M+1},\ldots,x_{N-1})\cr
&&\hskip 2.7in \times\cd(x_1)\cdots\cd(x_{M-1})\cu(x_{M+1})\cdots\cu(x_{N-1})
d^\dd_\ua d^\dd_\da .
\end{eqnarray}
\end{widetext}
We will look for eigenstates with total energy $E=\sum^N_{i=1}q_i$ where the $q_i$ are single particle
energies.  We will then take the coefficients $g(x_i)$,$e_\da (x_i)$, $e_\ua (x_i)$, and $f(x_i)$ to be of
the general form
\begin{widetext}
\begin{eqnarray}\label{eIIxxx}
g(x_1,\ldots,x_N) &=& \sum_{\sigma\in S_N}\prod g_{q_{\sigma (i)}}(x_i) \cdot A_{N,M}(x_i|\sigma);\cr\cr
e_\da(x_1,\ldots,x_{M-1},x_{M+1},\ldots,x_N) &=& M\sum_{\sigma\in S_N}e_{q_{\sigma(m)}}
\prod^N_{i=1\atop i\neq M}g_{q_{\sigma (i)}}(x_i) 
\cdot B_{\da N,M}(x_i,x_M=0|\sigma);\cr\cr
e_\ua(x_1,\ldots,x_{M},x_{M+1},\ldots,x_{N-1}) &=& (N-M)\sum_{\sigma\in S_N}e_{q_{\sigma(m)}}
\prod^{N-1}_{i=1}g_{q_{\sigma (i)}}(x_i) 
\cdot B_{\ua N,M}(x_i,x_N=0|\sigma);\cr\cr
f(x_1,\ldots,x_{M-1},x_{M+1},\ldots,x_{N-1}) &=& (N-M)M\sum_{\sigma\in S_N}e_{q_{\sigma(M)}}e_{q_{\sigma(N)}}
\prod^{N-1}_{i=1\atop i\neq M}g_{q_{\sigma (i)}}(x_i) \cr
&& \hskip 1.5in \times C_{N,M}(x_i,x_M=x_N=0|\sigma)
\end{eqnarray}
\end{widetext}
Here $\sum_{\sigma \in S_N}$ 
is a sum over the permutations of $N$.  $g_q(x)$ is the electron portion of the single
particle eigenstate defined in (\ref{eIIx}) with energy $q$. Similarly $e_q$ is the dot 
portion of this same wavefunction.  These are the sole ingredients of the wavefunction with an explicit
dependence upon $V_p$.  $A_{\da^M\ua^{N-M}}(x_i|\sigma)$ are coefficients which depend upon both $\sigma$
and the relative ordering of the $x_i$ (i.e. $A_{\da^M\ua^{N-M}}$ only changes value as the $x_i$ cross
one another).  Their forms are analogous to those of the Hubbard model detailed in Ref. \onlinecite{hubbard}. 
Moreover and most importantly,
the $A_{\da^M\ua^{N-M}}$ have no explicit dependence upon $V_p$ and are identical in form to
those of the $V_p=0$ Anderson model.  

To see this, we analyze the $N$-particle Schr\"odinger equation.  In doing so we find pieces of the form
\begin{widetext}
\begin{eqnarray}\label{eIIxxxi}
0 &=& \sum^N_{i=1}\bigg(-i\partial_{x_i} + V_p\delta (x_i)\bigg) g(x_1,\cdots,x_N) + \cdots ;\cr\cr
0 &=& \sum^N_{{i=1\atop i\neq M}}\bigg(-i\partial_{x_i} + V_p\delta (x_i)\bigg) 
e_\da(x_1,\cdots,x_{M-1},x_{M+1},\cdots,x_N) + \cdots ;\cr\cr
0 &=& \sum^N_{{i=1\atop i\neq N}}\bigg(-i\partial_{x_i} + V_p\delta (x_i)\bigg) 
e_\ua(x_1,\cdots,x_{M},x_{M+1},\cdots,x_{N-1}) + \cdots ;\cr\cr
0 &=& \sum^N_{{i=1\atop i\neq M,N}}\bigg(-i\partial_{x_i} + V_p\delta (x_i)\bigg) 
f(x_1,\cdots,x_{M-1},x_{M+1},\cdots,x_{N-1}) + \cdots .
\end{eqnarray}
\end{widetext}
But because we have constructed $g$, $e_\da$, $e_\ua$, and $f$ out of products of
single particle eigenfunctions and $g_q$(x) has the property
\begin{equation}\label{eIIxxxii}
\partial_x g_q(x) = V_p g_q(0) \delta (x) + \cdots ,
\end{equation}
all terms proportional to $V_p$ cancel straightforwardly.  The equations (or pieces thereof) that remain to be
solved are then equivalent to the ordinary Anderson model.\cite{kao,wie}  Although a $V_p$ dependence in
these equations lurks in $g_q(x)$ and $e_q$, it never has to be explicitly exhibited in their solution.

\subsection{Bethe Ansatz Equations: Analysis for $\ed \geq -U/2$}

In the previous section, we argued $N-$particle wavefunctions of energy $E=\sum_i q_i$
could be explicitly constructed.
Moreover the S-matrix describing the scattering of two electrons was shown to be identical to that
of the $V_p=0$ Anderson model.  $V_p$ appears only in the impurity scattering phase.  This implies that
the quantization conditions (the Bethe ansatz) for the $q_i$ when the system is placed in a system of finite
length, $L$, are near identical to those of the usual Anderson model.

On the basis of the above discussion, the Bethe ansatz equations for this system, mimicking 
Ref. \onlinecite{wie}, take the form
\begin{eqnarray}\label{eIIxxxiii}
e^{iq_jL+i\delta(q_j,V_p)} &=& \prod^M_{\alpha = 1}{g(q_j)-\lambda_\alpha+i/2 \over g(q_j)-\lambda_\alpha-i/2};\cr\cr
\prod^N_{j=1}{\lambda_\alpha-g(q_j)+i/2 \over \lambda_\alpha-g(q_j)-i/2} 
&=& -\prod^M_{\beta=1}{\lambda_\alpha-\lambda_\beta+i \over \lambda_\alpha-\lambda_\beta-i};\cr\cr
\delta (q,V_p) &=& -2\tan^{-1}({\Gamma \over q-\ed}+V_p);\cr\cr
g(q) &=& {(q-\ed-U/2)^2 \over 2U\G }.
\end{eqnarray}
These equations depend upon $V_p$ only through $\delta (q,V_p)$.   In fixing the allowed values of $q_j$, they
depend upon a set of $M$ rapidities, $\{\lambda_\alpha\}$ equaling the number of spin $\da$ electrons in the
system.  The total spin, $S_z$, of the wavefunction is $2S_z = N-2M$.

These equations have a rich structure of solutions for $\{q_j\}$ and $\{\lambda_\alpha\}$.  But at $T=0$ only
two types of solutions are important:
\vskip .2in

i) $q_j$ a real number;

\vskip .1in 

and

\vskip .1in 

ii) $\lambda_\alpha$ a real number to which are associated two 

complex q's, $q^\alpha_\pm$, given by
\begin{eqnarray}\nonumber
g(q^\alpha_\pm) &=& g(x(\lambda_\alpha)\mp iy(\lambda_\alpha )) = \lambda_\alpha \pm i/2;\cr
x(\lambda ) &=& U/2 + \ed - \sqrt{U\G }(\lambda+\sqrt{\lambda^2+1/4})^{1/2};\cr
y(\lambda ) &=& \sqrt{U\G }(-\lambda+\sqrt{\lambda^2+1/4})^{1/2}.
\end{eqnarray}
These two solutions are the only ones appearing in the ground state.  $x(\la )$ and $y(\la )$ defined above
satisfy $\Gamma U + 2(x-U/2-\ed )y$, the condition identified in the construction of the bound state
wave function in Section II.B.3.

\subsection{The Ground State}

The ground state of the finite $V_p$ Anderson model consists of the same set of solutions as at $V_p=0$.  
As such a ground state with spin $2S_z = N-2M$ consists of 
\vskip .1in
i) N-2M real $q_j$'s;
\vskip .05in
ii) M real $\lambda_\alpha$'s and the 2M associated $q_{\alpha\pm}$.
\vskip .1in
\noindent The construction of this ground state holds for $\ed \geq -U/2$.  We can understand
the case $\ed < -U/2$ through an explicit construction focusing upon holes rather than
particles.  There will be call to do so.  We can alternatively perform a particle-hole transformation.
This transformation however changes the sign of $V_p$.  We will find reason to examine both approaches.

\newcommand{\il}{\int^{\tilde{Q}}_Q d\la~}
\newcommand{\ilp}{\int^{\tilde{Q}}_Q d\la '}
\newcommand{\ik}{\int^{B}_{-D} dq~}

In analyzing the ground state we want to consider the thermodynamic limit.  Thus we no longer treat 
$q_j$ and $\lambda_\alpha$ as discrete but rather derive smooth distributions (per unit length), $\rho(q)$
for the $q_j$'s and $\sigma (\lambda )$ for the $\lambda_\alpha$'s and their associated complex $q$'s.  These
distributions, $\rho (q)$ and $\sigma (\lambda)$, mark out how the solutions of the Bethe ansatz equations are distributed
when the system is in its ground state.  As each $\lambda$ in the ground state reduces the ground state spin
by 1 (i.e. $2S_z = N-2M$), a $\lambda$ is known as a spin excitation and $\sigma (\lambda )$ the spin
distribution.  As each $q$ in the ground state at fixed particle number does not alter the spin,
$q$ is known as a charge excitation and $\rho (q)$ the charge distribution.
They are derived by taking $\log$'s of the Bethe ansatz equations,
\begin{eqnarray}\label{eIIxxxiv}
q_j L + \de (q_j,V_p) &=& 2\pi N_j - \sum^M_{\beta = 1} 
\theta_1 (g(q_j)-\lambda_\beta);\cr\cr
2\pi J_\alpha + \sum^M_{\beta=1} \theta_2 (\la_\al-\la_\be) &+&  
\sum^{N-2M}_{j=1} \theta_1 (\la_\al-g(q_j)) \cr\cr
&&\hskip -1in = -2Lx(\la_\al) -2{\rm Re}\de(x(\la_\al ) + iy(\la_\al),V_p);\cr\cr
\theta_n (x ) &=& 2\tan^{-1} ({2\over n}x)  -\pi .
\end{eqnarray}
where $N_j$ and $J_\alpha$ are quantum numbers with the $J_\alpha$ satisfying
\begin{eqnarray*}
-{N-2M \over 2} < J_\alpha \leq {N-2M \over 2}.
\end{eqnarray*}
By taking appropriate derivatives in (\ref{eIIxxxiv}) 
\begin{eqnarray}\label{eIIxxxv}
\rho (q) &=& {1\over 2\pi} + {\Delta (q,V_p) \over L} + 
g'(q) \il a_1(g(q)-\la) \sig (\la); \cr\cr
\sig (\la ) &=& - {x'(\la)\over\pi} + {\tilde{\Delta}(\la)\over L}
- \ilp a_2(\la '-\la)\sig (\la ')\cr\cr
&& - \int^B_{-D}dq a_1(\la - g(q))\rho (q),
\end{eqnarray}
where
\begin{eqnarray}\label{eIIxxxvi}
\Delta (q) &=& {1\over 2\pi} \partial_q \delta (q);\cr\cr
\tilde{\Delta} (\la ) &=& -{1\over \pi} \del_\la 
{\rm Re}\delta (x(\la)+iy(\la),V_p);\cr\cr
a_n(x) &=& {1\over 2\pi}
\partial_x \theta_n (x) =  {2n\over\pi} {1\over (n^2 + 4x^2)}.
\end{eqnarray}
In the equations for the charge and spin distributions appear the Fermi surfaces, $Q$ and $B$, and ``band'' bottoms
$\tilde Q$ and $-D$.  These mark out the range of $q$ and $\lambda$ over which excitations appear in the ground
state. $-D$ is the lower band edge of the charge excitations.  As each $\lambda$ has two associated complex $q$'s,
we expect $q_{q_+}(\tilde Q) + q_{q_-}(\tilde Q) = 2x(\tilde Q) = -2D$, thus determining $\tilde{Q}$.\cite{wie}  The Fermi
surfaces $Q$ and $B$ are fixed by insisting that the overall spin and particle number,
\begin{eqnarray}\label{eIIxxxvii}
N-2M &=& L\int^B_{-D} dq \rho (q);\cr\cr
M &=& L\int^{\tilde Q}_{Q} d\lambda \sigma (\lambda ) ,
\end{eqnarray}
are reproduced.

\subsection{Scattering Phase at the Fermi Surface: $T=0$ Transport}

We are now in a position to access the quantity we are ultimately interested in, the zero temperature scattering
phase, $\delta_e$, of electrons at the Fermi surface.  To compute $\delta_e$ we employ an argument used 
by Andrei (Ref. \onlinecite{andrei}) 
in the computation of the magnetoresistance in the Kondo model.  We imagine adding
an electron to the system.  Its momentum will be quantized according to $p = 2\pi n/L$.  If the impurity 
was absent, the momentum would be determined by the bulk properties of the system and we would write
$p = p_{\rm bulk}$.  In the presence of the impurity, the total momentum of the electron, $p$, experiences a shift away
from $p_{\rm bulk}$, necessarily
scaling as $1/L$,
\nonumber
\begin{eqnarray*}
p = p_{\rm bulk} + p_{\rm imp}/L .
\end{eqnarray*}
$\delta_e$ is then identified with $p_{\rm imp}$.  In the context of the Bethe ansatz, $p_{\rm imp}$ is
readily computed.

As we are interested in quantities scaling as $1/L$, it is useful to isolate the portions of the densities,
$\rho (q)$ and $\sigma (\lambda )$, scaling as such.
With this in mind we write
\begin{eqnarray}\label{eIIxxxviii}
\rho (q) &=& \rho_{\rm bulk} (q) + {1\over L}\rho_{\rm imp}(q);\cr\cr
\sigma (\lambda ) &=& \sigma_{\rm bulk} (\lambda ) + {1\over L}\sigma_{\rm imp}(\lambda ).
\end{eqnarray}
Here $\rho_{\rm imp} (q)$ and $\sigma_{\rm imp} (\lambda )$ determine the total number of electrons displaced by
the impurity, $n_{\rm imp}$, that is the electrons sitting on the dot, $n_d$, plus
the deviation of electron density in the leads away from their equilibrium value of $\rho_{\rm bulk}$:
\begin{eqnarray}\label{eIIxxxix}
n_{\rm imp \ua} &=& n_{d\ua} + \int dx \bigg[ \langle c^\dd_{e\ua}(x) c_{e\ua}(x)\rangle - \rho_{\rm bulk}\bigg]\cr\cr
&=& \int^{\tilde Q}_Qd\lambda \sigma_{\rm imp} (\lambda ) + \int^B_{-D}dq \rho_{\rm imp}(q);\cr\cr
n_{\rm imp \da} &=& n_{d\da} + \int dx \bigg[ \langle c^\dd_{e\da}(x) c_{e\da}(x)\rangle - \rho_{\rm bulk}\bigg]\cr\cr
&=& \int^{\tilde Q}_Qd\lambda \sigma_{\rm imp} (\lambda ).
\end{eqnarray}
Typically the deviation of the electron density in the leads is taken to be zero.  However this is not
appropriate in a magnetic field nor will it be appropriate here.  Indeed this deviation will be of the same
magnitude as the dot electron number $n_{d\ua}/n_{d\da}$.  

\newcommand{\rh}{\rho_{\rm bulk}}
\newcommand{\ri}{\rho_{\rm imp}}
\newcommand{\sh}{\sig_{\rm bulk}}
\newcommand{\si}{\sig_{\rm imp}}
\newcommand{\rph}{\rho_{p/h}}
\newcommand{\sph}{\sig_{p/h}}
\newcommand{\rp}{\rho_{p}}

To derive equations governing the densities, $\rho_{\rm bulk}(q)/\rho_{\rm imp} (q)$ and 
$\sigma_{\rm bulk}(\lambda )/\sigma_{\rm imp}(\lambda)$, we substitute (\ref{eIIxxxviii}) into (\ref{eIIxxxv}) and
segregate the bulk terms from those scaling as $1/L$
\begin{eqnarray}\label{eIIxl}
\rh (q) &=& {1\over 2\pi} + 
g'(q) \il a_1(g(q)-\la) \sh (\la); \cr\cr
\sh (\la ) &=& - {x'(\la)\over\pi} 
- \ilp a_2(\la '-\la)\sh (\la ') \cr\cr
&& \hskip 0in - \int^B_{-D}dq a_1(\la-g(q))\rh (q),
\end{eqnarray}
and
\begin{eqnarray}\label{eIIxli}
\ri (q) &=& \Delta (q) + 
g'(q) \il a_1(g(q)-\la) \si (\la); \cr\cr
\si (\la ) &=& \tilde{\Delta}(\la)
- \ilp a_2(\la '-\la)\si (\la ') \cr\cr
&& \hskip -.1in - \int^B_{-D}dq a_1(\la-g(q))\ri (q).
\end{eqnarray}
We now relate the impurity density to the impurity momenta and so the impurity scattering phases.

The total momenta of the $q$ and $\lambda$ excitations are derivable from the continuum limit of logarithms
of the Bethe ansatz equations (up to possible terms independent of $\lambda$ and $q$; see Ref. \onlinecite{long}):
\begin{eqnarray}\label{eIIxlii}
p (q) &=& q + \delta (q) + \il \sigma (\la ) 
\theta_1(g(q) - \la) ;\cr\cr
p (\la ) &=&  2 x(\la) + 2 Re \delta ( x(\la) +iy(\la))\cr\cr
&& \hskip -.5in + \ilp \sigma (\la ') \theta_2 (\la - \la ') 
 + \int^B_{-D}dq \rho (q) \theta_1 (\la - g(q)) .\cr &&
\end{eqnarray}
The impurity momenta correspond to the portion of the total momenta
scaling as $1/L$:
\begin{eqnarray}\label{eIIxliii}
p_{\rm imp}(q) &=& \delta (q) + \il \si (\la ) 
\theta_1(g(q) - \la) ;\cr\cr
p_{\rm imp}(\la ) &=&  2 Re \delta ( x(\la) +iy(\la))\cr
&& \hskip -.8in + \ilp \si (\la ') \theta_2 (\la - \la ') 
+ \int^B_{-D}dq \ri (q) \theta_1 (\la - g(q)) .\cr &&
\end{eqnarray}
We note that
\begin{eqnarray}\label{eIIxliv}
p_{\rm imp}(q\rightarrow -\infty) &=& -2\tan^{-1}(V_p);\cr\cr
p_{\rm imp}(\lambda \rightarrow \infty) &=& -4\tan^{-1}(V_p).
\end{eqnarray}
and so at the band edges, the impurity momenta reduce to their bare values.

The impurity density and momentum are intimately related.  It is straightforwardly
checked that
\begin{eqnarray}\label{eIIxlv}
\partial_q p_{\rm imp} (q) = 2\pi \rho (q);~~
\partial_\lambda p_{\rm imp} (\lambda) = -2\pi \sigma (\lambda ).
\end{eqnarray}
These relations will ultimately allow us to express $\delta_{e\ua/\da}$ in terms of
$n_{\rm imp\ua}/n_{\rm imp\da}$; that is, verify a version of the Friedel sum holds
in this case.  We again stress that in this proof of the Friedel sum rule, $\delta_{e\ua/\da}$
is not related to merely the electron occupancy of the dot, but the total electron number
displaced by the impurity.

In this description of the system an electron is a composite object.  It arises as a gluing
together of a charge excitation, $q$, as well as a spin excitation, $\lambda$.  To see this,
imagine a spin $\ua$ electron being added to the system.  The number of electrons in the
system as a whole goes from $N$ to $N+1$ with the system spin increasing
by $1/2$.  This means we must add a single $q$ excitation to the system.  But in doing
so, the range of quantum numbers, $-{N-2M \over 2} < J_\alpha < {N-2M \over 2}$ available
to the $\lambda_\alpha$ increases by $1$.  This additional slot is in effect a hole in the
spin distribution.  Thus a spin $\ua$ electron is composed of a charge $k$-excitation
and a spin $\lambda$-hole.

The electron scattering phase is the sum of the scattering phase of these two 
elementary objects
\nonumber
\begin{eqnarray}\label{eIIxlvi}
\delta^\ua_{e} = p_{\rm imp}(q) + (p_{\rm imp}(\lambda )-p_{\rm imp}(\lambda = {\tilde Q})).
\end{eqnarray}
The additional constant, $p(\lambda = {\tilde Q})$, ensures that at the band edge ($q=-D$ and $\lambda = \tilde Q$)
the electron scattering phase equals its bare value of $-2\tan^{-1}(V_p)$.
We obviously have some degree of freedom in how we choose $q$ and $\lambda$.  But
we want to construct an excitation at the Fermi surface.  To do so, we must
choose $q$ and $\lambda$ to be at their respect Fermi surfaces, $q=B$ and $\lambda=Q$.
Using relations (\ref{eIIxlvi}), we can then write $\delta^\ua_e$ in
terms of the impurity densities:
\begin{eqnarray}\label{eIIxlvii}
\delta^\ua_e &=& 2\pi\int^B_{-D}dq \rho_{\rm imp}(q) + 2\pi\int^{\tilde Q}_Q d\lambda \sigma_{\rm imp}(\lambda )\cr\cr
&& + p_{\rm imp}(q=-D)\cr\cr
&=& 2\pi n_{\rm imp\ua} - 2\tan^{-1}(V_p).
\end{eqnarray}
We have thus demonstrated $\delta^\ua_e$ is (up to an additive constant)
proportional to the number of electrons displaced by the impurity.

To determine the scattering of a spin $\da$ electron we employ particle-hole symmetry.  A particle-hole
transformation is implemented via
\begin{eqnarray}\label{eIIxlviii}
c^\dagger_\ua (q) &\rightarrow& c_\da (-q) ;\cr\cr
c^\dagger_\da (q) &\rightarrow& c_\ua (-q) ;\cr\cr
d^\dagger_\ua &\rightarrow& d_\da ;\cr\cr
d^\dagger_\da &\rightarrow& d_\ua ;\cr\cr
\ep_d &\rightarrow& -U - \ep_d ;\cr\cr
V_p &\rightarrow& -V_p.
\end{eqnarray}
Thus the scattering phase of a spin $\da$ hole is related to that of a spin $\ua$ electron via
\begin{equation}\label{eIIxlix}
\delta^\da_{ho}(-U-\ed,-V_p) = \delta^\ua_{e}(\ed,V_p).
\end{equation}
Thus at the Fermi surface we have
\begin{eqnarray}\label{eIIl}
\delta^\da_{ho} (-U-\ed,-V_p) &=& 
2\pi\int^{\tilde{Q}}_Q d\la  \si (\la  ) \cr\cr
&& \hskip -.75in + 2\pi\int^B_{-D} dq \rho_{\rm imp} (q) - 2\tan^{-1}(V_p)\cr\cr
&& \hskip -.75in = 2\pi n_{\rm imp\ua}(\ed) - 2\tan^{-1}(V_p).
\end{eqnarray}
As $n_{\rm imp\ua} (\ed,V_p) =  - n_{\rm imp\da} (-U-\ed,-V_p )~{\rm mod~1}$, we have (up to a factor of $2\pi$)
\begin{eqnarray}\label{eIIli}
\delta^\da_{ho}(-U-\ed ,-V_p) &=& -2\pi n_{\rm imp\da}(-U-\ed,-\vp ) \cr\cr
&& \hskip .25in - 2\tan^{-1}(V_p).
\end{eqnarray}
At the Fermi surface $\delta^\da_{ho}(-U-\ed,-V_p) = -\delta^\da_{e}(-U-\ed,-V_p)$.
Hence we have shown the Friedel sum rule holds for spin $\da$ electrons.

Imagine now that we are at the point where $U+2\ed = 0$ and are working at zero field, $H=0$ (thus $N=2M$ and $B=0$).  Hence
spin $\ua$ and $\da$ electrons scattering identically.  Moreover we expect the phase to be
insensitive to the sign of $V_p$ due to particle-hole symmetry.
\begin{equation}\label{eIIlii}
\delta^\ua_{e}(U=-2\ed,V_p) ~{\rm mod} 2\pi  = \delta^\ua_e (U=-2\ed,-V_p).
\end{equation}
We will however find this to be not na\"\i vely the case.  This seeming violation of particle-hole
symmetry at the point $U+2\ed = 0$ is a result of the Bethe ansatz yielding a state other
than the ground state over a range of $V_p$ .  In fact we are able to construct two candidate ground states
at the point $U+2\ed$.  As we vary $V_p$, the true ground state varies between these two 
possibilities.  Once this is factored into the calculation, particle-hole symmetry is restored.
Which ground state to use will be discussed in some detail when we discuss the computation of the
linear response conductance.

To construct the second candidate ground state, we must begin with the Bethe ansatz solution from a different
starting point, one in which we focus on hole not particle excitations.  We do so now.

\subsection{Bethe Ansatz Equations: Analysis for $\ed \leq -U/2$}

In the previous sections we analyzed the Bethe ansatz equations and ultimately determined the scattering
of electrons at $T=0$ at the Fermi surface.  But this analysis was limited to the case of 
$U+2\ed \geq 0$.  Here we consider the case $U \leq -2\ed$.  Our primary interest in doing so is to
understand how particle-hole symmetry is preserved at the point $U-2\ed = 0$.  To consider $U$ strictly
smaller than $-2\ed$, we
could simply exploit a particle-hole transformation.

To analyze the case $U \leq -2\ed$, we recast the even Hamiltonian in (\ref{eIIv})
as
\begin{eqnarray}\label{eIIliii}
H_e &=& \sum_\sigma \bigg\{\int dx -i c_{e\sigma}(x)\partial_xc^\dd_{e\sigma}(x) \cr\cr
&& - \sqrt{2\Gamma}(d_\sigma c^\dagger_{e\sigma} + c_{e\sigma} d^\dagger_\sigma )|_{x=0}
+{\tilde V}_p c_{e\sigma} c^\dagger_{e\sigma}|_{x=0} \bigg\}\cr\cr
&& + {\tilde \epsilon}_d \sum_{\sigma = \uparrow,\downarrow}{\tilde n}_\sigma + 
U {\tilde n}_\uparrow {\tilde n}_\downarrow + V_p + 2\ed + U,
\end{eqnarray}
where
\begin{eqnarray}\label{eIIliv}
{\tilde n}_{d\sigma} &=& d_\sigma d^\dd_{\sigma};\cr\cr
{\tilde \ed} &=& -\ed - U ;\cr\cr
{\tilde V_p} &=& -V_p.
\end{eqnarray}
In this form we can construct hole states in exactly the same fashion as particle
states governed by the parameters $({\tilde V}_p,{\tilde \ed})$.  For example
the single hole states take the form
\begin{eqnarray}\label{eIIlv}
|\psi_\sigma\rangle^{\rm hole} = \bigg[ \int^\infty_{-\infty}dx \{ g_\sigma (x) c_\sigma(x) \} 
+ e_\sigma d_\sigma\bigg]|0\rangle ,
\end{eqnarray}
where here the vacuum state $|0\rangle$ is completely full, i.e. $c^\dagger_\sigma |0\rangle = 0$.
For a single hole of energy $q$ we find
\begin{eqnarray}\label{eIIlvi}
g_\sigma (x) &=& \theta (x) e^{iqx + i\delta/2} + \theta (-x) e^{iqx - i\delta/2};\cr\cr
\delta (q) &=& -2\tan^{-1}({{\Gamma} \over q - {\tilde\ed}} + {\tilde V}_p)\cr\cr
&=& -2\tan^{-1}({\Gamma \over q - {\tilde\ed}} - V_p).
\end{eqnarray}
The bulk S-matrix describing two-hole scattering is derivable in a similar fashion to as before and is given
by
\begin{eqnarray}\label{eIIlvii}
S^{\sigma_1'\sigma_2'}_{\sigma_1\sigma_2} &=& b(p,q) I^{\sigma_1'\sigma_2'}_{\sigma_1\sigma_2}
+ c(p,q) P^{\sigma_1'\sigma_2'}_{\sigma_1\sigma_2};\cr\cr
b(p,q) &=& {g(p)-g(q) \over g(p)-g(q) -i};\cr\cr  
c(p,q) &=& {-i \over g(p)-g(q) -i},
\end{eqnarray}
where $g(p)=(p-{\tilde\ed}-U/2)/(2U\G )$.

With such an S-matrix we can readily determine the Bethe ansatz equations together with the solutions of these
equations which populate the ground state.  In the absence of a magnetic field, these are simply sets of real
$\lambda$ and associated complex momenta, $q_\pm$, given by
\begin{eqnarray}\label{eIIlviii}
g(q_{\alpha \pm}) &=& g(x(\lambda_\alpha)\mp iy(\lambda_\alpha )) = \lambda_\alpha \pm i/2;\cr\cr
x(\lambda ) &=& U/2 + {\tilde\ed} - \sqrt{U\G }(\lambda+\sqrt{\lambda^2+1/4})^{1/2};\cr\cr
y(\lambda ) &=& \sqrt{U\G }(-\lambda+\sqrt{\lambda^2+1/4})^{1/2}.
\end{eqnarray}
This set of solutions is valid for $\tilde\ed + U/2 \geq 0$.  (But of course as $\tilde\ed = -\ed -U$,
this corresponds to the condition, $\epsilon_d < -U/2$, in which we are interested.)

The above construction indicates that a hole scattering at the Fermi surface in a system characterized
by $V,\ed,V_p,$ and $U$ is equivalent to a particle scattering at 
$\tilde V = -V,\tilde\ed = -\ed-U,\tilde V_p = -V_p,$ and $U$.
This is exactly what is to be expected from a particle-hole transformation.
And thus at the point $\ed = -U/2 = -\tilde\ed$, we obtain a second inequivalent description of scattering.
Which description to use and when will be discussed in Section III.

\section{Linear Response Conductance at $T=0$}

Here in this section we will compute the linear response conductance, $G$, at $T=0$.  To compute $G$ we employ the
Landauer-B\"uttiker formula,
\begin{equation}\label{eIIIi}
G = {e^2\over h}(T^2_\uparrow (\epsilon = \epsilon_F) + T^2_\downarrow(\epsilon=\epsilon_F)).
\end{equation}
Here $T$ is the transmission amplitude for electron scattering at the Fermi surface while $T$ is related
to $n_{\rm imp}$, the number of electrons displaced by the impurity, through the relationship
\begin{equation}\label{eIIIii}
T^2_{\uparrow/\downarrow} = \sin^2({\delta_{e\uparrow/\downarrow} - \delta_{o\uparrow/\downarrow} \over 2}).
\end{equation}
(if $V_L=V_R$).
Here $\delta_{o\uparrow/\downarrow}$, the scattering of electrons off the impurity in the 
odd sector is given trivially
by $\delta_{o\ua/\da} = 2\tan^{-1}(V_p)$.  In the even sector on the other hand, $\delta_{e\uparrow/\downarrow}$ is given 
by $\delta_{e\ua/\da} = 2\pi n_{\rm imp\uparrow/\downarrow} -2\tan^{-1}(V_p)$ as developed in the previous section.
We once again emphasize that $n_{\rm imp}$ is not necessarily equal to the number, $n_d$, of electrons sitting on
the dot but may include contributions from displaced electrons in the leads.
Langreth makes this precise point in his proof of the Friedel sum rule.\cite{langreth}  As we will see this point will
be crucial in interpreting our results.

\begin{figure}
\vskip .58in
\begin{center}
\noindent
\epsfysize=0.45\textwidth
\epsfbox{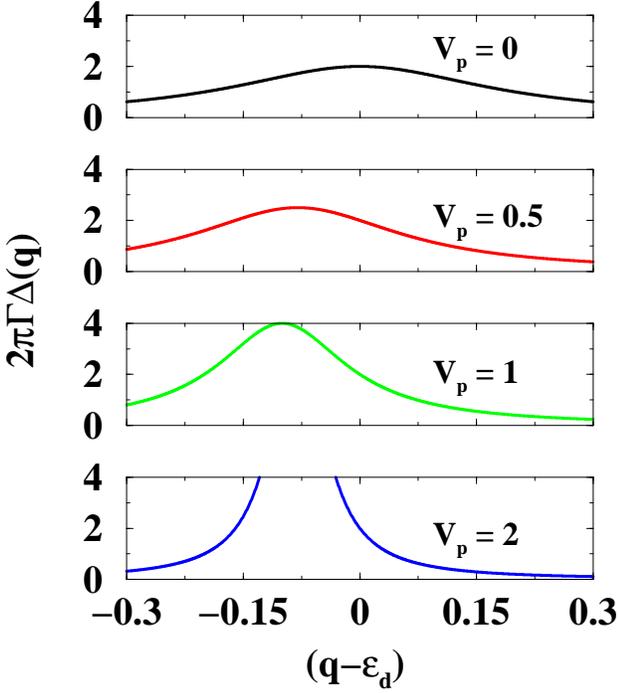}
\end{center}
\caption{\label{dqreal}Plots of $\Delta(q)$ for a number of different values of $V_p$.  Here $\Gamma =.1$}
\end{figure}

In the presence of finite $V_p$, the structure of $G$ is highly non-trivial.  This is readily seen.  Let us suppose
that $U/2+\epsilon_d > 0$.  In this case, 
the number of electrons displaced by the impurity is given by
\begin{eqnarray}\label{eIIIiii}
n_{\rm imp \uparrow} &=& \int^{\tilde Q}_Q d\lambda \sigma_{\rm imp} (\lambda ) 
+ \int^B_{-D}dq \rho_{\rm imp}(q) ;\cr\cr
n_{\rm imp \downarrow} &=& \int^{\tilde Q}_Q d\lambda \sigma_{\rm imp} (\lambda ).
\end{eqnarray}
The integral equations controlling $\sigma_{\rm imp}/\rho_{\rm imp}$ are
\begin{eqnarray}\label{eIIIiv}
\rho_{\rm imp}(q) &=& \Delta(q) + g'(q)\int^{\tilde Q}_Q d\lambda 
a_1(\lambda - g(q)) \sigma_{\rm imp}(\lambda);\cr\cr
\sigma_{\rm imp}(\lambda) &=& \tilde{\Delta} (\lambda) - \int^{\tilde Q}_Q d\lambda' a_2(\lambda - \lambda') 
\sigma_{\rm imp}(\lambda)\cr\cr
&& \hskip .2in  + \int^B_{-D}dq a_1(\lambda-g(q))\rho_{\rm imp} (q).
\end{eqnarray}
These equations cannot in general be solved in closed form.  However examining the source terms,
$\Delta(q)$ and $\tilde\Delta (\lambda)$, leads to significant insight.  In 
Figure \ref{dqreal} are found plots of $\Delta(q)$ for several values of $V_p$.  Recall $\Delta(q)$ is
related to the bare scattering phase of single particle eigenstates with energy $q$ via
$\Delta(q) = \del_q \Delta(q)/2\pi$. $\Delta (q)$ is continuous in $V_p$.  As $V_p$ is made finite, the
particle-hole breaking due to $V_p$ both shifts the peak away from $q=\epsilon_d$ (the direction of the
shift is dependent upon the sign of $V_p$) as well as making the peak narrower.  This is reflected in
the general analytic form for $\Delta (q)$,
\nonumber
\begin{eqnarray*}
\Delta (q) = {1\over \pi} {\bar\Gamma \over (q-\epsilon_d+V_p\bar\Gamma) + {\bar\Gamma}^2},
\end{eqnarray*}
where $\bar\Gamma = \Gamma/(1+V_p^2)$.

\begin{figure*}
\vskip .58in
\begin{center}
\noindent
\epsfysize=0.55\textwidth
\psfrag{y}{$\Gamma\tilde\Delta (\lambda)$}
\psfrag{x}{$\lambda$}
\epsfbox{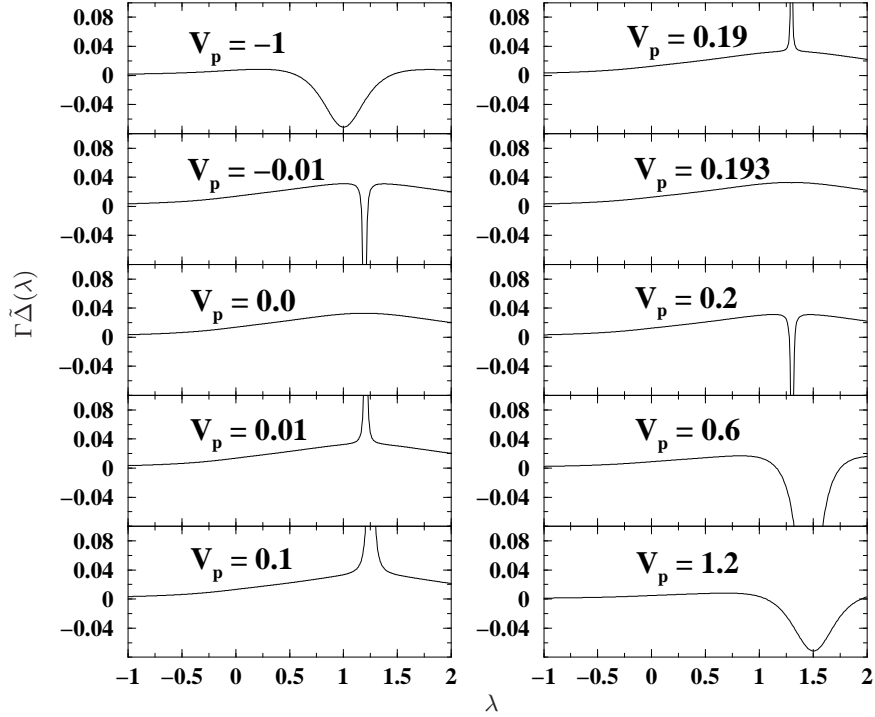}
\end{center}
\caption{\label{dqcomplex}Plots of $\Gamma\tilde\Delta(q)$ for a number of different values of $V_p$.  Here we
have chosen $U=1$ and $\Gamma = .1$.}
\end{figure*}

While $\Delta(q)$ is continuous with variations in $V_p$, $\tilde\Delta (\lambda)$ is not.  $\tilde\Delta (\lambda)$
is given in terms of the bare scattering phase, not of single particle eigenstates, but of two-particle
bound states.  We plot $\tilde\Delta (\lambda )$ in Figure \ref{dqcomplex} for a range of values of $V_p$.
Two marked transitions are apparent in this plot.

As $V_p$ is reduced from $-\infty$ a peak in $\tilde\Delta(\lambda)$ develops.  With $V_p$
approaching $0^-$, this peak develops into a $\delta$-function of weight $-1$.  At $V_p=0$, the
$\delta$-function disappears and $\tilde\Delta(\lambda)$ is smooth.  As $V_p$ enters the region of
positive values the delta function reappears but with weight 1.  As $V_p$ is increased from $0^+$
this $\delta$-function at first broadens out (see $V_p = .1$ in Figure \ref{dqcomplex}) but then
narrows again.  At some finite value of $V_p$, designated $V_p^{crit}$, 
the $\delta-$function disappears (in Figure \ref{dqcomplex} this occurs at $V_p \sim 0.2$ 
for the particular values of parameters chosen).  At $\vp = \vc + 0^+$,
the $\delta$-function reappears with weight -1.  Additional increases in $V_p$ serve to
broaden and reduce the peak.

To study further the linear response conductance (i.e. the behaviour of the equations (\ref{eIIIiv}))
we consider the cases of $H=0$ and $H\neq 0$ separately.  We first take up the zero field case.

\subsection{Zero Field Linear Response Conductance}

We will divide our discussion into three cases: 1) $U+2\epsilon_d>0$, 2) $U+2\epsilon_d<0$, and
3) $U+2\epsilon_d=0$.  The first two regions are studied through the two different developments of the
system's ground state found in Section II (i.e. in the first case, a `particle' basis was used
to develop the Bethe ansatz equations while in the second case we used a `hole' basis).  
In the third case, $U+2\epsilon_d =0$, the 
two developments are both ostensibly valid but give different results.  To distinguish which 
solution should be employed, we examine the energy of the two corresponding ground states, choosing the state with the
lower energy.

\subsubsection{$\bf U+2\epsilon_d >0$:}

\begin{figure}
\vskip .58in
\begin{center}
\noindent
\epsfysize=0.35\textwidth
\epsfbox{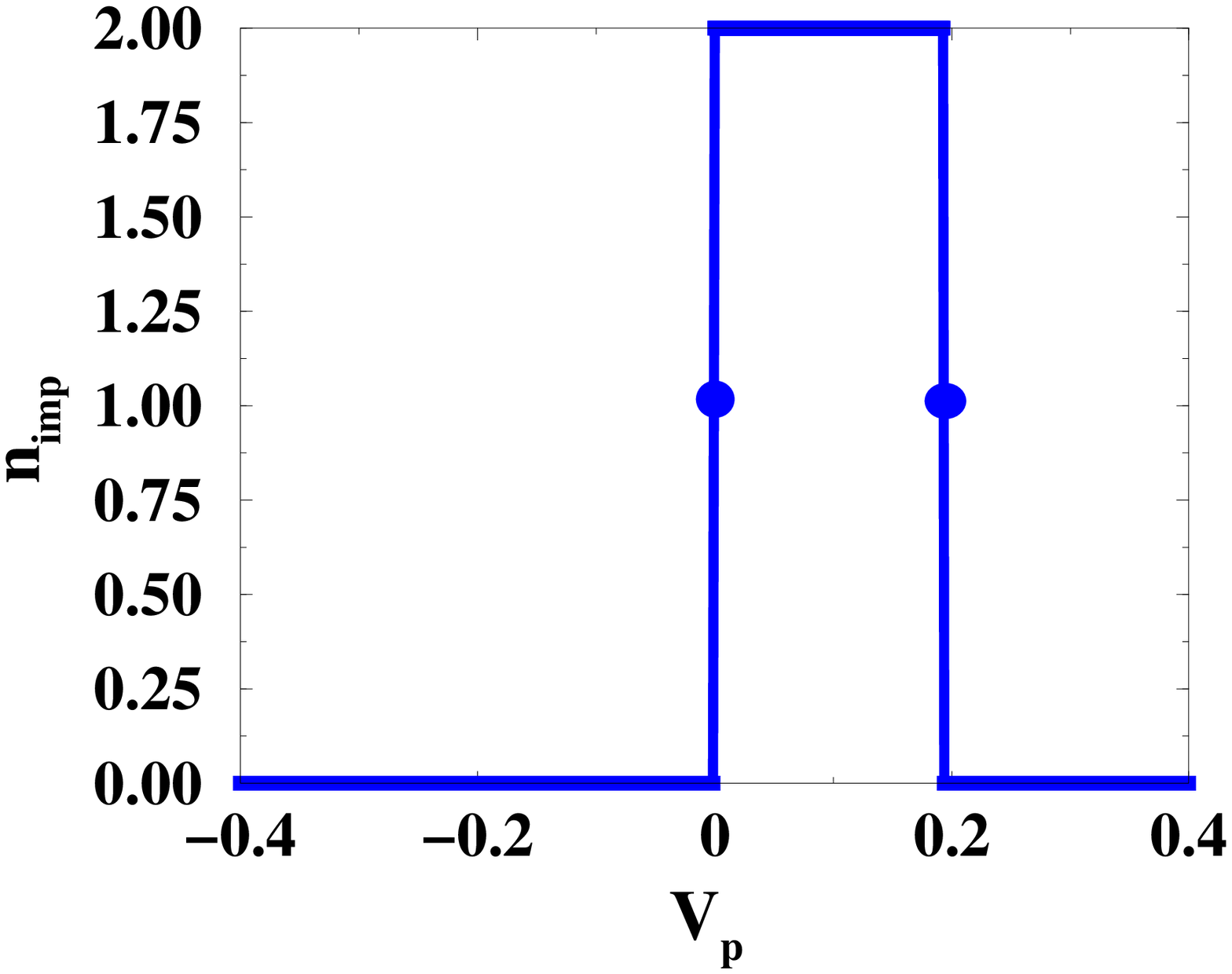}
\end{center}
\caption{\label{nimp}A plot of the total number of electrons (both spin species)
displaced by the presence of the impurity as
a function of $V_p$.  We have assumed $U+2\epsilon_d = 0^+$ and have chosen $U=1$ and $\Gamma = .1$.
The plotted filled circles indicate that at $V_p=0$ and $V_p = V_p^{\rm crit} \simeq 0.19$, $n_{\rm imp}$ is precisely 1.}
\end{figure}

\vskip .2in

\begin{center}
a) $U+2\epsilon_d = 0^+$
\end{center}

\vskip .2in

\begin{figure}
\vskip .58in
\begin{center}
\noindent
\epsfysize=0.35\textwidth
\epsfbox{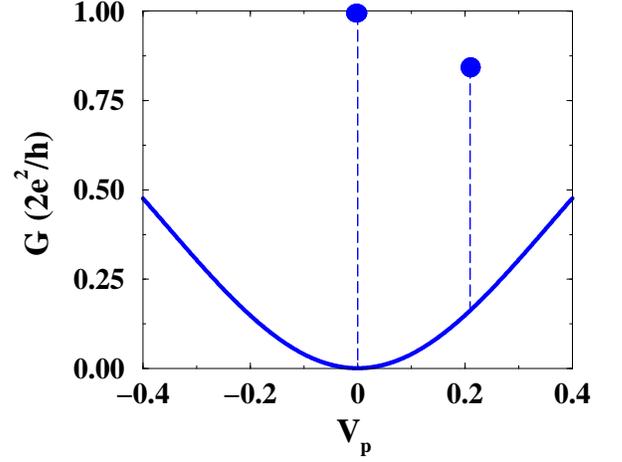}
\end{center}
\caption{\label{G}A plot of the conductance through the impurity
as a function of $V_p$.  We again take $U+2\epsilon_d = 0^+$ as well as $U=1$ and $\Gamma = .1$.
The plotted filled circles indicate the values of $G$ at $V_p=0$ and $V_p = V^{\rm crit}_p \simeq 0.19$.}
\end{figure}

As a starting point, we begin by computing the conductance at $U+2\epsilon_d = 0^+$, that is, at
a value of the dot chemical potential just in excess of the point
at which the model would possess particle-hole symmetry in the absence of a finite $V_p$.
At $U+2\epsilon_d = 0^+$, $Q=-\infty$, just as with the $V_p=0$ Anderson model, 
and it is straightforward to show that,
\begin{equation}\label{eIIIv}
n_{\rm imp \uparrow/\downarrow} = \int^{\infty}_{-\infty} d\lambda \sigma_{\rm imp} (\lambda) = 
{1\over 2}\int^{\infty}_{-\infty} d\lambda \tilde{\Delta}(\lambda,V_p).
\end{equation}
The discontinuities in $\tilde\Delta (\lambda)$ as a function of $V_p$ just discussed are then intimately related to
discontinuous changes in $n_{\rm imp}$ as a function of $V_p$.  We plot
$n_{\rm imp}=n_{\rm imp\ua}+n_{\rm imp\da}$ 
as function of $V_p$ in Figure \ref{nimp}.  We see for $V_p$ negative,
$n_{\rm imp} = 0$.  As $V_p$ is tuned to exactly zero, $n_{\rm imp} = 1$.
Upon further increasing $V_p$, $n_{\rm imp}$ changes to $2$.  This marks the
first transition in $\tilde\Delta (\lambda)$ as $V_p$ is varied from $-\infty$ to $+\infty$.  When $V_p$ is tuned through
its second transition, $n_{\rm imp}$ jumps back to $1$ and then to $0$.
Given that we have
\begin{eqnarray*}
G_{\uparrow/\downarrow} = {e^2\over h}\sin^2(\pi \nboth - 2\tan^{-1}(V_p)),
\end{eqnarray*}
$G$ experiences a similar
set of jumps as shown in Figure \ref{G}.

The jump in $n_{\rm imp}$ and $G$ as $V_p$ transits $V_p=0$ is the easiest to understand.  As $V_p$ is 
changed away from $0$, the particle-hole symmetry present is broken.  With this symmetry present, $n_{\rm imp}$
necessarily is $1$.  This change by $\pm 1$ in $n_{\rm imp}$ is mimicked by an exact diagonalization calculation
of the ground state occupancy for a lattice model equivalent to the continuum Anderson model.
To be specific, we consider the lattice model 
\begin{eqnarray}\label{eIIIvi}
{\cal H}_{\rm lattice} &=& -t\sum_{i=-\frac{N-1}{2}+1,\sigma}^{-1}
(c^\dagger_{i\sigma}c_{i-1\sigma}+{\rm h.c.})\cr\cr
&& -t\sum^{(N-1)/2-1}_{i=1,\sigma}
(c^\dagger_{i\sigma}c_{i+1\sigma}+{\rm h.c.})\cr\cr
&& + V \sum_\sigma( c^\dagger_{1\sigma}c_{0\sigma} + c^\dagger_{-1\sigma}c_{0\sigma} + {\rm h.c.})\cr
&& + V_{LR} \sum_\sigma (c^\dagger_{1\sigma}c_{-1\sigma} + {\rm h.c.})\cr\cr
&& + \epsilon_d \sum_\sigma c^\dagger_{0\sigma} c_{0\sigma} 
+Uc^\dagger_{0\uparrow}c_{0\uparrow}c^\dagger_{0\downarrow}c_{0\downarrow}.
\end{eqnarray}
On the lattice, the dot sits at $i=0$.
Here $t$ is the hopping amplitude between the ``lead'' (i.e. non-dot) lattice 
sites and is roughly equivalent to the bandwidth. 
$V$ is the hopping amplitude between the dot site and the neighboring lead sites, $i=\pm 1$.
The hopping amplitude, $V_{LR}$, allows the electron to bypass the dot, hopping from $i=1$ to $i=-1$
and vice versa.  The dot parameters $U$ and $\ed$ are the same as in the continuum Hamiltonian
(\ref{eIIi}).

We have constructed the ground state of this lattice model for a number of small lattice sites $N\leq 13$,
$N$ odd, at the point $U+2\ed = 0$.  With $V_{LR}=0$ the ground state is degenerate, one state with
$N-1$ electrons, one state with $N+1$ electrons.  Averaging over these states, each site on the lattice contains
exactly one electron.  Turning on $V_{LR}$ breaks the degeneracy of these two states.  For $V_{LR} > 0$,
the system prefers to be in a state with $N-1$ electrons, while for $V_{LR} < 0$ the system prefers
to be in a state with $N+1$ electrons.  Thus as $V_{LR}$ is swept through $0$ the ground state occupancy
changes by 2.  We suggest this is the lattice equivalent of the change by $2$ as seen in $n_{\rm imp}$ as computed
by the Bethe ansatz.

\begin{figure}
\vskip .58in
\begin{center}
\noindent
\epsfysize=0.35\textwidth
\epsfbox{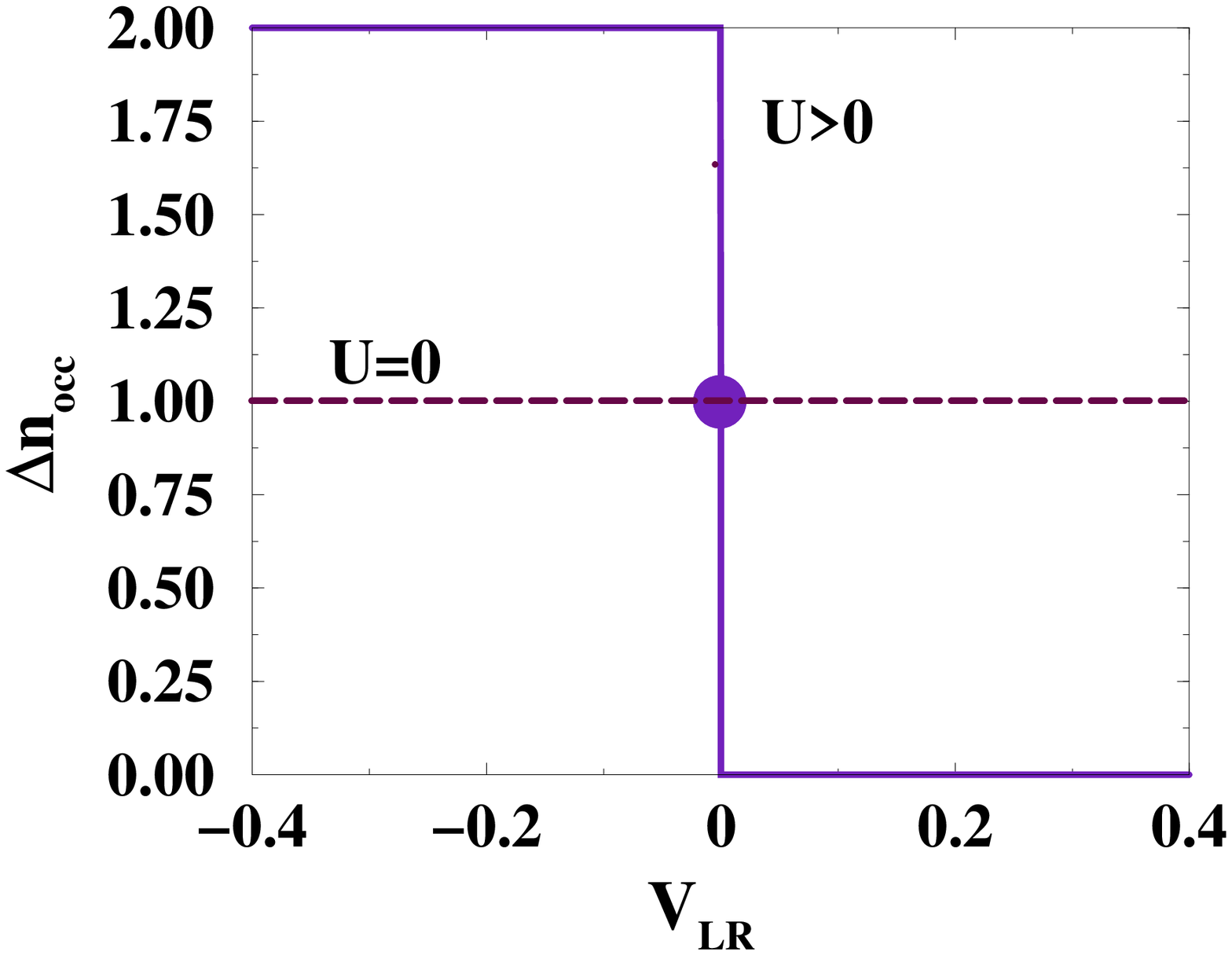}
\end{center}
\caption{\label{nocc}Plots as a function of $V_{LR}$ of the deviation in the ground state occupancy of an N-site Anderson
lattice model from a state with occupancy $N-1$. 
The exact diagonalization calculation of the model
takes $U+2\epsilon_d = 0$. 
The plotted filled circle indicates that at $V_{LR} = 0$, $\Delta n^{\rm ED}_{\rm occ} =1 $ corresponding
to exactly one electron living on the dot site $i=0$.}
\end{figure}

Now we point out, as is obvious from Fig. \ref{nimp} and Fig. \ref{nocc}, that the changes
in $n_{\rm imp}$ and $n^{\rm ED}_{\rm occ}$ as $V_p$ passes through zero are opposite in sign.  In the context
of the Bethe ansatz computation this is readily understandable.  Consider $V_p$ negative.  The exact
diagonalization computation indicates for $V_{LR}$ negative the system should increase its occupancy
by one electron.  For the Bethe ansatz to increase the
occupancy of the system, it must add a two-particle bound state to the bulk distribution.  
For an overall change of one electron, the impurity occupancy must then compensate by decreasing by one.
A similar argument applies to $V_p$ positive.  The opposing changes in $n_{\rm imp}$ and $n^{\rm ED}_{\rm occ}$
are then ultimately a consequence of the Bethe ansatz constructing the ground state out of two-particle
bound states.

This change in lattice occupancy can be characterized as a many body effect.  If $N=2R+1$ with $R$ even,
the change in ground state occupancy is present only for finite $U$.  If $U=0$ instead, the total ground state
occupancy is precisely $N$ regardless of the value of $V_{LR}$ (see Figure 11).  $R$ must be even for only then does the
non-interacting lattice system have zero energy single particle excitations.  With $R$ odd (and
no zero energy single particle excitation), the
ground state occupancy jumps as $V_{LR}$ passes through zero regardless of the value of $U$.  But
in choosing $R$ for taking the continuum limit, $R$ even is appropriate as the
continuum limit possesses such a zero energy excitation.

Although the changes in the ground state occupancy were computed for small lattice sizes, the nature
of the results are robust and will persist to arbitrarily large lattices.

The origin of the second transition in $n_{\rm imp}$ occurring at finite positive $V_p$ is less transparent.  The
transition at $V_p=0$, on the basis of the exact diagonalization calculations, can be interpreted as one
electron on average being added or removed from the ground state.  However at the second transition no
obvious change in the ground state occupancy occurs in the exact diagonalization of the lattice.
A clue to the nature of the second transition can be found in the value of $V_p$ at which it occurs.

To determine this value of $V_p$ we need to first determine the value of $\lambda$ 
at which $\tilde\Delta (\lambda )$ develops a singularity when $V_p$ passes through the
second transition.  At this value of $\lambda$, the argument of the inverse tangent appearing in 
the scattering phase of the two-particle bound state, i.e.
\begin{eqnarray*}
\delta_{\rm bd.~state} (\lambda ) = -2{\rm Re}\tan^{-1}\bigg[ {\G \over x(\lambda )-iy(\lambda)-\ed}+V_p\bigg],
\end{eqnarray*}
satisfies 
\nonumber
\begin{eqnarray}\label{eIIIvii}
{\rm Im} \bigg[{\G \over x(\lambda )-iy(\lambda)-\ed}+V_p\bigg] = 1.
\end{eqnarray}
Solving this equation for large $U/\G$ we find
\nonumber
\begin{eqnarray*}
\lambda = {U\over 8\G } \pm {\Gamma \over 2 U}.
\end{eqnarray*}
The $V_p=0$ singularity appears at $\lambda = {U\over 8\G } - {\Gamma \over 2 U}$ while the $V_p=V^{\rm crit}_p$
singularity is found at $\lambda = {U\over 8\G } + {\Gamma \over 2 U}$.
To determine then $V^{\rm crit}_p$, we insist that the argument of the inverse tangent in $\delta_{\rm bd.~state}$
satisfies 
\nonumber
\begin{eqnarray*}
{\rm Re} \bigg[{\G \over x(\lambda )-iy(\lambda)-\ed}+V_p\bigg] = 0.
\end{eqnarray*}
Solving we find the critical value of $V_p$ is
\begin{eqnarray}\label{eIIIviii}
V^{\rm crit}_p = {2\G \over U} - 8{\Gamma^3\over U^3} + {\cal O}({\Gamma^5\over U^5}) .
\end{eqnarray}
Thus the value of $V_p$ at which the second transition occurs is roughly $2\G/U$.  To interpret this
we suppose that at large $U/\G$, the Coulomb repulsion $U$ serves as a rough cutoff for the theory.
With $V_p$ a dimensionless parameter, the energy scale corresponding to $V_p^{\rm crit}$
is then $2\G$ or roughly the level broadening of the dot.  Thus the second transition occurs
when the tunneling from one lead to the other induced by $V_p$ has grown to be the same magnitude
as tunneling through the dot.

This transition only occurs at $V_p$ positive.  At $V_pU\sim \Gamma$, the potential scattering term is
strong enough to remove the electron from the impurity itself.  That is, the electron that must be
ejected from the system (at $V_p >0$ as indicated by the exact diagonalization computation) is now removed
directly from impurity.  We no longer have the construction by which $V_p >0$ induces the removal of a
two-electron bound state in the bulk and a concomitant addition of an electron to the integrated impurity density
of states.

For $U/2+\epsilon_d >0$, this suggests an inequivalent position for particle vs. hole scattering.  There is no corresponding
transition at $V_pU \sim -\Gamma$ where the impurity occupancy is increased by one.  We however find such a transition
for $U/2+\epsilon_d <0$ where the roles of particles and holes are reversed.  Exactly at $U/2+\epsilon_d=0$, both
transitions, $V_p U\sim \pm \Gamma$ appear, as they must by the demands of a particle-hole transformation which
leaves $U/2+\epsilon_d$ invariant but changes the sign of $V_p$.

\vskip .2in

\begin{center}
b) $U/2+\ed$ finite and positive:
\end{center}

\vskip .2in

\begin{figure*}
\begin{center}
\noindent
\epsfysize=0.6\textwidth
\epsfbox{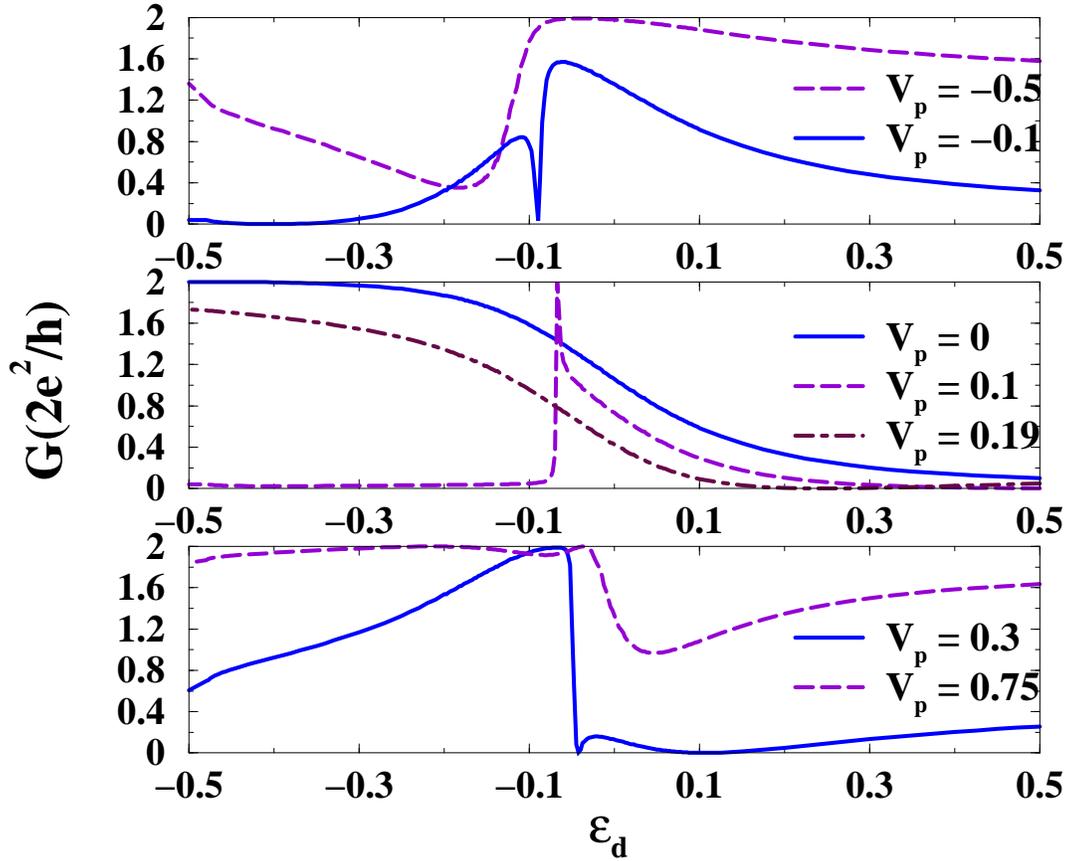}
\end{center}
\caption{Plots of the $T=0$ linear response conductance for $\ed > -U/2$ for a variety of values of $V_p$.
Here we have taken $U=1$ and $\Gamma =.1$}
\end{figure*}

We now move on to consider how the linear response conductance varies as $\epsilon_d$ is increased from the
value $-U/2+0^+$.  $G$ evolves according to the value of $\nboth$ at $\ed = -U/2+0^+$.  If $\nboth =0$ (as will
be the case for $V_p<0$ as well as for $V_p>\vc$), as $\ed$ tends to large values, the conductance, $G$,
will return to the value seen at $\ed = -U/2+0^+$.  We have that $G$ depends upon $\nboth$ via
\begin{equation}\label{eIIIix}
G = 2{e^2\over h} \sin^2(\pi\nboth-2\tan^{-1}(V_p)).
\end{equation}
$\nboth$ is necessarily $\sim 0$ at $\ed/\Gamma \gg 1$ for in this limit the Fermi surface, $Q$, of the
$\lambda$-excitations tends to $\infty$ and so
\begin{eqnarray*}
\lim_{\ed \rightarrow \infty}\nboth = \lim_{\ed\rightarrow \infty} \int^\infty_Q d\lambda \si (\lambda ) = 0.
\end{eqnarray*}
Thus if $\nboth = 0$ at $\ed = -U/2+0^+$, we necessarily have
\begin{eqnarray}\label{eIIIx}
G(\ed = -U/2 + 0^+) &=& G(\ed = \infty) \cr\cr
&=& 2{e^2\over h}\sin^2(2\tan^{-1}(V_p))\cr\cr
&=& 2{e^2\over h}{4V_p^2 \over (1 + V_p^2)^2}.
\end{eqnarray}
The value of $G$ at these two values of $\ed$ is then solely a function of $V_p$.

There will of course be significant variation in $G$ as $\ed$ changes from $-U/2+0^+$ to $+\infty$.
This variation will occur at the value of $\ed$ at which the $\lambda$-Fermi surface, $Q$, reaches
the position of the largest variation in $\tilde\Delta (\lambda )$.  As $\vp$ is tuned away from $0$, the
position at which the non-analyticity in $\tilde\Delta (\lambda )$ opens up is $\lambda \sim U/8\Gamma $.  
It is thus reasonable to locate the largest variation in $\tilde\Delta (\lambda)$ at this value
of $\lambda$.  The behaviour of $Q$ as a function of $\ed$ has been determined in Ref. \onlinecite{wie}.
Provided $Q>0$, $Q$ can be written in terms of $\ed$, $U$, and $\Gamma$ as follows
\begin{eqnarray}\label{eIIIxi}
Q &=& q^* + {1\over 2\pi}\log (2\pi e q^*);\cr\cr
\sqrt{q^*} &=& {\ed + U/2 \over \sqrt{2U\Gamma}}.
\end{eqnarray}
Thus the critical value of $\ed$ at which $Q=U/8\Gamma $ and $G$ sees a large variation
is
\begin{equation}\label{eIIIxii}
\ed^{\rm crit} = -{U\over 2} + \sqrt{U\over \Gamma}{U\over 2^{5/2}} - {\sqrt{2U\Gamma} \over 2\pi}
\log (2\pi e{U\over 8\Gamma}),
\end{equation}
provided $U/\Gamma \gg 1$.

The size of the variation in $G$ occurring at $\ec$ can be estimated.  From 
(\ref{eIIIix}), the variation in $G$ can be related to the variation in $\nt=n_{\rm imp \uparrow}+n_{\rm imp \da}$:
\begin{eqnarray}\label{eIIIxiii}
\delta G &=& 2{e^2 \over h}\bigg(2\pi {\vp(1-\vp^2) \over (1+\vp^2)^2}\delta\nt\cr\cr
&+& {\pi^2\over 2}{1-3\vp^2+\vp^4 \over (1+\vp^2)^2}(\delta\nt)^2\bigg)
\end{eqnarray}
$\delta \nt$ can in turn be estimated to be
\begin{eqnarray}\label{eIIIxiv}
\delta \nt &=& \delta \bigg( {1\over \pi} {\rm Re}\tan^{-1}\big({\Gamma \over x(\lambda ) -iy(\lambda )-\ed} 
+ \vp\big)\bigg|_{\lambda=Q}\bigg)\cr\cr
&=& {1\over \pi}{1\over 1+ \vp^2} \delta \bigg({\rm Re}\big({\Gamma \over x(\lambda ) -iy(\lambda )-\ed}\big)
\bigg|_{\lambda=Q}\bigg)\cr\cr
&=& {1\over \pi}{1\over 1+ \vp^2}.
\end{eqnarray}
The first estimate in (\ref{eIIIxiv}) follows as we expect $\nt$ obeys
\begin{eqnarray}\label{eIIIxv}
\nt &=& 2\int^{\tilde Q}_Q d\lambda \sigma_{\rm imp}(\lambda )
= \int^{\tilde Q}_Q d\lambda \tilde \Delta(\lambda )\cr\cr
&=& {1\over \pi} {\rm Re}\delta(x(\lambda )+iy(\lambda )).
\end{eqnarray}
The second equality in (\ref{eIIIxiv}) follows provided $V_p$ is large while the third and final estimate
holds as ${\rm Re}(\Gamma /(x(\lambda)-iy(\lambda)-\ed))$ varies between $-1/2$ and $1/2$ in the vicinity
of $\lambda =Q$.  Hence $\delta G$ is given by
\begin{eqnarray}\label{eIIIxvi}
\delta G = 2{e^2 \over h}\bigg(4 {\vp(1-\vp^2) \over (1+\vp^2)^3}
+ {1-3\vp^2+\vp^4 \over (1+\vp^2)^4}\bigg).
\end{eqnarray}
According to this expression, near $\vp = 0$, the sign of the variation changes.

In Figure 12 these various analytic formulae are borne out.  At $\vp = -0.5, -0.1, 0.3,$ and $0.75$, we have
$\nboth(U/2+\ed=0^+)=0$.  We see that the value $G$ asymptotes to as $\ed\rightarrow \infty$ equals the value
of $G$ at $U/2+\ed=0^+$.  The variation in $G$ takes place at $\ed \approx -0.1$ in accordance with
(\ref{eIIIxii}).  Finally the size of the variation in $G$ clearly depends upon $\vp$ with the variation
decreasing in size as $\vp$ increases.  We also see that variations in $G$ on opposite sides of $\vp=0$ are
of different signs in accordance with (\ref{eIIIxvi}).

We now consider the behaviour of $G$ as a function of $\ed$ in the region of $\vp$, $0 < \vp < \vc$, where
$\nboth(\ed+U/2=0^+) = 1$.  By virtue of (\ref{eIIIix}), we again find that
\begin{eqnarray}\label{eIIIxvii}
G(\ed = -U/2 + 0^+) &=& G(\ed = \infty) \cr\cr
&=& 2{e^2\over h}{4V_p^2 \over (1 + V_p^2)^2}.
\end{eqnarray}
The variation in $G$ as $\ed$ is increased from $-U/2$ still takes place at $\ed = \ec$ (\ref{eIIIxii}) as the
primary variation in $\tilde\Delta (\lambda)$ still occurs at $\lambda = U/8\Gamma $.  However the size
of the variation in $G$ differs from the case $\nboth(U/2+\ed = 0^+)=0$.  Here $G$ will take on its
maximal ($2e^2/h$) and minimal ($0e^2/h$) values as $\ed$ is increased upwards from $-U/2$.  This is
a consequence of (\ref{eIIIix}) and the fact that $\nboth$ will decrease monotonically from $1$ to $0$ as $\ed$
is made large and positive.

This behaviour in again seen in the conductance plot for $\vp = 0.1$ in Figure 12.  
$G$ there is $0.08e^2/h$ at $\ed = -U/2+0^+$, varies
rapidly in the region of $\ed = \ec = -0.1$, taking on values of both $2e^2/h$ and $0e^2/h$, then proceeds to
decay back to $0.08e^2/h$ for larger values of $\ed$.

\newcommand{\nbh}{n^{\rm hole}_{\rm imp\ua/\da}}

\newcommand{\nht}{n^{\rm hole}_{\rm imp}}

\begin{figure}
\vskip .58in
\begin{center}
\noindent
\psfrag{y}{$\bf n_{\rm imp}^{\rm hole}$}
\epsfysize=0.35\textwidth
\epsfbox{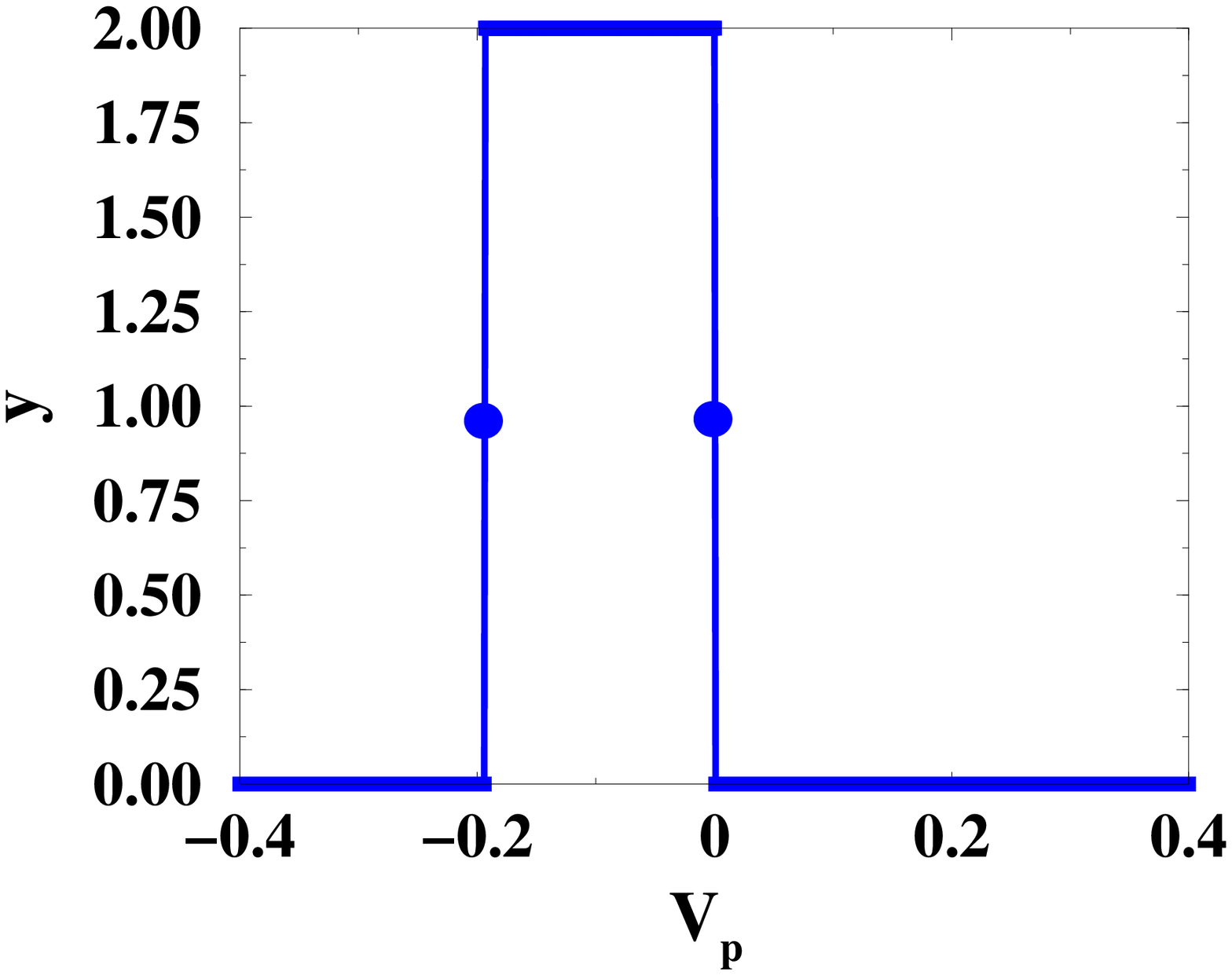}
\end{center}
\caption{\label{Ghole}A plot of the total number of holes (of both spin species)
displaced by the presence of the impurity as
a function of $V_p$.  We have taken $U+2\epsilon_d = 0^-$, $U=1$ and $\Gamma = .1$.
The plotted filled circles indicate that at $V_p=0$ and $V_p = \vc \simeq -0.19$, $\nht$ is precisely 1.}
\end{figure}

Having considered values of $\vp$ where $\nboth (\ed=-U/2+0^+)=0$ and $1$, we now turn to the two remaining
undiscussed values of $\vp$, $\vp = 0$ and $\vc$, where here $\nboth (\ed=-U/2+0^+)=1/2$.  For these two values of $\vp$,
the two limiting values of $G$ are 
\begin{eqnarray}\label{eIIIxviii}
G(\ed = -U/2+0^+) &=& {2e^2\over h} \cos^2(2\tan^{-1}(\vp ))\cr\cr 
&=& 2{e^2\over h}\bigg({1-\vp^2 \over 1+\vp^2}\bigg)^2;\cr\cr
G(\ed = \infty ) &=& {2e^2\over h}{4\vp^2 \over (1+\vp^2)^2}.
\end{eqnarray}
For $\vp = 0$ and $\vc$, $\tilde\Delta(\lambda )$ does not possess large variations at $\lambda = U/8\Gamma$.
Thus $G$ sees no large variations.  At $\vp=0$, $G$ decreases monotonically from $2e^2/h$ as $\ed$ is increased
from $-U/2$.  If $U/\Gamma \gg 1$
and correspondingly $\vc \ll 1$, $G(\vp = \vc )$ will also decrease monotonically as $\ed$ is varied upwards from $-U/2$.
The plots of $G$ for $\vp=0$ and $\vc$ in Figure 12 bear this out.

\begin{figure}
\vskip .58in
\begin{center}
\noindent
\epsfysize=0.35\textwidth
\epsfbox{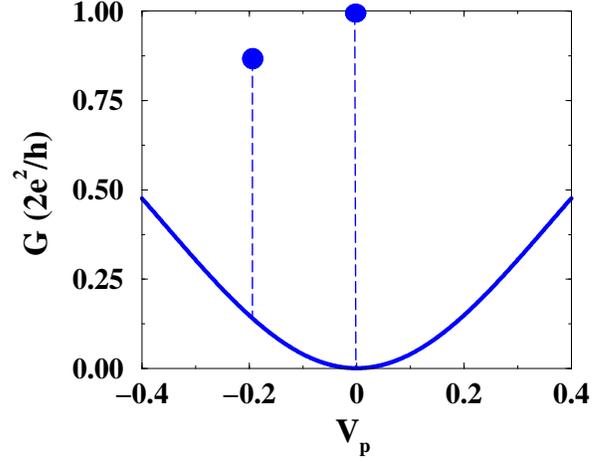}
\end{center}
\caption{\label{nimphole}A plot of the conductance through the impurity
as a function of $V_p$.  Here $U+2\epsilon_d = 0^-$ while $U=1$ and $\Gamma = .1$.
The plotted filled circles indicate the values of $G$ at $V_p=0$ and $V_p \simeq -0.19$.}
\end{figure}

\begin{figure*}
\begin{center}
\noindent
\epsfysize=0.65\textwidth
\epsfbox{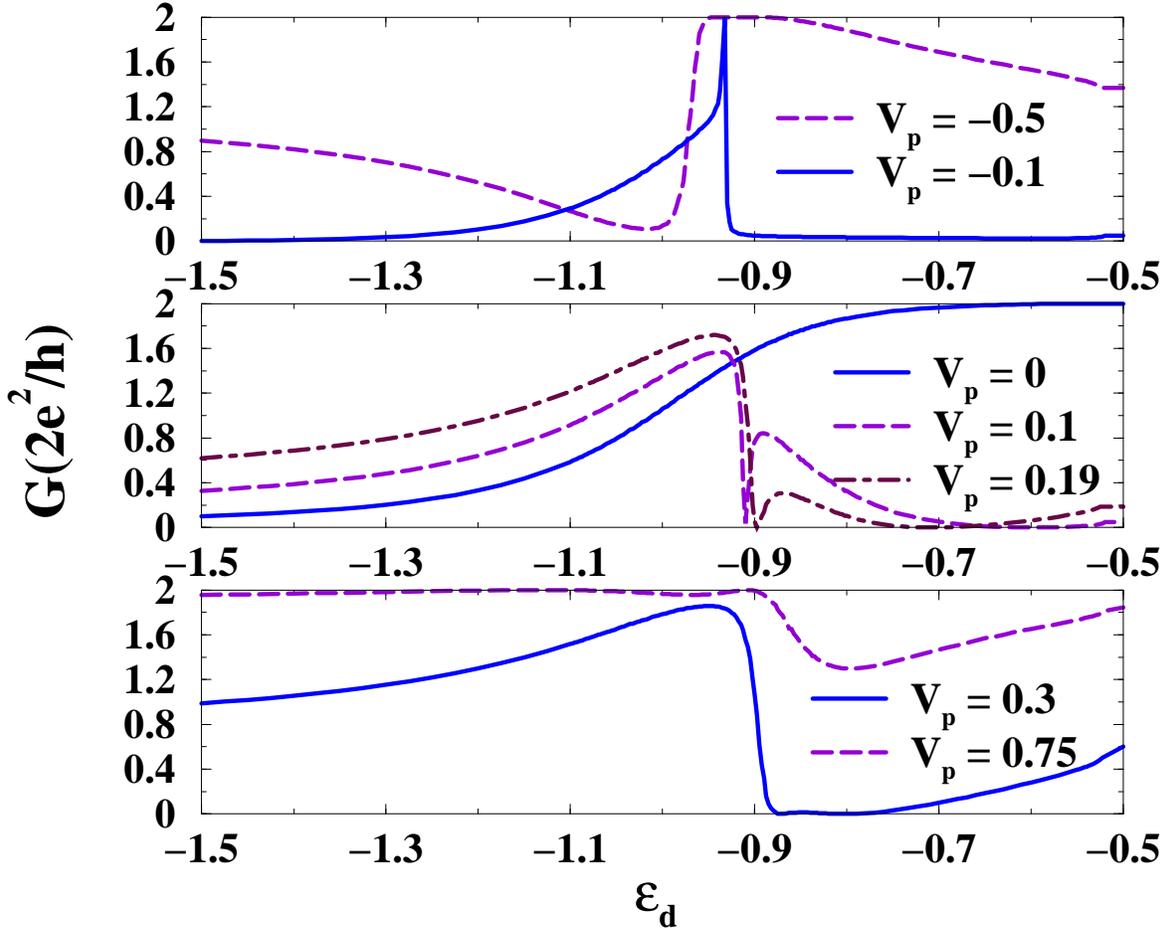}
\end{center}
\caption{Plots of the $T=0$ linear response conductance for $\ed < -U/2$ for a variety of values of $V_p$.}
\end{figure*}

\subsubsection{$\bf U+2\epsilon_d <0$}

To treat the case $U+2\ed < 0$, we need to use the development of the Bethe ansatz equations
detailed in Section II.F.  Here everything is cast in terms of holes not
particles.  The conductance is given by
\begin{equation}\label{eIIIxix}
G(\ed ,\vp) = 2{e^2\over h}\sin^2(\pi\nbh(\ed,\vp) + 2\tan^{-1}(\vp))
\end{equation}
where $\nbh$ is the number of holes displaced by the impurity.  $\nbh$ is given by (at zero field)
by
\begin{equation}\label{eIIIxx}
\nbh = \int^{\tilde Q}_{Q}d\lambda \si (\la,-\vp,-U-\ed),
\end{equation}
where $\si (\la,-\vp,-U-\ed )$ is governed by (\ref{eIIIiv}).

\vskip .2in

\begin{center}
a) $U/2+\ed = 0^-$:
\end{center}

\vskip .15in

Akin to the treatment of $U+2\ed > 0$ in Section III.A.1, we first examine the case $U+2\ed = 0^-$.  At
$U+2\ed = 0^-$, the number of holes displaced by the impurity is
\begin{equation}\label{eIIIxxi}
\nbh (U+2\ed = 0^-,\vp) = {1\over 2}\int^\infty_{-\infty} d\la \tilde\Delta (\lambda,-\vp).
\end{equation}
Thus $\nbh(U+2\ed=0^-,\vp) = \nboth(U+2\ed=0^+,-\vp)$ as would be expected from a particle-hole
transformation.  Plotted in Figure 13 is $\nht(U+2\ed = 0^-)$ as a function of $\vp$.  It is a mirror
image of Figure 9.

We ascribe the jumps in $\nht(U+2\ed=0^-,\vp)$ at $\vp = 0$ and $\vp = -\vc$ to similar
mechanisms as the jumps seen in $\nt(U+2\ed=0^+)$.  The jump in $\nht (U+2\ed = 0^-)$ as $\vp$
moves away from 0 is caused by the breaking of particle-hole symmetry by finite $\vp$.  For $\vp = 0^\pm$,
the system wants to have one more/one less hole in the ground state.  It thus adds/subtracts one two-hole
bound state to/from the bulk occupancy and subtracts/adds one hole from/to the impurity occupancy.

Again the jump in $\nht$ at $\vp = -\vc$ can be thought to be due to the rough equivalence occurring at this point of the two tunneling
paths through the impurity.  Here the transition occurs at $\vp$ negative.  The hole
that must be removed from the system at $\vp <0$ now is removed directly from the impurity occupancy as
opposed to being removed from the bulk.

As in the case $U/2+\ed = 0^-$, the conductance at $U/2+\ed = 0^-$ experiences a set of discontinuities.
This is plotted in Figure 14, a mirror image of Figure 10.

\vskip .2in 

\begin{center}
b) $U/2+\ed$ finite and negative:
\end{center}

\vskip .10in

The behaviour of $G$ as $\ed$ is decreased from $-U/2$ is determined by the value of $\nbh$ at $\ed = -U/2+0^-$.
As we have already carried out an analogous discussion of $G$'s evolution as $\ed$ is increased from $-U/2$ and its
dependence upon $\nboth(\ed=-U/2+0^+)$, we merely summarize in point form the results in this case.

\vskip .3in

\noindent {$\bf a) ~~\nbh(\ed = -U/2+0^-) = 0$, ($\bf \vp > 0$ and $\bf\vp < -\vc$)}:

\vskip .2in

\noindent $\bullet$ The conductance at $\ed=-U/2+0^-$ equals $G$ at $\ed=-\infty$, namely
\begin{equation}\label{eIIIxxii}
G(\ed-U/2+0^-)=G(\ed=-\infty)= {4\vp^2 \over (1+\vp^2)^2}.
\end{equation}

\vskip .2in 

\noindent $\bullet$ The largest variation in $G$ as $\ed$ varies between $-\infty$ to $-U/2+0^-$ will occur at
\begin{eqnarray}\label{eIIIxxiii}
\tilde{\epsilon}^{\rm crit}_d &=& -U-\ec \cr\cr
&=& -{U\over 2} - \sqrt{U \over \Gamma}{U \over 2^{5/2}} 
+ {\sqrt{2U\Gamma}\over 2\pi}\log (2\pi e {U\over 8\Gamma}).\cr &&
\end{eqnarray}

\vskip .2in

\noindent $\bullet$ The variation in $G$ at $\ed = \tilde\ec$ is given by
\begin{equation}\label{eIIIxxiv}
\delta G = 2{e^2\over h} \bigg(-4{\vp(1-\vp^2 ) \over (1+ \vp^2)^2} + {1-3\vp^2+\vp^4 \over (1+\vp^2)^4}\bigg).
\end{equation}

\vskip .2in

\noindent $\bullet$ These features are seen in Figure 15 in the plots of G for $\vp = .75,.3,.19,.1,$ and $-.5$
(where $\tilde \epsilon^{\rm crit}_d \simeq -0.9$).

\vskip .3in

\noindent {$\bf b) ~~\nbh(\ed = -U/2+0^-) = 0$, ($\bf 0 > \vp > -\vc$)}:

\vskip .2in

\noindent $\bullet$ At the two limits, $\ed = -U/2+0^-$ and $\ed=-\infty$, the conductance satisfies
\begin{equation}\label{eIIIxxv}
G(\ed-U/2+0^-)=G(\ed=-\infty)= 2{e^2\over h}{4\vp^2 \over (1+\vp^2)^2}.
\end{equation}

\vskip .2in

\noindent $\bullet$ The variation in G occurs again at $\tilde\ec = -U-\ec$.

\vskip .2in

\noindent $\bullet$ The variation in $G$ as $\ed$ is varied from $-\infty$ to $-U/2+0^-$ is $2e^2/h$.  That
is, the minimal value of $G$ is $0e^2/h$ while its maximal value is $2e^2/h$.  

\vskip .2in

\noindent $\bullet$ These features are seen in
Figure 15 for the conductance plot for $\vp = -0.1$.

\vskip .3in

\noindent {$\bf c) ~~\nbh(\ed = -U/2+0^-) = 1/2$, ($\vp = 0 , -\vc$)}:

\vskip .2in

\noindent $\bullet$ 
At the two values of the dot chemical potential, $\ed = -U/2+0^-$ and $\ed=-\infty$, the conductance satisfies
\begin{eqnarray}\label{eIIIxxvi}
G(\ed-U/2+0^-) &=& 2{e^2\over h}{(1-\vp^2) \over 1+\vp^2};\cr\cr
G(\ed=-\infty) &=& 2{e^2\over h}{4\vp^2 \over (1+\vp^2)^2}.
\end{eqnarray}

\vskip .2in

\noindent $\bullet$ G varies smoothly and monotonically as $\ed$ is varied between $-\infty$ and $-U/2+0^-$.

\vskip .2in

\noindent $\bullet$ 
Comparing Figure 12 and Figure 15, we see the conductance is continuous as $\ed$ crosses $-U/2$ for all values
of $\vp$ but for $\vp=\vc$.  This jump is of magnitude
\begin{eqnarray}\label{eIIIxxvii}
\Delta G &=& G(\ed=-U/2+0^+)-G(\ed=-U/2+0^-)\cr\cr
&=& 2{e^2\over h}{1+{\vc}^4-6{\vc}^2 \over (1+{\vc}^2)^2}.
\end{eqnarray}
A similar zero temperature discontinuity in $G$ is seen at $\vp = -\vc$.
This discontinuity gives rise to the question of what is the actual value of $G$ when $\ed$ is precisely $-U/2$.  
We resolve this question in the next section.

\subsubsection{$\bf U+2\epsilon_d =0$}

\begin{figure}
\vskip .58in
\begin{center}
\noindent
\epsfysize=0.35\textwidth
\epsfbox{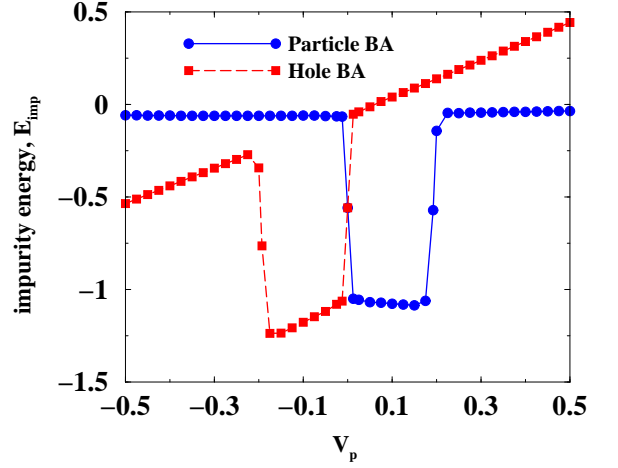}
\end{center}
\caption{A plot of the impurity energy at $U+2\epsilon_d = 0$ as a function of $V_p$
with $U=1$ and $\Gamma = 0.1$ as computed with both the particle and hole Bethe ansatz solutions.}
\end{figure}

At $\ed = -U/2$ (and $\ed = -U/2$ alone), two different developments of the Bethe ansatz equations are available:
one involving a basis of single {\it particle} states and one involving a basis of single {\it hole} states.
These two treatments are not equivalent for all values of $\vp$ as the discontinuity at $\ed=-U/2$ in $G$ at
$\vp = \pm\vc$ shows.  This discontinuity is a reflection of our ability to construct two not one putative
ground states.  To determine the true ground state at a given $\vp$, we compute the energy of the two competing states.

The energy of the particle Bethe ansatz state (for $H=0$) equals
\begin{equation}\label{eIIIxxviii}
E_{\rm part}(\ed ,\vp) = L\int^{\tilde Q}_Q d\la \sigma (\la,\vp,\ed)2x(\lambda) ,
\end{equation}
where $\sigma(\la )$, the density of bound states at a given $\lambda$, satisfies (\ref{eIIxxxv}) and 
$2x(\lambda )$ is the (bare) energy of a two-electron bound state.  Decomposing $\sigma(\la )$ into its bulk
and impurity contributions, $\sigma = \sigma_{\rm bulk}+\sigma_{\rm imp}/L$, allows us to
similarly decompose $E_{\rm part}$:
\begin{eqnarray}\label{eIIIxxix}
E_{\rm part}(\ed,\vp) &=& E^{\rm bulk}_{\rm part}(\ed,\vp) + {1\over L}E^{\rm imp}_{\rm part} ;\cr\cr
E^{\rm imp}_{\rm part}(\ed,\vp) &=& \int d\la \sigma_{\rm imp} (\la,\vp,\ed) 2x(\lambda,\ed,U );\cr\cr
E^{\rm bulk}_{\rm part}(\ed,\vp) &=& L \int d\la \sigma_{\rm bulk} (\la ,\ed)2x(\lambda,\ed,U ).\cr &&
\end{eqnarray}
$\sigma_{\rm bulk}(\lambda )$ shares its independence from $\vp$ with $E^{\rm bulk}_{\rm part}$.  The energy
of the hole Bethe ansatz state is given by
\begin{eqnarray}\label{eIIIxxx}
E_{\rm hole}(\ed,\vp) &=& L \int d\la \sigma_{\rm hole}(\la, \vp,\ed)2x(\la,-U-\ed,U )\cr\cr
&& \hskip .2in + \vp + U + 2\ed\cr\cr
&=& L \int d\la \sigma(\la, -\vp,-U-\ed)\cr
&& \hskip .5in \times 2x(\la,-U-\ed,U)\cr\cr
&& \hskip .2in  + \vp + U + 2\ed .
\end{eqnarray}
We have used the fact that the density of two-hole bound states in the hole Bethe ansatz state, 
$\sigma_{\rm hole}(\la,\vp,\ed)$, is equal to the density, $\sigma(\la,-\vp,-U-\ed)$, of two-electron bound
states as computed using the particle Bethe ansatz up to a change in sign of $-\vp$.
We can again perform a decomposition into bulk and impurity pieces:
\begin{eqnarray}\label{eIIIxxxi}
E_{\rm hole}(\ed,\vp) &=& E^{\rm bulk}_{\rm hole}(\ed,\vp) + {1\over L}E^{\rm imp}_{\rm hole} ;\cr\cr
E^{\rm imp}_{\rm hole}(\ed,\vp) &=& \vp + U +2\ed \cr\cr
&& \hskip -.75in +\int d\la \sigma_{\rm imp}(\la,-U-\ed,-\vp) 2x(\lambda,-U-\ed,U) \cr\cr
E^{\rm bulk}_{\rm hole}(\ed,\vp) &=& \cr\cr
&& \hskip -.95in L \int d\la \sigma_{\rm bulk}(\la , -U-\ed) 2x(\lambda,-U-\ed,U).
\end{eqnarray}
So while $E^{\rm bulk}_{\rm part}$ and $E^{\rm bulk}_{\rm part}$ are equal at $\ed=-U/2$, 
the impurity contributions are not.
By determining these impurity contributions then we can determine the state of lowest energy, the true 
ground state.

\begin{figure}
\vskip .58in
\begin{center}
\noindent
\epsfysize=0.35\textwidth
\epsfbox{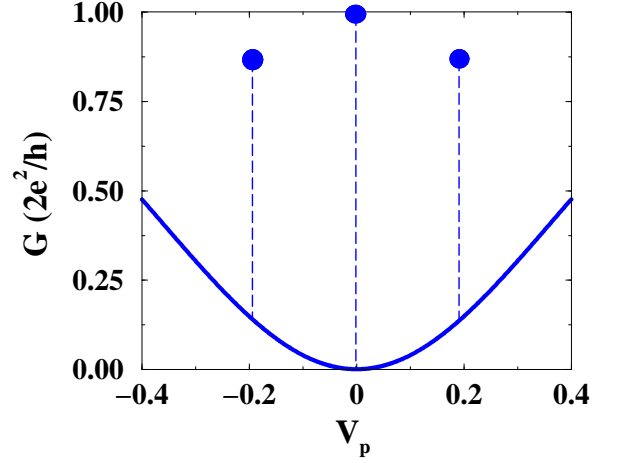}
\end{center}
\caption{A plot of the conductance as a function of $V_p$ 
for $U+2\epsilon_d = 0$, $U=1$ and $\Gamma = 0.1$.
The filled circles mark the discontinuous values of the conductance 
at $V_p=0$ and $V_p = \vc \simeq \pm 0.19$.}
\end{figure}

Plotted in Figure 16 are the impurity contributions to the energy arising from both the hole and particle
Bethe ansatz states.  $\sigma_{\rm imp}$, necessary to compute $E^{\rm imp}_{\rm part, hole}$ is determined
by equations (\ref{eIIxxxvii}) and (\ref{eIIxli}).  For $\vp >0$, we see the particle Bethe ansatz state
forms the true ground state while for $\vp < 0$, it is the hole Bethe ansatz state that gives the
ground state.  At $\vp=0$, the two states become equivalent.

Having determined the true ground state as a function of $\vp$, we can determine $G$ at $U=2\ed$.  We plot
$G(\ed = -U/2)$ in Figure 17.  As a function of $\vp$, it sees three discontinuities: two at $\vp = \pm\vp$ and
one at $\vp =0$.  Despite these discontinuities, $G$ is particle-hole symmetric, i.e.
$$
G(\ed=-U/2,\vp) = G(\ed=-U/2,-\vp),
$$
as it must be.

The determination of the true ground state as a function of $\vp$ thus eliminates a potential problem.
If one were to look at $G(\ed = -U/2)$ on the basis of either the particle Bethe ansatz state or the
hole Bethe ansatz state, $G$ would not satisfy particle-hole symmetry (see for example Figures 
10 and 14).  Only by taking into account both solutions, do we establish a conductance consistent 
with particle-hole symmetry.

\begin{figure*}
\vskip .58in
\begin{center}
\noindent
\epsfysize=0.35\textwidth
\psfrag{y}{$\hskip -.1in \rho_{\rm imp}/\partial_q\epsilon$}
\psfrag{x}{$\hskip .9in$ energy, $\epsilon /T_k$}
\epsfbox{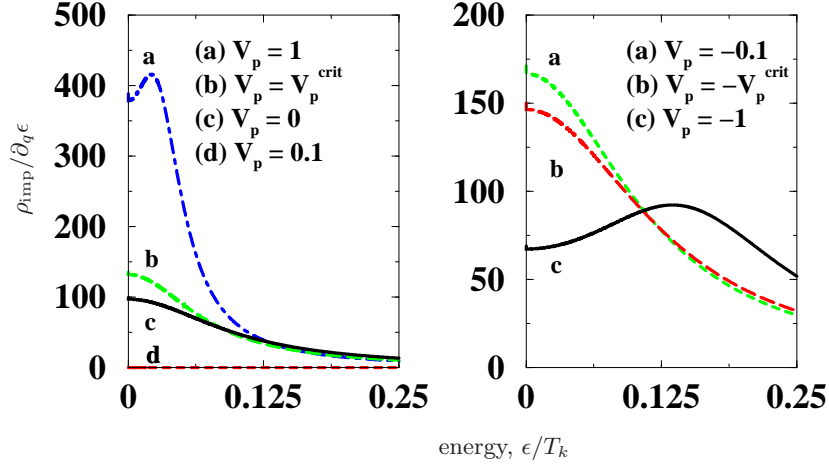}
\end{center}
\caption{Plots of the impurity density of states $\rho_{\rm imp}/\partial_q\epsilon$ 
as a function of energy for a number of a values of $V_p$.  Here we have taken $U=1$ and $\Gamma = 0.1$.}
\end{figure*}

\subsection{Finite Field Conductance}

In this section we will consider how $G$ behaves as a function of $H$.  Making $H$ finite changes the ground
state of the system.  From consisting solely of two-electron (or hole) bound states, the ground state begins to admit
the presence of single particle states with real wavevectors, $q$.  The scattering phase is still
given by (\ref{eIIxlvii}) and (\ref{eIIli}) but now the number of electrons displaced by the impurity is
given by (\ref{eIIxxxix}) with $B>-D$:
\begin{eqnarray}\label{eIIIxxxii}
n_{\rm imp \ua} &=& \int^{\tilde Q}_Qd\lambda \sigma_{\rm imp} (\lambda ) + \int^B_{-D}dq \rho_{\rm imp}(q);\cr\cr
n_{\rm imp \da} &=& \int^{\tilde Q}_Qd\lambda \sigma_{\rm imp} (\lambda ).
\end{eqnarray}
Thus the behaviour of $\rho_{\rm imp}(q)$ is dispositive of how $G$ will vary as $H$ changes from zero
(as $H$ is increased, $B$ increases from $-D$).

To obtain a feel for how $G$ evolves as a function of $H$, we examine $\rho_{\rm imp}(q)$ at both $H=0$
and $\ed = -U/2+0^+$.  If there is low energy spectral weight in $\rho_{\rm imp}(q)$, we expect $G$ to vary
rapidly with $H$.  If not, $G$ will be relatively constant as a function of $H$.  We plot 
$\rho_{\rm imp}(q)/\partial_q\epsilon (q)$ vs $\epsilon(q)$ for a number of representative values of $\vp$ in 
Figure 18.  $\epsilon (q)$ is the energy needed to add a $q$-excitation to the system.  As there are no $q$-excitations
in the groundstate at $H=0$, $\epsilon (q)$ is necessarily positive.  The derivation of $\epsilon (q)$ may be found
in Section II of Ref. \onlinecite{long} (applicable to the case here as $\epsilon (q)$ does not depend on $\vp$).

The behaviour of $\rho_{\rm imp}(q)$ at $H=0$ and $U/2+\ed =0^+$ as a function of $\vp$ can be put into
three categories.  If $\nboth(\vp,\ed=-U/2+0^+)=1/2$ (i.e. $\vp = 0,\vc$) there is low lying spectral weight.
With $\vp=0$, this spectral weight is associated with (but not identical to) the Kondo peak in the dot
density of states.  This low-lying weight ensures $G$ varies rapidly as a function of $H$.  If 
$\nboth(\vp,\ed=-U/2+0^+)=0$, i.e. $\vp<0$ or $\vp > \vc$, the low-lying spectral weight doubles
approximately in size.  It is as if spectral weight has shifted from being below the Fermi energy
and encoded in $\nboth$ at $\vp = 0$ to being above the Fermi energy and encoded in $\rho_{\rm imp}(q)$.
Again this spectral weight guarantees $G$ will change rapidly with the introduction of finite $H$.  Finally
if $\nboth (\vp,\ed = -U/2+0^+) = 1$, (i.e. $0<\vp <\vc$), the low-lying spectral weight in $\rho_{\rm imp}(q)$ is
absent.  Here it is as if the spectral weight has shifted from $\rho_{\rm imp}(q)$ to below the Fermi
surface increasing $\nboth(\vp,\ed=-U/2+0^+)$ from $1/2$ to $1$.  In this case $G$ will be unresponsive
to small magnetic fields ($H\sim T_K$).

\subsubsection{Conductance at $\ed=-U/2+0^+$}

We first turn to computing the conductance at the point $\ed=-U/2+0^+$.  In Figure 19, we plot the conductance
as a function of $H$ for $\ed=-U/2+0^+$.  We see that there is significant variation in the conductance for $\vp$
outside the range $(0,\vc )$.  In contrast for $\vp$ inside this range, $G(\ed=-U/2+0^+)$ is nearly
constant.  This is in accordance with the presence or absence of low-lying excitations as parameterized 
by $\rho_{\rm imp}(q)$.

Figure 19 also encodes information on the finite field conductance for $\ed = -U/2+0^-$.  Taking $\vp \rightarrow -\vp$
in Figure 19 gives the behaviour of $G(\ed-U/2+0^-)$.  We thus realize $G$ experiences generically a discontinuous
jump at finite $\vp$ and finite $H$ as $\ed$ crosses $-U/2$.  We plot the size of this jump as a function of $H$
for a set of representative values of $\vp \geq 0$ in Figure 20.  The corresponding discontinuities
for $\vp < 0$ can be obtained via
$$
\Delta G (\vp ) = \Delta G (-\vp ).
$$
While only at $\vp = \vc$ is $\Delta G$ non-zero at $H=0$, as $H$ is turned on, $\Delta G$ rapidly becomes finite
and large.  Although not plotted, these discontinuities will persist to large values of $H$ and will not disappear
until $H\gg \sqrt{U\Gamma }$.

We are able to derive analytically some results at $\ed = -U/2 + 0^\pm$.  We will focus upon $\ed = -U/2+0^+$ as results
from $\ed = -U/2+0^-$ can be obtained via a particle-hole transformation, i.e. $\vp \rightarrow -\vp$.  With
$\ed=-U/2+0^+$, $\tilde\Delta (\lambda )$, the bare impurity density of states can be approximated via
\begin{equation}\label{eIIIxxxiii}
\tilde\Delta(\lambda ,\vp) = \tilde\Delta(\lambda,\vp =0) + S(\vp )\delta(\lambda - \lambda_0),
\end{equation}
provided $|\vp |$ is not much greater than $\vc$.  If in addition $U\gg \Gamma$, this approximation works for the 
entire range, $0 \leq \vp \leq \vc$, as $\vc$ is small.  We will work under this assumption.  Then from our analysis of the $H=0$
conductance we can write
\begin{equation}\label{eIIIxxxiv}
\lambda_0 \simeq {U \over 8\Gamma} - {\Gamma \over 2U}(1 - 2{\vp \over \vc});~~~0\leq\vp\leq\vc ,
\end{equation}
and
\begin{equation}\label{eIIIxxxv}
S(\vp ) = \cases{
-1, & for $\vp = 0^-$;\cr
0,  & for $\vp = 0$;\cr
1,  & for $0<\vp<\vc$;\cr
0,  & for $\vp = \vc$;\cr
-1, & for $\vp = \vc + 0^+$.}
\end{equation}
In the expression for $\lambda_0$, we have used a linear extrapolation in $\vp$ between the value of $\lambda_0$
at $\vp =0$ and $\lambda_0$ at $\vp =\vc$.

\begin{figure*}
\vskip .58in
\begin{center}
\noindent
\psfrag{y}{$G(H)$}
\psfrag{x}{$\hskip .9in H/T_k$}
\epsfysize=0.35\textwidth
\epsfbox{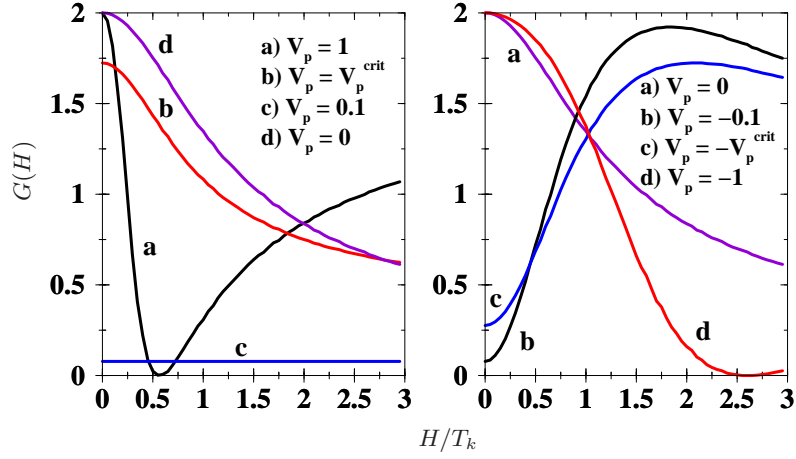}
\end{center}
\caption{Plots of the magnetoconductance at $\epsilon_d = -U/2$  
as a function of $H$ for a number of a values of $V_p$.  In these plots, we have taken
$U=1$ and $\Gamma = 0.1$.}
\end{figure*}

To compute the magnetoconductance at $H\neq 0$ and $\ed = -H/2$, we need to solve ultimately the integral
equations
\begin{eqnarray}\label{eIIIxxxvi}
\ri (q) &=& \Delta (q) + 
g'(q) \int^\infty_{-\infty} d\la  a_1(g(q)-\la) \si (\la); \cr\cr
\si (\la ) &=& \tilde{\Delta}(\la)
- \int^\infty_{-\infty} d\la 
a_2(\la '-\la)\si (\la ') \cr\cr
&& \hskip .25in - \int^B_{-D} dq a_1(\la-g(q))\ri (q).
\end{eqnarray}
As the limits of the integrals involving $\si (\la )$ run from $-\infty$ to $+\infty$, the
above equations can be recast into a single equation:
\begin{eqnarray}\label{eIIIxxxvii}
\ri (q) = \Delta (q) &+& g'(q)\int^\infty_{-\infty}d\la \tilde\Delta(\la ) s(g(q)-\la )\cr\cr
&&\hskip -.8in - g'(q)\int^B_{-D}dq R(g(q)-g(q'))\rho_{\rm imp}(q),
\end{eqnarray}
where
\begin{figure}[b]
\vskip .58in
\begin{center}
\noindent
\psfrag{y}{$\Delta G (H)$}
\psfrag{x}{$H/T_k$}
\epsfysize=0.35\textwidth
\epsfbox{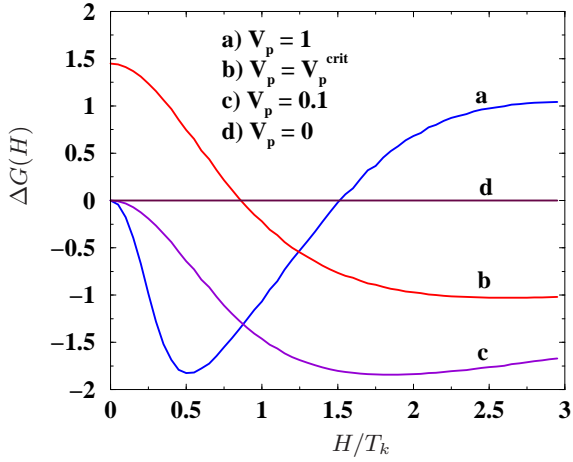}
\end{center}
\caption{Plots of the discontinuity in the magnetoconductance at $\epsilon_d = -U/2$  
as a function of $H$ for a number of a values of $V_p$.  Here the discontinuity is
defined as $\Delta G(H) = G(\epsilon_d = -U/2 + 0^+,H) - G(\epsilon_d = -U/2 + 0^-,H)$.
In these plots, we have taken $U=1$ and $\Gamma = 0.1$.}
\end{figure}
\begin{eqnarray}\label{eIIIxxxviii}
s(\la ) &=& {1\over 2}{1\over \cosh (\pi\la )};\cr\cr
R(\la ) &=& {1\over 2\pi}\int d\omega e^{i\omega\lambda }{1\over 1 + e^{|\omega |}}.
\end{eqnarray}
Needed information on $\si (\la)$ can extracted from the relation
\begin{equation}\label{eIIIxxxix}
\int^\infty_{-\infty} d\la \si (\la ) = {1\over 2}\int^\infty_{-\infty}\tilde\Delta (\la) -{1\over 2}\int^B_{-D}\ri (q).
\end{equation}
To solve the integral equation (\ref{eIIIxxxvii}), we note that using (\ref{eIIIxxxiii}) and the results
borrowed from Ref. \onlinecite{wie}, the source term of (\ref{eIIIxxxvii}) can be recast as 
\begin{eqnarray}\label{eIIIxl}
\Delta (q) &+& g'(q)\int^\infty_{-\infty}d\la \tilde\Delta(\la ) s(g(q)-\la ) \cr\cr
&&\hskip -.8in = g'(q)\big(S(\vp )s(g(q)-\lambda_0)-s(g(q)-I^{-1})\big).
\end{eqnarray}
where 
$$
I^{-1}= {U \over 8\Gamma} - {\Gamma \over 2U},
$$
and again we assume $\vp$ to be small.
With this observation, the integral equation can be solved and $\int^B_{-D}dq\rho_{\rm imp}(q)$ determined
from Wiener-Hopf techniques on exactly the same lines as in Ref. \onlinecite{wie}.  The result is
\begin{eqnarray}\label{eIIIxli}
\int^B_{-D}dq\rho_{\rm imp}(q) &=&
{\sqrt{2} \over \pi} \sum^\infty_{n=0} 
{G_+(i\pi (2n+1))\over (2n+1)} e^{-b\pi(2n+1)}\cr\cr
&& \hskip -.5in  \times (-1)^n\bigg(e^{\pi (2n+1)/I)} - S(\vp )e^{\pi\lambda_0(2n+1)}\bigg);\cr\cr
b &=& {1\over\pi}\log ({2\over H}\sqrt{U\Gamma\over\pi e}).
\end{eqnarray}
We point out that for $U/\Gamma \gg 1$, $I^{-1} = \lambda_0$.  Thus when $0 < \vp < \vc$ and $S(\vp ) =1$,
$\int^B_{-D}dq\rho_{\rm imp}(q) \simeq 0$.

\begin{figure*}
\vskip .58in
\begin{center}
\subfigure[$\vp=0$]{
\epsfysize=0.85\textwidth
\includegraphics[height=7.5cm]{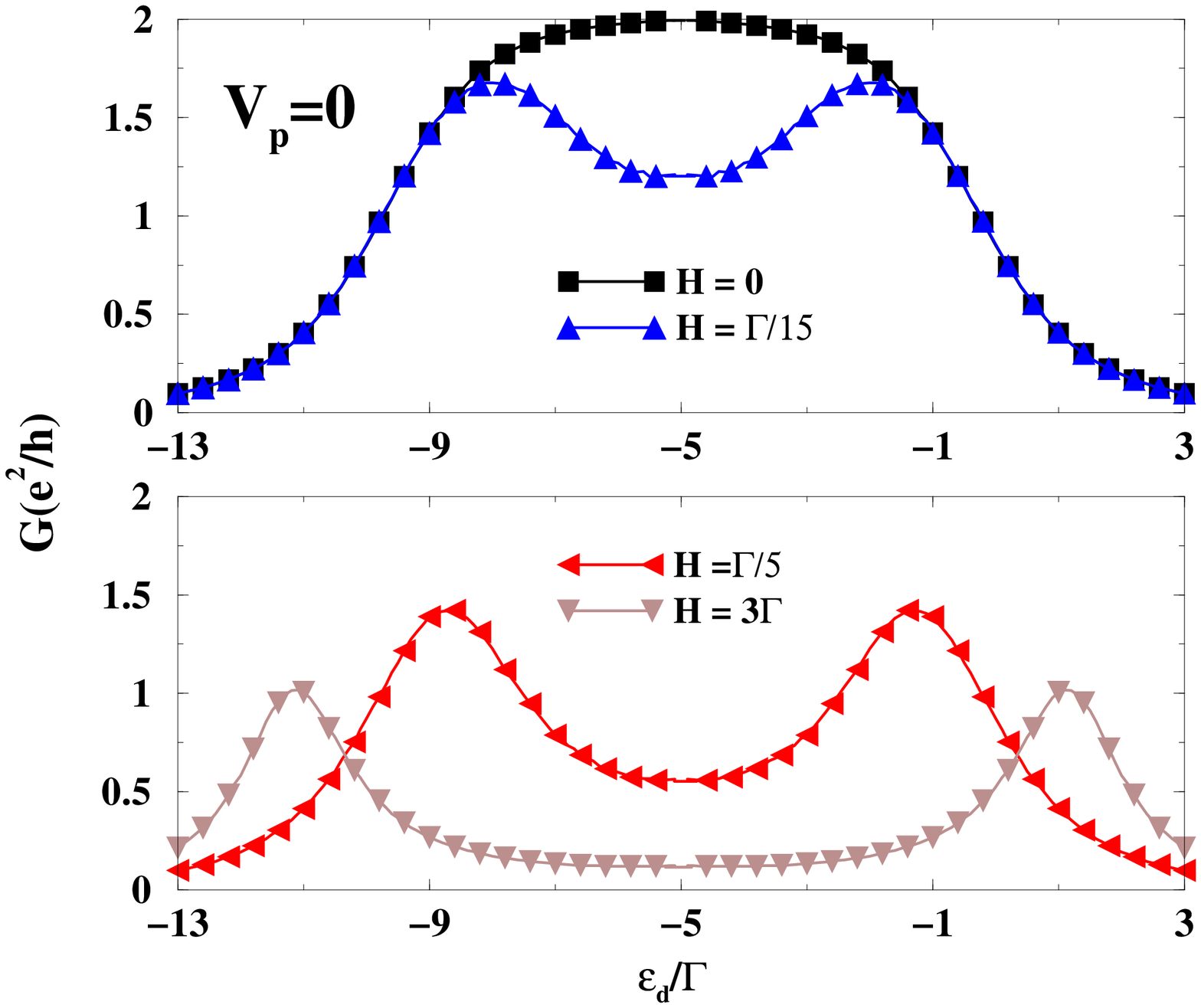}}
\subfigure[$\vp=0.1$]{
\epsfysize=0.85\textwidth
\includegraphics[height=7.5cm]{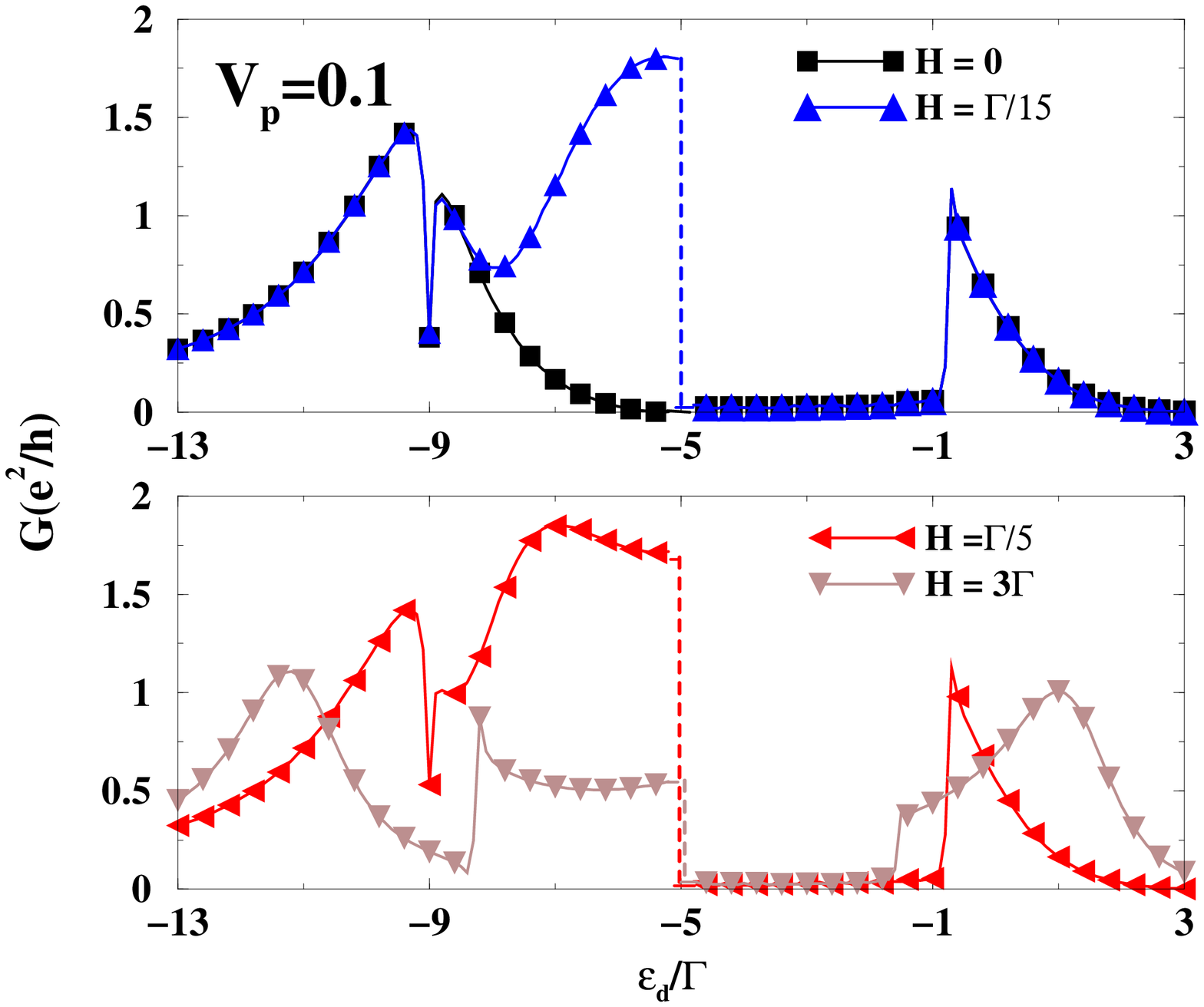}}
\end{center}
\begin{center}
\subfigure[$\vp=\vc\simeq 0.19$]{
\epsfysize=0.85\textwidth
\includegraphics[height=7.5cm]{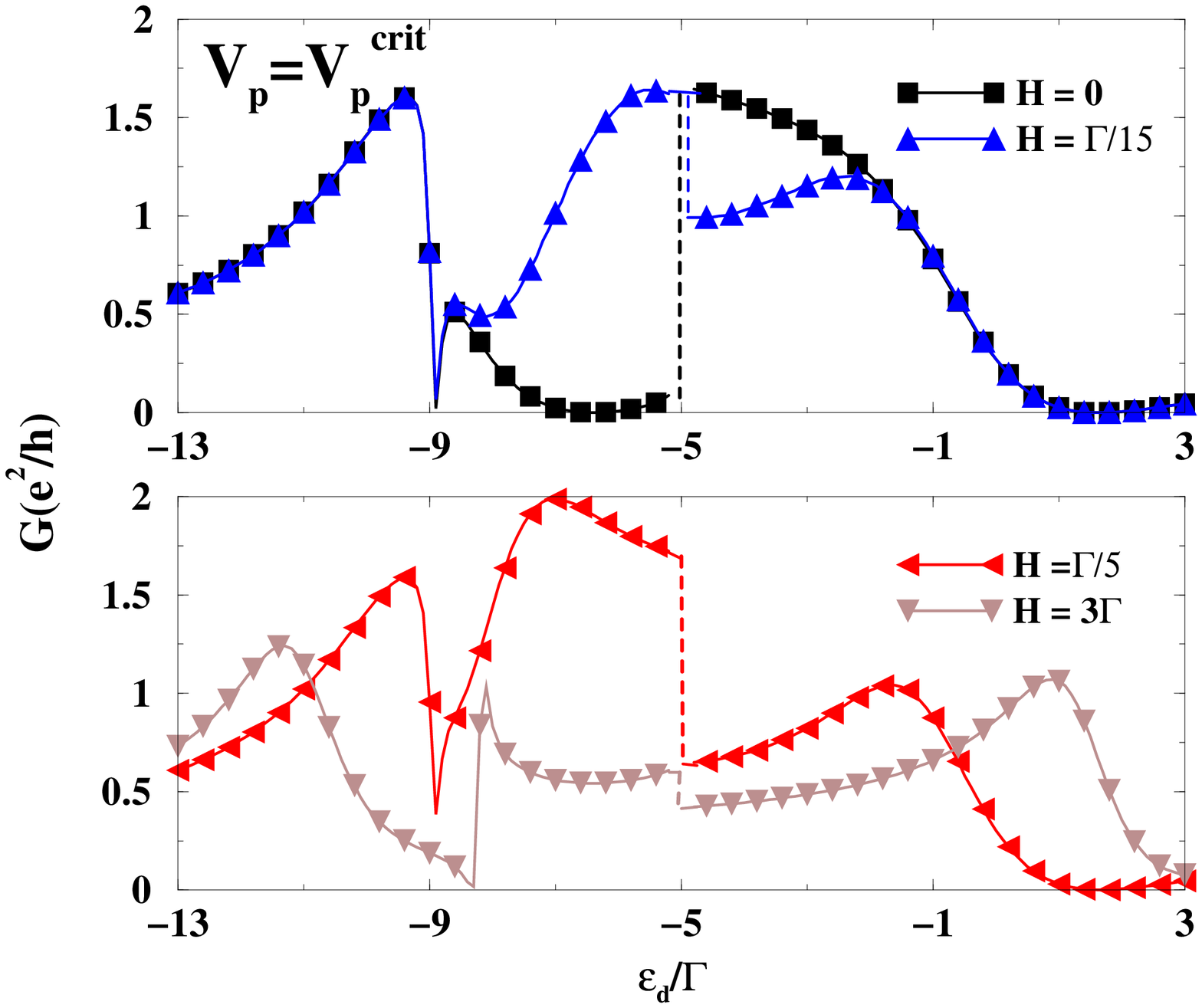}}
\subfigure[$\vp=1$]{
\epsfysize=0.85\textwidth
\includegraphics[height=7.5cm]{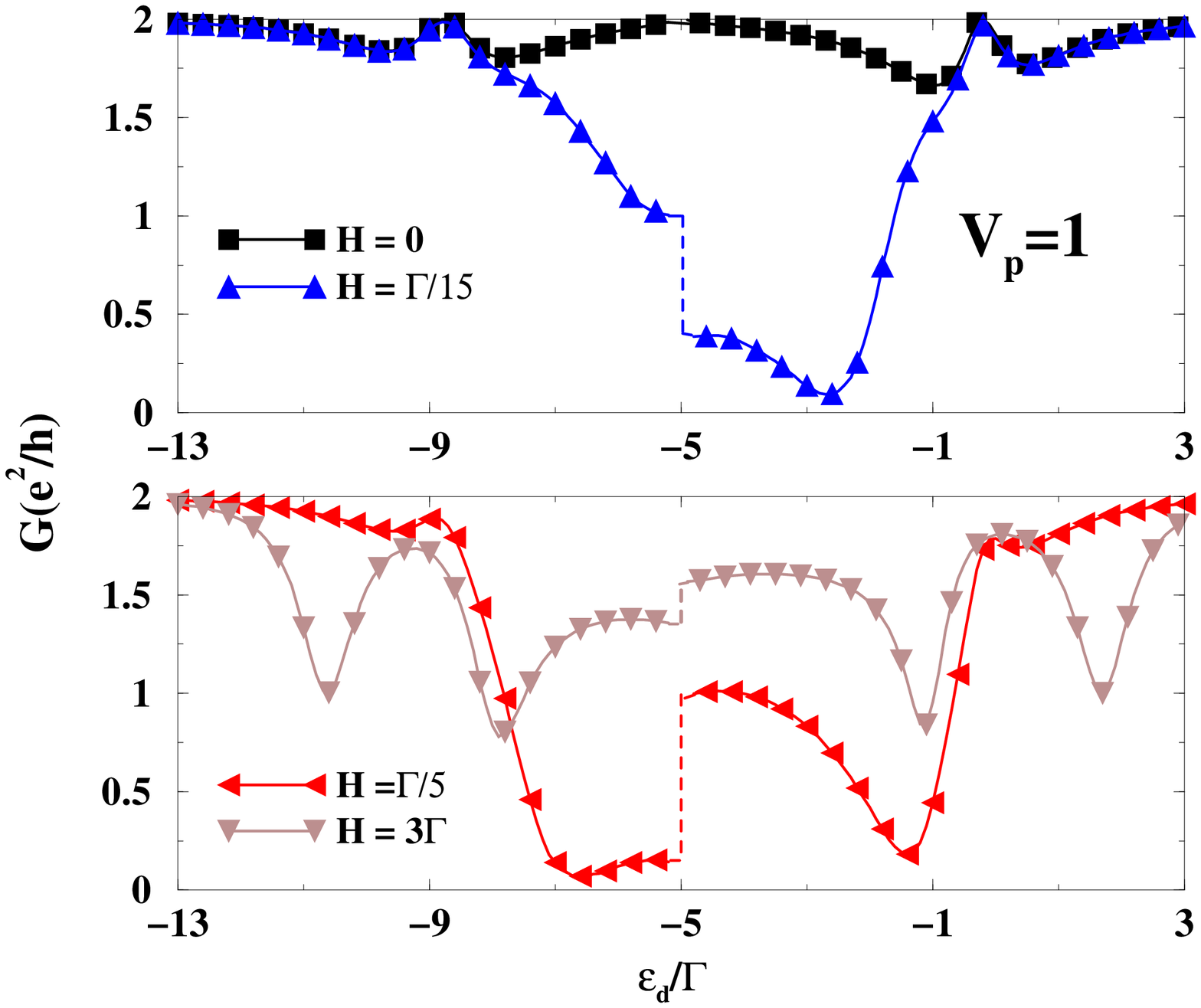}}
\end{center}
\caption{Plots of the $\vp = 0,0.1,0.19,$ and $1$ magnetoconductance as a function of $\epsilon_d$  
for a number of values of $H$.  In these plots, we have taken $U=1$ and $\Gamma = 0.1$.}
\end{figure*}

For small $H/T_k$, where the Kondo temperature $T_k$ (at $\ed = -U/2$) is defined by\cite{haldane,wie}
\begin{equation}\label{eIIIxlii}
T_k = \sqrt{U\Gamma \over 2}e^{-\pi I^{-1}},
\end{equation}
we can arrive at explicit expression for $G(\vp,\ed=-U/2+0^+)$.  With small $H/T_k$, the sum in
(\ref{eIIIxli}) can be approximated by its first term.  So using (\ref{eIIix}), (\ref{eIIxlvii}), (\ref{eIIIxxxix}),
and (\ref{eIIIxli}), we can write down the following
\begin{widetext}
\begin{equation}\label{eIIIxliii}
G(\vp,\ed=-{U\over 2}+0^+,{H \over T_k} \ll 1) 
= \cases{ 2{e^2 \over h} \big({\pi^2 H^2 \over 4T_k^2}{1+\vp^4-6\vp^2 \over (1+\vp^2)^2} 
+ {4\vp^2 \over (1+\vp^2)^2}\big) , & for $\vp < 0, \vp > \vc + 0^+$;\cr
2{e^2 \over h}\big (1 - {\pi^2 \over 16}{H\over T_k}^2\big), & for $\vp = 0$;\cr
2{e^2 \over h}{4\vp^2 \over (1+\vp^2)^2}, & for $0<\vp < \vc$;\cr
2{e^2 \over h}\big(1 - {\pi^2 H^2 \over 16 T_k^2} + {\pi^2H^2 \over 2T_k^2}{\vp^2 \over (1+\vp^2)^2} 
- {4\vp^2 \over (1+\vp^2)^2}\big), & for $\vp = \vc$.}
\end{equation}
\end{widetext}
We see the deviations in $G$ from its zero field value are quadratic in $H/T_k$.  This indicates that regardless
of the value of $\vp$, the ground state is Fermi liquid-like at $\ed = -U/2+0^+$.  Thus the discontinuous changes
as $\vp$ is varied mark first order transitions between Fermi liquid states.

On the basis of these expressions, we are able to write down analytic expressions for the discontinuity
in the conductance occurring at small but finite $H$ and $\vp$
\begin{widetext}
\begin{eqnarray}\label{eIIIxliv}
\Delta G &=& G(\vp,\ed=-{U\over 2}+0^+) - G(\vp,\ed=-{U\over 2}+0^-) \cr\cr
&=& G(\vp,\ed=-{U\over 2}+0^+) - G(-\vp,\ed=-{U\over 2}+0^+) \cr\cr
&=& \cases{0{e^2 \over h} , & for $\vp = 0$;\cr
-2{e^2 \over h}\big( {\pi^2 H^2 \over 4T_k^2}{1+\vp^4-6\vp^2 \over (1+\vp^2)^2} \big), & for $0<\vp < \vc$;\cr
2{e^2 \over h}\bigg(1 - {8\vp^2 \over (1+\vp^2)^2} + {\pi^2H^2 \over T_k^2}\big({\vp^2 \over 2(1+\vp^2)^2} - {1\over 16}
- {1\over 4}{1+\vp^4-6\vp^2 \over (1+\vp^2)^2}\big)\bigg), & for $\vp = \vc$;\cr
0{e^2 \over h}, & for $\vp = \vc + 0^+$.}
\end{eqnarray}
\end{widetext}
For $0<\vp <\vc$, the jump in the conductance develops as $H^2/T_k^2$ from $0$.  At $\vp = \vc$ the magnetic
field decreases the size of the jump.  At $\vp = \vc + 0^+$ the discontinuity disappears for small $H/T_k$.  But
for larger values of $\vp$, as evidenced by Figure 20, the jump reappears.

\subsubsection{Finite H Linear Response Conductance Away from $\ed = -U/2$}

Away from $\ed = -U/2$, the transmission amplitude of the electrons is determined by 
(\ref{eIIxlvii}) where
now $Q \neq -\infty$.  In general these equations do not admit analytic solutions.  Nonetheless we can solve them
numerically.  In Figure 21 we plot the linear response conductance for a variety of values of $H$ and $\vp$ as
a function of $\ed$.  We take $\vp > 0$ (again $\vp < 0$ can be obtained via a particle-hole 
transformation, $G(\ed,\vp ) = G(-U/2-\ed,-\vp)$).

In Figure 20a as a reference we have plotted
$G$ as a function of $\ed$ at $\vp=0$ for successive
values of $H$.  We see that each plot is symmetric about $\ed = -U/2$ (as it must be for $\vp=0$ by 
particle-hole symmetry).  As $H$ is increased, the conductance peak at $\ed = -U/2$ splits into
two smaller peaks.  For large $H$, these peaks occur at $\ed = H/2$ and $\ed =-U-H/2$ with a conductance
$G \simeq e^2/h$.  The peaks may be modeled by Breit-Wigner resonances, i.e.
\begin{eqnarray}\label{eIIIxlv}
G(\ed, H \gg T_k) &=& {e^2 \over h} \bigg(\sin^2\big(\tan^{-1}({\Gamma \over \ed-H/2})\big)\cr\cr
&& + \sin^2\big(\tan^{-1}({\Gamma \over \ed+U+H/2})\big)\bigg)\cr\cr
&=& {e^2 \over h} {\Gamma^2 \over (\ed-H/2)^2+\Gamma^2}\cr\cr
&& \hskip -.3in + {e^2 \over h} {\Gamma^2 \over (\ed+U+H/2)^2+\Gamma^2}.
\end{eqnarray}
Thus $G$ appears as if due to two species of spinless fermions, one with a resonant level
at $\ed = H/2$ and one with a resonant level at $\ed = -H/2-U$.

Moving to finite $\vp$, the behaviour of $G$ as a function of $\ed$ radically changes.  No longer
symmetric about $\ed = -U/2$, $G$ now sees sharp variations.  These variations arise for the same reason
variations at $H=0$ happen.  They thus occur at gate voltages corresponding to
\begin{eqnarray*}\label{eIIIxlvi}
\ed = \ec = -U/2 + \sqrt{U\over \Gamma}{U\over 2^{5/2}} - {\sqrt{2U\Gamma} \over 2\pi}\log(2\pi e {U\over 8\Gamma})
\end{eqnarray*}
and
\begin{equation}
\ed = -{U\over 2}-\ec .
\end{equation}
These variations are absent only for $\vp = \vc, \ed > -U/2$ and $\vp=-\vc, \ed < -U/2$.  
As discussed previously, $\vp=\vc$ is the only
non-zero value of $\vp$ at which $\tilde\Delta(\lambda ,\vp)$ is relatively smoothly varying.

From Figure 21, we see the discontinuities in the magnetoconductance at $\ed=-U/2$ discussed previously.
As $H$ is increased, the discontinuities grow smaller.  In fact one can check (numerically at least)
the discontinuities disappear in the regime $H\gg \Gamma$.  This suggests that large $H$ destroys
all non-perturbative physics.  Indeed for $H\gg \Gamma$, $G(\ed,H\gg T_k)$ behaves as
\begin{eqnarray}\label{eIIIxlvii}
G(\ed,H\gg T_k) &=& \cr\cr
&& \hskip -1.1in {e^2\over h}\bigg(\sin^2\big(\tan^{-1}({\Gamma \over \ed-H/2}+\vp) + \tan^{-1}(\vp )\big)\cr\cr
&& \hskip -1.35in + \sin^2\big(\tan^{-1}({\Gamma \over \ed+H/2+U}+\vp) + \tan^{-1}(\vp )\big)\bigg),\cr
&&
\end{eqnarray}
in analogy with (\ref{eIIIxlv}).

\section{Attractive U Analysis}

In this section we will consider the $U<0$ Anderson model.  In doing so we will demonstrate that the
non-analyticities present at $U>0$ and $\vp\neq 0$ disappear.  This indicates the problem
is non-perturbative in $U$.

The Bethe ansatz equations describing the quantization of momentum are the same as in (\ref{eIIxxxiii}).  However
for $U<0$, the solutions of the equations are of a different character.  At $U>0$, the ground state consists
of bound states characterized by complex pairs of wavevectors, $q_\pm$, tied to particular values of $\lambda$ (the spin momenta).  
But at $U<0$, the ground
state is composed of real values of $q$ independent (nominally) of any specific $\lambda$.
As in the $U>0$ case, we can write down a set of equations describing
the distributions, $\rho (q)$ and $\sigma (\lambda )$, of $k$'s and $\lambda$'s in the continuum limit of
the ground state:
\begin{eqnarray}\label{eIVi}
\rho (q) &=& {1\over 2\pi} + {\Delta (q) \over L} \cr\cr
&+& g'(q)\int^Q_{\tilde Q}d\la a_1(\la - g(q))\sigma (\la );\cr\cr
\sigma(\la ) &=& -\int^Q_{\tilde Q}d\la' \sigma (\la' )a_2(\la -\la')\cr\cr
&& \hskip .4in +\int^B_{-D}dq\rho (q)a_1(\la - g(q)).
\end{eqnarray}
As with $U>0$, the limits $-D$ and $\tilde Q$ mark the bandwidths of the $k$ and the $\lambda$-distributions.
But now $\tilde Q$ occurs at $\lambda << 0$ -- in fact it can be shown to be given by
solving
$$
|U|/2 +\ed-\sqrt{|U|\Gamma}(-\tilde Q + (\tilde Q^2+1/4)^{1/2})^{1/2} = -D.
$$  
The remaining two limits, $B$ and $Q$,
mark the Fermi surfaces of the two distributions and so are characterized by
\begin{eqnarray}\label{eIVii}
\int^B_{-D} dq \rho (q) &=& {D+\mu \over \pi};\cr\cr
\int^Q_{\tilde Q}d\la \sigma (\la ) &=& {D+\mu \over 2\pi} - S_z = {D+\mu \over 2\pi} - {H\over 2\pi}.
\end{eqnarray}
Here $\mu$ is the chemical potential.  We introduce $\mu$ for technical reasons.  If one wants to explore
the effects of varying $\ed$, one can alternatively vary $\mu$ keeping $\ed$ fixed as it is only the relative
distance between the dot energy level and the lead-electron Fermi surface that is relevant to the physics.
As such for the purposes of any actual computations, we fix $\ed$ at $-U/2$ and simply vary $\mu$.

As before we can divide these densities into a bulk and an impurity piece via writing
$\rho (q) = \rho_{\rm bulk}(q)+\rho_{\rm imp}(q)/L$ and $\sigma (\la) = \sigma_{\rm bulk}(\la ) + \sigma_{\rm imp}(\la )/L$:
\begin{eqnarray}\label{eIViii}
\rho_{\rm bulk} (q) &=& {1\over 2\pi} + g'(q)\int^Q_{\tilde Q}d\la a_1(\la - g(q))\sigma_{\rm bulk} (\la );\cr\cr
\sigma_{\rm bulk}(\la ) &=& -\int^Q_{\tilde Q}d\la' \sigma_{\rm bulk} (\la' )a_2(\la -\la')\cr\cr
&& \hskip 0.in + \int^B_{-D}dq\rho_{\rm bulk} (q)a_1(\la - g(q)),
\end{eqnarray}
and
\begin{eqnarray}\label{eIViv}
\rho_{\rm imp} (q) &=& \Delta (q) + g'(q)\int^Q_{\tilde Q}d\la a_1(\la - g(q))\sigma_{\rm imp} (\la );\cr\cr
\sigma_{\rm imp}(\la ) &=& -\int^Q_{\tilde Q}d\la' \sigma_{\rm imp} (\la' )a_2(\la -\la')\cr\cr
&& \hskip 0.in + \int^B_{-D}dq\rho_{\rm imp} (q)a_1(\la - g(q)).
\end{eqnarray}
The impurity spin and occupancy are given by
\begin{eqnarray}\label{eIVv}
n_{\rm imp\ua}+n_{\rm imp\da} &=& \int^B_{-D}dq\rho_{\rm imp}(q);\cr\cr
&& \hskip -1.1in  n_{\rm imp\ua} - n_{\rm imp\da} \equiv 2S^z_{\rm imp} =\cr\cr 
&& \hskip -.7in = -2\int^Q_{\tilde Q}d\la \sigma_{\rm imp}(\la ) 
+ \int^B_{-D}dq\rho_{\rm imp}(q).
\end{eqnarray}
The bulk densities, as was the case for $U>0$, are independent of $\vp$.  As such the bulk densities
can be related to those for $U>0$ via a spin-$\da$ particle-hole transformation.  This will not be
exploited here but is discussed in some detail in Ref. \onlinecite{wie}.

We can conclude on the basis of this description of the $U<0$ ground state alone that there are
no non-analyticities as a function of $\vp$ in the various associated impurity quantities.  $\vp$
appears solely in $\Delta (q)$ and $\Delta (q)$ is a smooth function of $\vp$.  Nonetheless
we will compute the scattering phases of electrons at the Fermi surface and show they have a similar
smooth dependence upon $\vp$.  This will also allow a comparison between our and the Dyson equation
approach of Section V.

To determine the scattering phases of electrons off the impurity, it is necessary to isolate the impurity
portion of the momentum.  The total momentum is available from taking logarithms of the Bethe ansatz
equations (\ref{eIIxxxiii}) (up to possible constant terms)
\begin{eqnarray}\label{eIVvi}
p(q) &=& {2\pi N \over L} = q + {\delta (q) \over L} +{1\over L}\sum^M_{\beta =1}\theta_1(g(q)-\lambda_\beta)\cr\cr
&=& q + {\delta (q)\over L} + \int^Q_{\tilde Q} \sigma (\la ) \theta_1(g(q) - \la) ;\cr\cr
p(\la ) &=& {2\pi J \over L} = -{1\over L}\sum^M_{\beta =1}\big(\theta_2(\la - \la')+2\pi\big) \cr\cr
&& \hskip .4in + {1\over L}\sum^N_{\alpha =1}\big(\theta_1(\la - g(q_\alpha)) + 2\pi\big)\cr\cr
&& \hskip -.4in = -\int^Q_{\tilde Q} \sigma (\la ') \big(\theta_2 (\la - \la ') + 2\pi\big) \cr\cr
&& \hskip .1in + \int^B_{-D}dq \rho (q) \big(\theta_1 (\la - g(q)) + 2\pi\big) .
\end{eqnarray}
The impurity momenta are then found by isolating the $1/L$ pieces:
\begin{eqnarray}\label{eIVvii}
p_{\rm imp}(q) &=& \delta (q) + \int^Q_{\tilde Q} \si (\la ) 
\theta_1(g(q) - \la) ;\cr\cr
p_{\rm imp}(\la ) &=&  - \int^Q_{\tilde Q} \si (\la ') \big(\theta_2 (\la - \la ') + 2\pi\big)\cr\cr
&& \hskip -.3in + \int^B_{-D}dq \ri (q) \big(\theta_1 (\la - g(q)) + 2\pi\big).
\end{eqnarray}
We have chosen the branch cuts of the arctans here in order to ensure that
the impurity momenta take on their bare values at the band edges, i.e.
$p_{\rm imp}(q=-\infty ) = -2\tan^{-1}(\vp )$ and $ p_{\rm imp}(\lambda = -\infty) = 0$.

We again have a relationship between the impurity momenta and the impurity density of states:
\begin{eqnarray}\label{eIVviii}
\sigma_{\rm imp}(\la ) &=& {1\over 2\pi}\partial_\lambda p_{\rm imp} (\la );\cr\cr
\rho_{\rm imp}(q) &=& {1\over 2\pi}\partial_q p_{\rm imp} (q).
\end{eqnarray}
These relations will allow us to express again the scattering phase of an electron off the impurity in
terms of the impurity occupancy so verifying a variant of the Friedel sum rule holds for $U<0$ and $\vp$
finite.

To compute the scattering phase of an electron one must specify the gluing rules between the charge, $q$,
and spin, $\lambda$, excitations.  For simplicity let us focus on the scattering phase of a spin $\ua$ electron.
In adding a spin $\ua$ electron to the system, we increase the particle number, $N$, by 1 thus forcing an additional
real $q$ to be added to the ground state.  We, in addition, open up a hole in the $\lambda$-distribution.
A spin-$\ua$ electron is then composed of a charge $q$-excitation and a spin $\la$-hole.

The momentum of the added electron is then a sum of the momentum of its component pieces, $p_{\rm el} = p(k) - p(\lambda )$,
and so the corresponding scattering phase (i.e. the impurity portion of $p_{\rm el}$) is
\begin{equation}\label{eIVix}
\delta_{\rm el\ua} = p_{\rm imp}(k) - p_{\rm imp}(\la ).
\end{equation}
By virtue of the relations (\ref{eIVviii}), this can be recast as
\begin{eqnarray}\label{eIVx}
\delta_{\rm el\ua} &=& \int^B_{-D} dq \partial_q p_{\rm imp} + p_{\rm imp}(\lambda = -D )
- \int^Q_{\tilde Q} \partial_Q p_{\rm imp}(\la ) \cr\cr
&=& 2\pi \int^B_{-D} dq \rho_{\rm imp}(q) - 2\pi\int^Q_{\tilde Q} d\la \sigma_{\rm imp} (\la ) \cr 
&& \hskip 1.9in -2\tan^{-1}(\vp)\cr\cr
&=& 2\pi n_{\rm imp \ua}  -2\tan^{-1}(\vp) .
\end{eqnarray}
Thus we have established a relationship between the scattering phase and the number of electrons
displaced by the impurity.

In the case $U>0$, we argued that a discontinuous change in the impurity occupancy was mirrored in a discontinuous
change in the number of electrons occupying the ground state as revealed by exact diagonalization computations.
We also argued that the discontinuity in $n_{\rm imp}$ as a function of $\vp$ was intimately related
to a $U>0$ ground state being composed of two-particle bound states.
Given that the $U<0$ ground state is composed of a sea of single particle states, how is the lack of a discontinuous change
in $n_{\rm imp}(U<0)$ to be understood?

Regardless of the sign of $U$, discontinuous changes in the ground state occupancy, $n^{\rm ED}_{\rm occ}$, 
as a function of $\vp$
are seen in the exact diagonalization studies.  With the $U<0$ ground state being composed of a sea
of single particle states, changes in $n^{\rm ED}_{\rm occ}$ are able to be reflected solely in 
$n^{\rm BA}_{\rm bulk} \equiv n_{\rm bulk}$,
leaving $n^{\rm BA}_{\rm imp} \equiv n_{\rm imp}$ untouched.  If $\Delta n^{\rm ED}_{\rm occ} = \pm 1$ then
$\Delta n^{\rm BA}_{\rm bulk}$ can by adjusted by $1$ as the composition of the ground state permits such
changes, in contrast to the $U>0$ ground state composed of two-particle bound states only permitting changes
satisfying $\Delta n^{\rm BA}_{\rm bulk} = \pm 2$.  In this latter case we needed to alter $\Delta n^{\rm BA}_{\rm imp}$ by $\mp 1$
in order to satisfy $\Delta n^{\rm BA}_{\rm bulk} + \Delta n^{\rm BA}_{\rm imp} = \Delta n^{\rm ED}_{\rm occ} = \pm 1$.

\section{Conductance from a Dyson Equation Analysis}

In this section we consider an alternative approach relying on Dyson equations to compute the linear
response conductance.  This approach leads to results for the conductance, $G$, considerably different
from those of the Bethe ansatz presented in this article.
We will show that this approach, used in Refs. \onlinecite{bulka} and \onlinecite{hofstetter}, 
is equivalent to employing the transformation in (\ref{eIxvi}) by which the potential scattering
in the Hamiltonian is gauged away.  We then argue that because the gauge transformation incorrectly
computes certain finite $U$ quantities, so must the Dyson equations.
We will further argue that the methodology
assumes implicitly
both $\vp$ and $U$ are perturbative quantities and so cannot possibly
capture the non-perturbative physics we believe present in this problem.

\subsection{Overview of Approach}
We will begin by giving a review of the methodology used in Refs. \onlinecite{bulka} and \onlinecite{hofstetter}
as few details appeared in the two letters.
In this approach the current is written as
\begin{eqnarray}\label{eVi}
I &=& {1\over 2}e\partial_t\langle n_R-n_L\rangle\cr\cr
&=& i{e\over 2h}\langle [{\cal H},n_R]-[{\cal H},n_L]\rangle,
\end{eqnarray}
where $n_{R/L}=\sum_\sigma\int dx c^\dagger_{\sigma R/L}(x)c_{\sigma R/L}(x)$ and the
Hamiltonian, $\cal H$, is given by (\ref{eIIi}).  For the purpose of simplicity, we take
$V_L=V_R=V$ and $V_{LL}=V_{RR}=0$.  Thus $I$ can be written as
\begin{eqnarray}\label{eVii}
I &=& -{e\over 2h}V\sum_\sigma \int d\omega 
\bigg(G^{-+}_{Rd}(\om)-G^{-+}_{dR}(\om)\cr\cr
&&\hskip .4in -G^{-+}_{Ld}(\om)+G^{-+}_{dL}(\om)\bigg)\cr\cr
&& -{e\over h}V_{LR}\sum_\sigma \int d\om \bigg(G^{-+}_{RL}(\om)-G^{-+}_{LR}(\om)\bigg),
\end{eqnarray}
where $G^{+-}$ are the standard Keldysh correlators, for example,
\begin{eqnarray}\label{eViii}
G^{-+}_{Rd}(t) &=& i\langle c^\dagger_R(x=0,t=0)d(t)\rangle;\cr\cr
G^{-+}_{RL}(t) &=& i\langle c^\dagger_R(x=0,t=0)c_L(x=0,t)\rangle.
\end{eqnarray}
The remaining ($RL$) Keldysh correlators are given by (so as to fix conventions),
\begin{eqnarray}\label{eViv}
G^{--}_{RL}(t) &=& -i\theta(t)\langle c_L(0,t)c^\dagger_R(0,0)\rangle \cr 
&& + i\theta(-t)\langle c^\dagger_R(0,0)c_L(0,t)\rangle;\cr\cr
G^{++}_{RL}(t) &=& -i\theta(-t)\langle c_L(0,t)c^\dagger_R(0,0)\rangle \cr 
&& + i\theta(t)\langle c^\dagger_R(0,0)c_L(0,t)\rangle;\cr\cr
G^{+-}_{RL}(t) &=& -i\langle c_L(0,t)c^\dagger_R(0,0)\rangle .
\end{eqnarray}
As here we are interested in the zero field transport, we suppress all spin indices appearing in
the correlators.

The next step in the development of this approach is to recast all of the correlators
appearing in (\ref{eVii}) in terms of full interacting correlators involving only dot degrees
of freedom ($d/d^\dagger$) together with non-interacting (i.e. $V=0$) correlators of the lead
fermions.  To this end we can write
\begin{widetext}
\begin{eqnarray}\label{eVv}
G^{-+}_{dR}(V,\om ) - G^{-+}_{Rd}(V,\om ) &=& V\bigg(
G^{-+}_{RR}(0,\om )\big(G^{--}_{dd}(V,\om )+G^{++}_{dd}(V,\om )\big)
-G^{-+}_{dd}(V,\om )\big(G^{++}_{RR}(0,\om )+G^{--}_{RR}(0,\om )\big)\bigg)\cr\cr
&& \hskip -.7in + V\bigg(
G^{-+}_{LR}(0,\om )G^{--}_{dd}(V,\om )+G^{-+}_{RL}(0,\om )G^{++}_{dd}(V,\om )
-G^{-+}_{dd}(V,\om )\big(G^{++}_{LR}(0,\om )+G^{--}_{RL}(0,\om )\big)\bigg);\cr\cr\cr
G^{-+}_{RL}(V,\om ) - G^{-+}_{LR}(V,\om ) &=& G^{-+}_{RL}(0,\om ) - G^{-+}_{LR}(0,\om )\cr\cr
&& \hskip -.3in + V^2G^{++}_{dd}(V,\om )\bigg[\big(G^{-+}_{LL}(0,\om ) + G^{-+}_{RL}(0,\om )\big)
\big(G^{++}_{RL}(0,\om ) + G^{++}_{RR}(0,\om )\big) - \big(R\leftrightarrow L\big)\bigg]\cr\cr
&& \hskip -.3in - V^2G^{+-}_{dd}(V,\om )\bigg[\big(G^{-+}_{LL}(0,\om ) + G^{-+}_{RL}(0,\om )\big)
\big(G^{-+}_{RL}(0,\om ) + G^{-+}_{RR}(0,\om )\big) - \big(R\leftrightarrow L\big)\bigg]\cr\cr
&& \hskip -.3in - V^2G^{-+}_{dd}(V,\om )\bigg[\big(G^{--}_{LL}(0,\om ) + G^{--}_{RL}(0,\om )\big)
\big(G^{++}_{RL}(0,\om ) + G^{++}_{RR}(0,\om )\big) - \big(R\leftrightarrow L\big)\bigg]\cr\cr
&& \hskip -.3in + V^2G^{++}_{dd}(V,\om )\bigg[\big(G^{--}_{LL}(0,\om ) + G^{--}_{RL}(0,\om )\big)
\big(G^{-+}_{RL}(0,\om ) + G^{-+}_{RR}(0,\om )\big) - \big(R\leftrightarrow L\big)\bigg].
\end{eqnarray}
\end{widetext}
We now must compute the set of lead electron correlators at $V=0$ (but finite $\vp$).

To compute such correlators, we combine the equations of motion,
\begin{eqnarray}\label{eVvi}
\partial_t c_L &=& i[{\cal H},c_L] = -\partial_x c_L -iV_{LR}c_R\delta (x=0);\cr\cr
\partial_t c_R &=& -\partial_x c_R -iV_{LR}c_L\delta (x=0),
\end{eqnarray}
together with the mode expansions
\begin{eqnarray}\label{eVvii}
c_{L/R}(x,t) &=& \int {dk \over 2\pi}\bigg(c_{+kL/R}e^{ik(x-t)}\theta(x) \cr
&& \hskip .5in + c_{-kL/R}e^{ik(x-t)}\theta (-x)\bigg);\cr\cr
c_{L/R}(x=0,t) &=& {1\over 2}\big(c_{L/R}(x=0^+,t)+c_{L/R}(x=0^-,t)\big) .\cr
&&
\end{eqnarray}
$c_{+k}$ and $c_{-k}$ mark modes to the right and left of the impurity at $x=0$
and are governed by
$$
\langle c^\dagger_{-kL/R}c_{-k'L/R}\rangle = \delta(k-k')f_{L/R}(\om ).
$$
Here $f_L$ and $f_R$ are the Fermi
functions in the left and right leads.
Using these mode expansions together with the constraints introduced by the equations of motion, we
find for the $G_{RR}$ correlators,
\begin{eqnarray}\label{eVviii}
G^{-+}_{RR}(0,\om ) &=& {i \over (1+\vp^2)^2}\big(f_R(\om ) + \vp^2f_L(\om )\big);\cr\cr
G^{+-}_{RR}(0,\om ) &=& \cr 
&& \hskip -.6in -{i \over (1+\vp^2)^2}\big(1+\vp^2 - f_R(\om ) - \vp^2f_L(\om )\big);\cr\cr
G^{++}_{RR}(0,\om ) &=& \cr
&& \hskip -.6in {i \over 2(1+\vp^2)^2}\big(2 f_R(\om ) - 1 + \vp^2(2f_L(\om )-1)\big);\cr\cr
G^{--}_{RR}(0,\om ) &=& \cr
&& \hskip -.6in {i \over 2(1+\vp^2)^2}\big(2 f_R(\om ) - 1 + \vp^2(2f_L(\om )-1)\big),\cr
&& 
\end{eqnarray}
while we find for the $G_{RL}$ correlators,
\begin{eqnarray}\label{eVix}
&& \hskip -.5in G^{-+}_{RL}(0,\om ) = G^{+-}_{RL}(0,\om ) = G^{++}_{RL}(0,\om ) = G^{--}_{RL}(0,\om )  \cr\cr
&& \hskip .15in = -{\vp \over (1+\vp^2)^2}\big(f_L(\om ) - f_R(\om )\big).
\end{eqnarray}
Here $\vp=V_{LR}/2$, consistent with the notation used in developing the Bethe ansatz solution.
The correlators $G_{LL}$ and $G_{LR}$ can be obtained from the above by making the substitution $L\leftrightarrow R$.

There is an ambiguity in the result for the $G_{RL}$ correlators.  The $G_{RL}$ correlators
satisfy the Dyson relation
\begin{eqnarray}\label{eVx}
G^{ab}_{RL}(\om ) &=& -V_{LR}\sum_{\mu=\pm}\bigg(\mu G^{a\mu}_{LL}(\vp ,\om)G^{\mu b}_{RR}(\vp=0,\om )\bigg),\cr &&
\end{eqnarray}
with $a,b=\pm$.
Consistency with these Dyson relations demands that $G^{--}_{RL}$ and $G^{++}_{RL}$
are given not by (\ref{eVix}), but as follows
\begin{eqnarray}\label{eVxi}
G^{++\atop --}_{RL}(0,\om ) &=& -{\vp \over (1+\vp^2)^2}(f_L(\om )-f_R(\om )) \cr\cr
&& \hskip .3in \pm {1\over 2}{\vp \over (1+\vp^2)}.
\end{eqnarray}
This ambiguity is perhaps related to the fact that the potential scattering term acts at all energies
equally and has no natural cutoff.

Substituting these $V=0$ correlators ((\ref{eVviii}), (\ref{eVix}),
and (\ref{eVxi})), we arrive at an expression for the current, $I$,
given solely in terms of the retarded correlator of the dot, $G^{\rm ret}_{dd}=G^{--}_{dd}-G^{-+}_{dd}$:
\begin{eqnarray}\label{eVxii}
I &=& {2e\over h}\int d\om T(\om )\big(f_L(\om ) - f_R(\om )\big);\cr\cr
T(\om ) &=& {4\vp^2 \over (1+\vp^2)^2} + 4\vp V^2{1-\vp^2 \over (1+\vp^2)^3} {\rm Re}G^{\rm ret}_{dd}(\om )\cr\cr
&& -{V^2 \over 1+\vp^2}\big(1-{8\vp^2\over (1+\vp^2)^2}\big){\rm Im}G^{\rm ret}_{dd}(\om ).
\end{eqnarray}
This agrees with Ref. \onlinecite{hofstetter}.  By taking $f_{L/R}(\om) = f(\om\pm\mu/2 )$, the $T=0$
linear response conductance is then given by 
$$G= {2e^2 \over h}T(\om = 0).
$$

\subsection{Dyson Equation Approach and the Bethe Ansatz}

In the non-interacting case ($U=0$), we can connect the expression for the current given in (\ref{eVxii})
with the scattering phase approach
adopted in this paper.  At $U=0$, $G^{\rm ret}_{dd}(\om )$ is given by
\begin{eqnarray}\label{eVxiii}
D(\om ) &=& ((1+\vp^2)(\om -\ed)^2+2V^2(\om -\ed )\vp + V^4);\cr\cr
G^{\rm ret}_{dd}(\om ) &=& {(\om -\ed)(1+\vp^2) + V^2\vp - iV^2 \over D(\om )}.
\end{eqnarray}
And so the conductance takes the form 
\begin{eqnarray}\label{eVxiv}
G &=& 2{e^2\over h} {1\over 1+\vp^2}{1\over D(\om = 0)}\cr\cr
&& \hskip .3in \times\big( V^4+4\vp^2\ed^2-4\vp V^2\ed \big) .
\end{eqnarray}
This, as is easily checked, equals the $U=0$ conductance computed on the basis of scattering phases
\begin{eqnarray}\label{eVxv}
G &=& 2{e^2\over h}\sin^2({1\over 2}(\delta_e-\delta_o)) \cr\cr
&=& 2{e^2\over h}\sin^2\bigg(\tan^{-1}\big(-{V^2\over\ed}+\vp\big)+\tan^{-1}(\vp)\bigg).\cr &&
\end{eqnarray}
We can also check that a direct computation of the $U=0$ correlators in (\ref{eVii}) 
(bypassing the Dyson relations in (\ref{eVv})) yields an expression for the current consistent with the above.  We have
\begin{eqnarray}\label{eVxvi}
G^{-+}_{RL}(\om ) - G^{-+}_{LR}(\om ) &=& \cr
&& \hskip -2in {1\over D(\om )}
{1\over 1+\vp^2}\bigg(-2\vp(\om -\ed)^2 - V^2(\om -\ed)\bigg).\cr\cr
{1\over 2}\big(G^{-+}_{Rd}(\om ) - G^{-+}_{dR}(\om ) -(L\leftrightarrow R)\big) &=& \cr
&& \hskip -2in -{1\over D(\om )}{1\over 1+\vp^2}
\big(V^3+2\vp(\om-\ed)V\big).
\end{eqnarray}
Substituting the above into (\ref{eVii}), we again find the same expression for the conductance (\ref{eVxiv}).

Thus the two approaches can be made to agree at $U=0$.  Moreover we see that at $U=0$
computing correlators via the Dyson equations is consistent
with a direct computation (provided we insist on using the $V=0$ correlators found in (\ref{eVxi})).
But what of $U \neq 0$.
To answer this question, we recast the two predicted forms of the conductance.  If we identify
\begin{eqnarray}\label{eVxvii}
e &=& {1\over \bar{V}^2}\big(\ed + {\rm Re}\Sigma (\om =0)\big);\cr\cr
\bar{V}^2 &=& {V^2 \over 1 + \vp^2},
\end{eqnarray}
where $\Sigma (\om )$ is the self energy appearing in the full dot correlator,
\begin{eqnarray}\label{eVxviii}
G^{\rm ret}_{dd}(\om ) &=& {1\over w-\ed-\Sigma (\om )}\cr\cr
&=& {1\over w-\ed-{\rm Re}\Sigma (\om ) +i\bar{V}^2},
\end{eqnarray}
we can recast the expression for the linear response conductance, $G^{\rm Dyson}$, arising from the Dyson equations, as follows,
\begin{eqnarray}\label{eVxix}
G^{\rm Dyson} &=& 2{e^2 \over h}{4\vp^2\over 1+\vp^2}{(e+q)^2\over e^2+1};\cr\cr
q &=& -{1\over 2\vp}(1-\vp^2).
\end{eqnarray}
We point out that $e$, depending only upon the self energy of the dot correlator, is actually a quantity computable using
the Bethe ansatz for the $\vp =0$ Anderson model.  
The effect of $\vp$ upon the dot correlator is to renormalize the bare values of $\Gamma$ and $\ed$:
\begin{eqnarray}\label{eVxx}
\Gamma^{\rm eff} &=& {\bar V}^2;\cr\cr
\ed^{\rm eff} &=& \ed - {\vp \Gamma^{\rm eff}}.
\end{eqnarray}
Once this renormalization is taken into account, $\vp$ can henceforth be ignored.  Invoking the Friedel sum rule,
$e$ can be expressed as 
\begin{eqnarray}\label{eVxxi}
e &=& \cot\big({\pi\over 2}n_d(U,\ed^{\rm eff},V^{\rm eff},\vp = 0)\big),
\end{eqnarray}
where $n_d$ is the number of electrons sitting on a dot with effective parameters, $\ed^{\rm eff}$ and $V^{\rm eff}$.

We can recast the conductance arising from the Bethe ansatz calculation,
\begin{equation}\label{eVxxii}
G = 2{e^2\over h}\sin^2({1\over 2}(\delta_e-\delta_o)),
\end{equation}
in a similar form as (\ref{eVxix}) with the identification
\begin{eqnarray}\label{eVxxiii}
\tilde e &=& \cot({1\over 2}(\delta_e+\delta_o));\cr\cr
\tilde q &=& -\cot(2\tan^{-1}(\vp ) ).
\end{eqnarray}
$G$ then becomes
\begin{equation}\label{eVxxiv}
G = 2{e^2 \over h}{4\vp^2\over 1+\vp^2}{(\tilde e+\tilde q)^2\over \tilde e^2+1}.
\end{equation}
Given $\delta_e = \pi n_{\rm imp} - 2\tan^{-1}(\vp )$ and $\delta_o = 2\tan^{-1}(\vp )$, we see that
$\delta_e+\delta_o = \pi n_{\rm imp}$ where again $n_{\rm imp}$ is the total number of electrons displaced
by the impurity.

For the two approaches to agree we require
\begin{eqnarray*}\label{eVxxv}
e &=& \tilde e
\end{eqnarray*}
or
\begin{eqnarray}
\delta_e+\delta_o &=& \pi n_{\rm imp} = \pi n_d.
\end{eqnarray}
Thus we ask that the number of electrons displaced by the impurity equal the number of electrons
sitting on the dot.  As we have argued in both Section II and III, this is clearly not the case with the Bethe ansatz
computation.
The number of electrons on the dot is a well-behaved quantity of both $\vp$ and $\ed$ whereas $n_{\rm imp}$
sees numerous discontinuities.

The relations in (\ref{eVxxv}) make it possible to show that the Dyson equation approach is equivalent to gauging away
the potential scattering term as is done in (\ref{eIxvi}).  With potential scattering gone, the transformed scattering phase,
$\tilde\delta_e$, is simply equal to $\pi n_d$ by the Friedel sum rule.  To find the scattering phase
of the original electron, we undo the gauge transformation giving us
\begin{eqnarray}\label{eVxxvi}
\delta_e &=& \tilde\delta_e - 2\tan^{-1}(\vp )\cr\cr
&=& \pi n_d - 2\tan^{-1}(\vp ).
\end{eqnarray}
Comparing (\ref{eVxxv}) and (\ref{eVxxvi}) we see by identifying $\delta_e + \delta_o$ with $\pi n_d$,
the Dyson equations are implicitly carrying out
this gauge transformation.

We now argue that this gauge transformation incorrectly computes the scattering phase at finite $U$.
(On the other hand, if one was interested in quantities involving the
dot degrees of freedom alone, this gauge transformation is more than adequate.)  We are able to demonstrate
this at the level of quantum mechanics.  Consider
the two-particle eigenstate constructed in Section II.B.2.  This eigenstate is composed of two
electrons of energy $q$ and $p$.  The scattering phase, $\delta_q$, 
the electron with energy $q$ as it travels from $x=-\infty$ to $x=\infty$ is given by
\begin{eqnarray}\label{eVxxvii}
\delta_q (p,\ed,\vp ,\Gamma) &=& 
{1\over i} \log {\phi (x>0,q,p,\ed,\Gamma) \over \phi (x < 0,q,p,\ed,\Gamma)} \cr\cr 
&& + ~\delta (q,\ed,\Gamma,\vp),
\end{eqnarray}
where $\phi (x) $ is defined in (\ref{eIIxix}) and (\ref{eIIxx}) while $\delta (q)$ is given (\ref{eIIxiii}).
The first term in the above involving $\phi (x)$ is the phase due to the interaction between the two electrons.
It disappears if $U=0$.
If the gauge transformation led to equivalent results we would find that the scattering phase can be written as 
\begin{eqnarray}\label{eVxxviii}
\delta_{q} (q,\ed,\vp ,\Gamma) &=& {1\over i} 
\log {\phi (x>0,q,p,\ed^{\rm eff},\Gamma^{\rm eff}) \over \phi (x < 0,q,p,\ed^{\rm eff},\Gamma^{\rm eff})} \cr\cr
&& \hskip -.8in + ~\delta (q,\ed^{\rm eff},\Gamma^{\rm eff},\vp=0) - 2\tan^{-1}(\vp ),
\end{eqnarray}
where the final contribution to the phase, $-2\tan^{-1}(\vp )$, is added in undoing the transformation.
As is easily checked, the two expressions are not equivalent.  In the presence of finite $U$,
the gauge transformation thus does not lead to the correct form of the scattering phase.

While this is our most definite critique of the Dyson equations, we can draw attention
to two other more amorphous difficulties with the approach.  The first stems from discrepancies between
the direct computation of a correlator and computing it using the Dyson equation.  We have already
seen this issue arise at $V=0$ in computing $G^{++\atop --}(V=0,\om )$.  And we find it persists to finite dot-lead
coupling, $V$.  To this end, consider the Dyson relation for $G^{\rm ret}_{RL}(V,\om )$:
\begin{eqnarray}\label{eVxxix}
G^{\rm ret}_{RL}(V,\om ) &=& G^{\rm ret}_{RL}(0,\om ) + V^2\big(G^{\rm ret}_{RR}(0,\om)+G^{\rm ret}_{RL}(0,\om)\big)\cr\cr
&& \hskip -.7in \times G^{\rm ret}_{dd}(V,\om )
\big(G^{\rm ret}_{LL}(0,\om)+G^{\rm ret}_{LR}(0,\om)\big).
\end{eqnarray}
If we then employ the correlators in (\ref{eVxi}) to determine the $V=0$ retarded lead-lead correlators appearing
in (\ref{eVxxix}), we find
\begin{eqnarray}\label{eVxxx}
G^{\rm ret}_{RL}(V,\om ) &=& -{1\over 2}{\vp \over (1+\vp^2)} \cr\cr 
&& \hskip -.7in + {V^2\over 4(1+\vp^2)}{1\over D(\om )}
\big((\om-\ed)(i+\vp)^2+V^2(i+\vp)\big). \cr&&
\end{eqnarray}
But if we compute $G^{\rm ret}_{RL}(V,\om )$ directly using the equations of motion (\ref{eVvi}) and mode
expansions (\ref{eVvii}), we find instead
\begin{eqnarray}\label{eVxxxi}
G^{\rm ret}_{RL}(V,\om ) &=& 
{V^2\over 4(1+\vp^2)}{1\over D(\om )}\cr
&&\hskip -.5in \times \big((\om-\ed)(i+\vp)^2+V^2(i+\vp)\big).
\end{eqnarray}
We see that even at $V\neq 0$, the two methods of computing $G^{\rm ret}_{RL}(\om )$ give a
discrepancy of $-\vp/2(1+\vp^2)$.  Agreement between the two computations moves much further apart
if we instead take $G^{\rm ret}_{RL}(0,\om )$ to vanish identically as was originally indicated by
the equations of motion, i.e. as in (\ref{eVix}).

Thus we generally find discrepancies between the computation of correlators using the Dyson relations 
and a direct computation using the equations of motion.  We could eliminate the discrepancy between (\ref{eVxxx})
and (\ref{eVxxxi}) by instead (again) computing $G^{\rm ret}_{RR/LL}(V,\om )$ and then using the Dyson
relations in (\ref{eVx}) to compute $G^{\rm ret}_{RL}$, but this is not overly satisfying.  The problematic
nature of the Dyson's relations at $U=0$ does not instill confidence in their use at finite U.

The second difficulty arises in that the Dyson relations express all quantities
in terms of the dot correlator, $G^{\rm ret}_{dd}(V,\om)$, and non-interacting ($V=0$) lead-lead
correlators.  Now we believe that it is correct to treat $G^{\rm ret}_{dd}(V,\om)$ as perturbative in both $U$ and $\vp$
(just as the gauge transformation in (\ref{eIxvi}) leads to correct results for dot quantities).
At $U=0$, the only effect of $\vp$ on $G^{\rm ret}_{dd}$ is to renormalize both $V$ and $\ed$.
Thus the perturbative series in $U$ for the $G^{\rm ret}_{dd}$ correlator must be well-behaved and convergent
for the same reason it is at arbitrary $V$ and $\ed$ but $\vp=0$.  But while $G^{\rm ret}_{dd}(\om )$
is perturbative in $U$ and $\vp$, $G^{\rm ret}_{RL}$ need not be (and must not be for consistency with our
solution).  Thus expressing $G^{\rm ret}_{RL}$ in terms of $U,V=0$ correlators and $G^{\rm ret}_{dd}(V,\om)$
is a mistake.  
It implicitly assumes the problem is 
perturbative in both $\vp$ and $U$.
We note that this critique applies only to the $\vp \neq 0$ case.  If $\vp = 0$,
the reduced Dyson relations (see for example Ref. \onlinecite{wingreen}) agree with the exact solution.

If correct, the origin of this difficulty lies in the manner the Dyson relations are established.  The Dyson
relations are not non-perturbative but rather are valid order by order in perturbation theory, representing
a regrouping of terms.  The Dyson relations needed at $\vp \neq 0$ (i.e. the second relation in (\ref{eVv}))
involve a regrouping of terms at ${\cal O}(V^2)$ whereas the Dyson relations needed at $\vp =0$ involve
a regrouping of terms at only ${\cal O}(V)$.  It would then seem that in this more extensive regrouping of terms
that non-perturbative information is lost.

\begin{figure}
\vskip .58in
\begin{center}
\noindent
\psfrag{y}{$\pi n_{\rm imp}/\pi n_d$}
\epsfysize=0.35\textwidth
\epsfbox{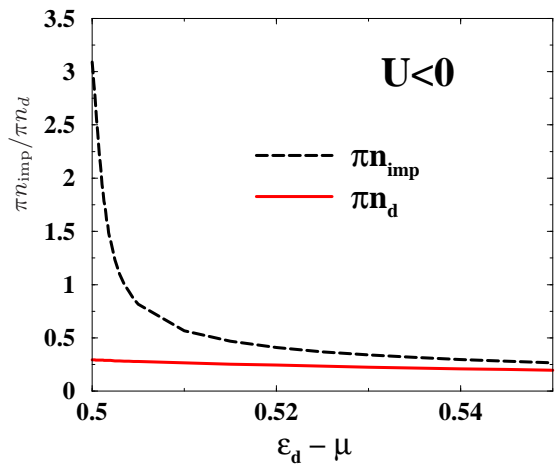}
\end{center}
\caption{$U<0$ plots of $\pi n_d$, with $n_d$ the number of electrons sitting on the dot, and  $\pi n_{\rm imp}$, with
$n_{\rm imp}$ the
number of electrons displaced by the impurity, as a function of $\ed-\mu$, where $\mu$ is the chemical potential.  
In these plots we have used $U=-1$ and $\Gamma = 0.1$.}
\end{figure}

\subsection{Dyson Equations and the $U<0$ Bethe Ansatz}

While it is obvious the $U>0$ Bethe ansatz and the Dyson equations 
give diverging answers for the conductance $G$, it is not {\it a priori} obvious the same is true at $U<0$.
At $U<0$, as detailed in Section IV, the conductance is a smooth function of $\vp$ and $\ed$.  The quantum
phase transitions that appear at $U>0$ vanish at $U<0$.
Might in this case then the two approaches agree?

We plot the two quantities, $(\delta_e + \delta_o) = \pi n_{\rm imp}$ and $\pi n_d$, in Figure 22
as a function of $\ed - \mu$.  We see that at $U<0$ the number of electrons displaced by the impurity is still different than
the number of electrons occupying the dot.  This suggests that recasting the correlators
appearing in (\ref{eVii}) in terms of the full dot correlators and $V,U=0$ lead-lead correlators remains
inappropriate.  As our demonstration in Section V.B of differing scattering phases 
between the two approaches was made without explicit recourse to the sign of $U$, this is
not surprising.

We can draw one additional conclusion from the $U<0$ analysis.
The lack of
discontinuities/non-analyticities in the transport for $U<0$ suggests that each term of a perturbative expansion of the
current in $U$ is well defined, but the series itself is not absolutely convergent.  If this is correct,
the Dyson equations are then attempting, in some sense, to manipulate the terms in a poorly behaved series.

\section{Comparison with G\"ores et al.}

In this section we study in some detail the experimental observations of Fano resonances
in the linear response conductance through a quantum dot
reported by G\"ores et al. (Ref. \onlinecite{gores}).  We will focus upon three aspects of their observations:
i) the general behaviour of the Fano resonances
as function of gate voltage; ii) their behaviour as a function of the dot-lead coupling;
and iii) their behaviour in a magnetic field.  Before considering any of this in
detail, one must treat the question of how appropriate it is to compare the predictions arising
from the model in (\ref{eIIi}) with the observations of G\"ores et al.

The presence of asymmetrically shaped peaks in the conductance reported in Ref. \onlinecite{gores}
suggests strongly that transport through the region containing the quantum dot occurs through
(at least) two distinct paths, one energy dependent and so resonant, one not.  
However the nature of the paths is equivocal.  In using the
Hamiltonian in (\ref{eIIi}), we model the resonant path as a tunneling through the quantum dot while we model 
the second tunneling path in the most minimal
way possible: as a contact term.  The appropriateness of this is not entirely clear.  The comparison
between theory and the experimental setup would be more robust if the geometry of scattering in
the dot region was more clear, for example if the observed Fano resonances arose from a dot embedded
in an Aharonov-Bohm ring.  Notwithstanding this uncertainty, we will find that a reasonable comparison
between our predictions and the observations of Ref. \onlinecite{gores} can be made.
This perhaps suggests that the physics is determined merely by the presence of two interfering paths and not
so much by the exact details of the paths.

In (\ref{eIIi}) we take
the dot to have a single spin degenerate level.
This requires the level broadening, $\Gamma \sim V^2$,
to be considerably less than $U+\Delta \epsilon$, where 
$\Delta \epsilon$ is the level spacing.  At least for a subset
of the data in Ref. \onlinecite{gores}, this condition is met
with $\Gamma/(U+\Delta \epsilon) \sim 1/35$.  We note that in
experimental measurements on dots with a single tunneling
path\cite{gold}, we have $\Gamma/(U+\Delta \epsilon) \sim 1/12$.  Here
the Anderson model does an excellent job of describing the
scaling behaviour of the reported finite temperature linear
response conductance.

An alternative approach to understanding the observations of Ref. \onlinecite{gores} 
was adopted in Clerk et al. (Ref. \onlinecite{clerk}).
There a combination of random matrix theory together with a non-interacting Green's function treatment modeling the size and
shape of the nano-junction (following Ref. \onlinecite{baranger}) was used to describe scattering through a dot coupled
to a non-resonant channel.  Aspects of the observations in Ref. \onlinecite{gores} were well
described by these methods.  In particular Ref. \onlinecite{clerk} was able to reproduce generic features
of the linear response conductance both in a magnetic field as well as its response to changes
to the dot-lead coupling, $\Gamma$.  However these approaches were unable
to fully explore the consequences of Coulomb interactions.  We, on the other hand, have argued that
Coulomb interactions lead to highly non-trivial physics non-perturbatively related to the non-interacting
case.  Nonetheless it is unclear whether this approach here is more appropriate than that adopted in Ref. \onlinecite{clerk}.
At least one can say our approaches share some of the same physics.  We will comment
upon this further in the context of actual comparison with experimental data.

We now turn to discussing the aforementioned observations of Ref. \onlinecite{gores}.

\begin{figure*}
\vskip .58in
\begin{center}
\noindent
\epsfysize=0.6\textwidth
\epsfbox{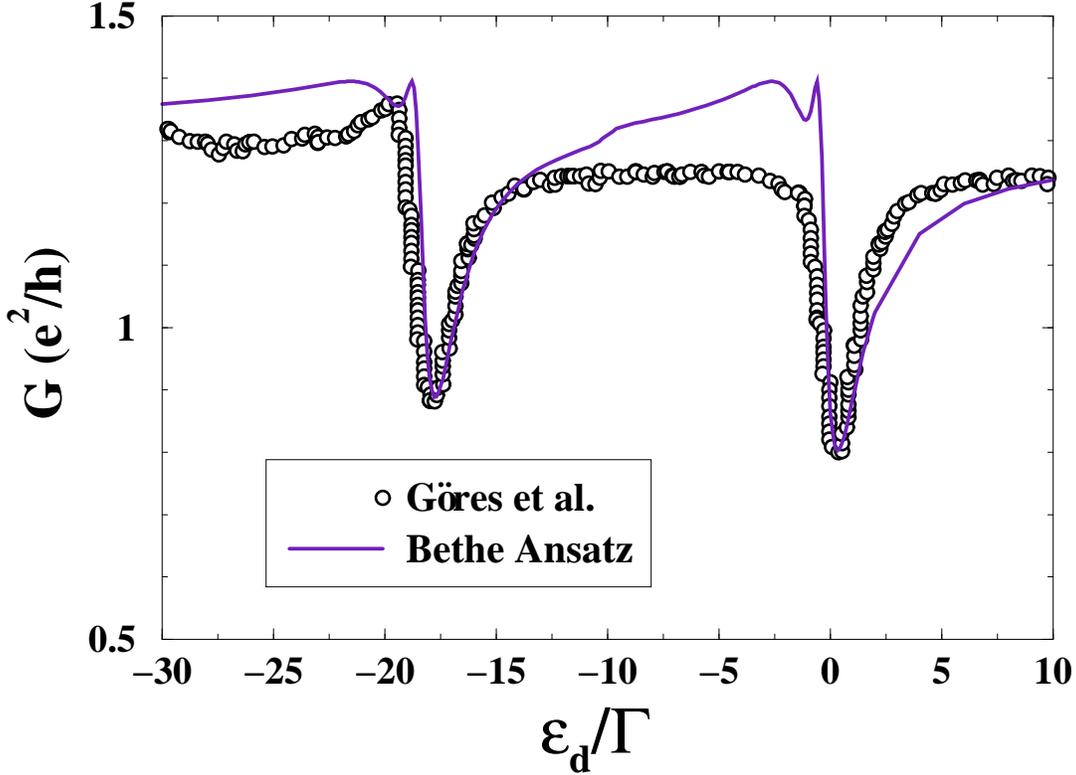}
\end{center}
\caption{A successive pair of Fano resonances observed in Ref. \onlinecite{gores}.  The black circles mark
the experimental data, the solid curve the computation from the Bethe ansatz.  For this computation, we take $U/\Gamma = 1/20$,
$\vp = 0.78$, and $V_R/V_L = 0.53$.}
\end{figure*}

\subsection{Behaviour of Fano resonances with gate voltage}

G\"ores et. al. report two well developed Fano resonances.  The resonances are plotted in Figure 23, appearing
as asymmetric dips.  These two dips are (slightly) differently shaped with different minima and maxima.

To model these resonances we must first take into account the possibility of an asymmetry between the dot-lead
couplings, i.e. $V_L \neq V_R$.  If $V_{LR}=0$, the only effect of $V_L \neq V_R$ is to reduce the maximal possible conductance:
\begin{equation}\label{eVIi}
G_{\rm max} = 2{e^2\over h} {4V_L^2V_R^2 \over (V_L^2+V_R^2)^2} \equiv 2{e^2\over h}\gamma.
\end{equation}
However with $V_{LR}$ finite, an asymmetry in $V_{L/R}$ produces a more complex effect.  As we indicated
in Section II, integrability in the presence of an asymmetry, $V_R\neq V_L$, requires an additional term in the
Hamiltonian, 
\begin{equation}\label{eVIii}
\delta {\cal H} = \bigg( V_{LL}c_L^\dagger c_L + V_{RR}c_R^\dagger c_R)\bigg|_{x=0},
\end{equation}
with
\begin{equation}\label{eVIiii}
V_{LL} = -V_{RR} = {V_{LR} \over 2V_LV_R}(V_L^2-V_R^2).
\end{equation}
Thus in order to take into account the reduction in $G_{\rm max}$ due to $V_L\neq V_R$ we must tolerate the presence of 
$V_{RR}/V_{LL}$ terms.  However we believe that $V_{RR}/V_{LL}$ will only
quantitatively change the predicted nature of transport through the impurity.  And a qualitative description is perhaps 
the best for which can be hoped given the overall simplicity
of the model together with the lack of complete understanding of the tunneling paths in the experiment.

We now consider how to fix the various parameters in the Hamiltonian.  We take the full width at half maximum of the resonances
to be indicative of $2\Gamma \simeq (V_L^2+V_R^2)$.  As such the ratio of the distance between the two 
resonances and $\Gamma$ is roughly $1/20$.  To determine then $U/\Gamma$ we note that the resonances occur
roughly at a dot chemical potential corresponding to $\ec$ and $\tilde\ec$.  Thus we must have
\begin{equation}\label{eVIiv}
{\ec - \tilde\ec \over \Gamma} \simeq 20 .
\end{equation}
Using (\ref{eIIIxii}) and (\ref{eIIIxxiii}) we then find that $U/\Gamma \simeq 20$ (and consequently the distance between
the two resonances corresponds to $U$).  This leaves us to determine $V_L/V_R$ and $V_p$.  We know from
the analysis of Section II that in general (unless $\vp=\pm\vc$), the conductance at $\ed =-U/2$  and at
$\ed =\infty$ is
\begin{equation}\label{eVIv}
G(\ed = -U/2) = G(\ed = \infty) = 2{e^2\over h} \gamma {4\vp^2 \over (1+\vp^2)^2}.
\end{equation}
For one of the resonances, we then determine the variation in the conductance, $\Delta G = G_{\rm res.~max}-G_{\rm res.~min}$ 
as a function of $\vp$ numerically.  Knowing $\Delta G (\vp )$ and $G(\ed=\infty)$ as a function
of $\vp$ permits $\vp$ to be then fixed.  We so find $\vp \simeq 0.78$.  Alternatively, and more crudely, we
could use the estimate for $\Delta G$ given in (\ref{eIIIxvi}).
If we do so we end up with an estimate of $\vp = 0.64$.  The difference between the two computations of
$\vp$ arises in that (\ref{eIIIxvi}) is only valid for $\vp$ large.
With $\vp$ in hand we can then fix the last unknown, the ratio of $V_L$ to $V_R$.  From (\ref{eVIv}) we
find $\gamma \simeq .66$ which then leads to $V_L/V_R \simeq .53$.

\begin{figure}
\vskip .5in
\begin{center}
\noindent
\epsfysize=.45\textwidth
\epsfbox{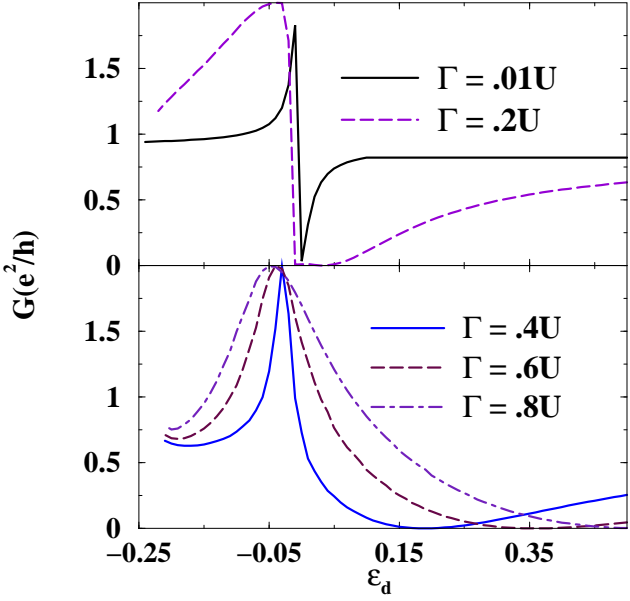}
\end{center}
\caption{A set of Fano resonances for differing values of $\Gamma$.
These curves are computed using $V_p = 0.4$ and with $V_L=V_R$.}
\end{figure}

A final consideration before a comparison can be made is the determination of how the gate voltage, $V_g$, applied
to the dot is related to the dot chemical potential, $\ed$.  For simplicity we assume $V_g \propto \ed$.  Midway
between the two resonance marks the point $\ed = -U/2$.  We can then fix the proportionality constant between $V_g$
and $\ed$ by insisting the positions of the minima of the rightmost resonance match.

Having done this, we are thus able to reproduce the behaviour of both Fano resonances in a reasonable
fashion.  We do note that the predicted conductance in the immediate vicinity of the rightmost resonance 
overshoots the measured conductance by 15\%.  The predicted conductance, G, also possesses a number of sharp features
not seen in the data.  However at the particular gate voltages at which these occur,
we are comparing a zero temperature computation with finite but low temperature
($T=100$mK) data.  As such some of the sharp features could be seen as a $T=0$ artifact.

A notable characteristic of this data set is that the resonances do not vanish.  One possibility is that there
is some incoherent background to the conductance, something assumed in Ref. \onlinecite{gores}.  An incoherent background
is necessary if one assumes a non-interacting problem.  In such a case we may write for the conductance 
(without a background contribution) the following,
\begin{equation}\label{eVIvi}
G = 2{e^2\over h}{(e+q)^2\over e^2 +1},
\end{equation}
where $e=\cot (\pi n_d/2)$ ($n_d$ is the total number of electrons sitting on the dot) and $q=-\cot (2\tan^{-1}(\vp ) )$.
Thus as $\ed$ is varied (so changing $n_d$ from $0$ to $2$), $G$ will necessarily vanish at some $\ed$.
In the Dyson equation approach (see Section V) at finite $U$, $G$ still vanishes for the same reason.
In either case a background contribution to the conductance is needed in order to be consistent
with the data set.

As an alternative to explicitly adding a background contribution, one can break time reversal symmetry.
With this symmetry broken, $q$ in (\ref{eVIvi}) is potentially complex and $G$ then need not vanish as $\ed$
is varied.  This is seen in the random matrix treatment of Ref. \onlinecite{clerk} as well as in
Refs. \onlinecite{bulka} and \onlinecite{hofstetter} where the introduction of an Aharonov-Bohm phase serves to break
time reversal symmetry.

The approach used in this article, on the other hand, does not necessitate either the breaking of time reversal symmetry nor the 
explicit introduction of a $G_{\rm background}$.  Rather $G$ remains finite through the non-perturbative effects of
$U$.  Although $G$ under the Bethe ansatz can be cast in the form (\ref{eVIvi}), $e$ in this case does not equal
$\cot (\pi n_d /2)$.  Instead $e=\cot (\pi n_{\rm imp}/2)$ where $n_{\rm imp}$ is
the total number of electrons displaced by the impurity.  This quantity, behaving considerably differently than $n_d$
as a function of $\ed$, does not force the vanishing of $G$.

G\"ores et al. report on the temperature dependence of the resonances in Ref. \onlinecite{gores}.  They observe
that the region in between the resonances is markedly insensitive to temperature while the resonances themselves
grow in a Kondo log fashion as the temperature is lowered from $T=1000$mK to $T=100$mK.  While we are not prepared
to go into an in-depth treatment of the temperature dependence here, we will make a few sketchy comments.  

\begin{figure*}
\vskip .5in
\begin{center}
\noindent
\epsfysize=.5\textwidth
\epsfbox{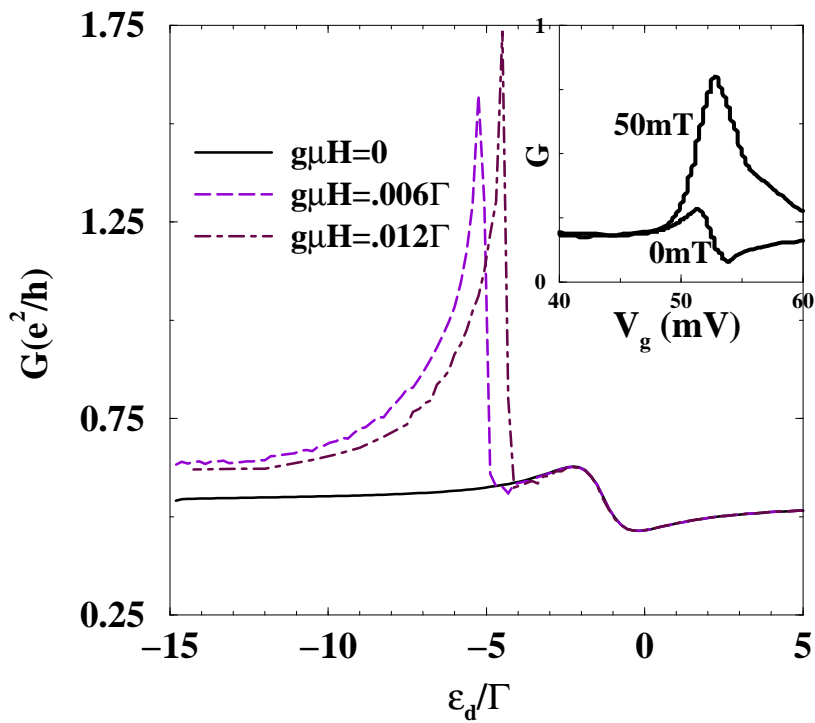}
\end{center}
\caption{The response of a bipolar Fano resonance to the application
of small magnetic fields, $H$.  To compute these curves
we employ $\Gamma/U = 1/30$ (with $V_L=V_R$) and $V_p = -3.6$.
Inset: Observed response in Ref. \onlinecite{gores}
of a Fano resonance to small applied fields.}
\end{figure*}

In our
discussion of the magnetoconductance in Section III we identified the ranges of values of $\vp$ at $\ed = -U/2$ where there would
be low-lying spectral weight present.  Our purpose there was to identify regimes of $\vp$ where small magnetic
fields would drastically change the conductance.  But what is true about a finite magnetic field is also true
of finite temperature.  At $\vp = 0.78$ and with the parameters, $U$ and $\Gamma$, identified as above,
we expect on the basis of Section III a surfeit of low lying spectral weight.  Thus as temperature is lowered
we expect the conductance to vary rapidly in the region of $\ed = -U/2$.  In  G\"ores et al. no
such variation is observed.  The cause of this is an extremely low Kondo temperature at $\ed=-U/2$.  Using\cite{haldane,wie}
\begin{equation}\label{eVIviii}
T_k = \sqrt{U\Gamma \over 2}e^{\pi\big(\ed (\ed+U)-\Gamma^2\big)/(2\Gamma U)},
\end{equation}
as the expression for the Kondo temperature, we find
\begin{eqnarray*}
T_k(\ed =-U/2) \sim 1 {\rm mK}.
\end{eqnarray*}
As such the temperature range studied in Ref. \onlinecite{gores} ($100-1000$mK) is decidedly in the high temperature
regime (in the region of $\ed \sim -U/2$).  Our zero temperature computation nonetheless finds reasonable agreement
because the low temperature behaviour, $T\ll T_k$, is equivalent to the high temperature behaviour, $T \gg T_k$.
This differs markedly from the Anderson model at $\vp=0$.  At $\vp = .78$ the low
lying spectral weight (as encoded in the $\rho_{\rm imp}(q)$ distribution) doubles.  Crudely speaking then
we expect the high temperature scattering phase, $\delta_e$, to behave as
\begin{eqnarray}\label{eVIvii}
\delta_e (T \gg T_k, \ed = -U/2) \sim \delta_e (T=0,\ed=-U/2) + 2\pi .\cr &&
\end{eqnarray}
The shift by $2\pi$ (as opposed to merely $\pi$ at $\vp =0$) thus implies a coincidence of the high and low
temperature conductance.  It also explains why our $T=0$ computation adequately conforms to the (high temperature) observations
(about $\ed \sim -U/2$).  

While at $\ed = -U/2$ the measurements in Ref. \onlinecite{gores} are high temperature, the corresponding observations
in the region of the resonances may be considered low temperature. 
In the region of the rightmost resonance, $\ed = 0$ and $T_k \sim 1 K$.  This is consistent with the observations
in Ref. \onlinecite{gores} of the resonance growing in depth as temperature is lowered.  Thus at the lowest temperature
at which measurements were taken, $T=100$mK, the system is in a low temperature regime.  And consequently we do not
the behaviour in the two regimes to coincide to obtain a reasonable match between data and theory.

\subsection{Dependence of width of Fano resonances upon $\Gamma$}

In Ref. \onlinecite{gores}, Fano resonances were studied as a function of
the total dot-lead coupling strength, $\Gamma$, where it was observed
that the width of Fano resonances exhibit a non-monotonic dependence upon
$\Gamma$.  (In a dot with a single tunneling path, the width of a resonance
merely increases with $\Gamma$.)  Together with this non-monotonicity,
the overall shape and amount of asymmetry in the Fano resonances was observed
to be sensitive to the strength of $\Gamma$.

We can reproduce this array of behaviour.  Plotted in Figure 24 is the
linear response conductance for a set of differing $\Gamma$'s.  For
$\Gamma$ small, a Fano resonance appears as a sharply peaked bipolar structure.
As $\Gamma$ is increased, as na{\"\i}vely expected, the bipolar peak
broadens.  However at some critical value of $\Gamma \sim (.3-.4)U$,
the bipolar resonance is replaced by a narrow unipolar one.  
With further increases in $\Gamma$,
this resonance broadens out.

We point out that the reported behaviour of the Fano resonances as a function of $\Gamma$ mandates 
against the form
$$
\delta_e(q) = -2\tan^{-1}({\Gamma \over q-\ed}) - 2\tan^{-1}(\vp ),
$$
of the single particle scattering phase (see (\ref{eIxiv})).  
This is to say, these results imply the scattering due to the
two paths should be treated as coincident (as contained in (\ref{eIxii})).
This is implicit in both the behaviour of the Fano resonances as a function of gate
voltage as well as the resonances' behaviour in a magnetic field in the following section.
But here we perhaps have a more explicit demonstration.
If the above was appropriate, variations in $\Gamma$ would not affect the asymmetry parameter, 
$q = -\cot (2\tan^{-1}(\vp ))$.  But as a measure of the shape of the resonances, $q$ must
and does vary (see Figure 3 of Ref. \onlinecite{gores} for explicit numbers).

In Ref. \onlinecite{clerk} a different approach is taken to describe this
phenomenon.  Using a specific microscopic picture of the dot, these authors consider the effect of modifying the second
tunneling path (as opposed to directly changing the strength of the dot-lead hopping of the first tunneling path).  
They find that the width of the resonances behaves non-monotonically as the strength of an impurity placed in the second
pathway is varied.  We might mimic this approach by varying $\vp$ instead of $\Gamma$.  If we did so, we would
certainly find the width of the Fano resonances varies non-linearly (see for example Figures 12 and 15).

\subsection{Linear response conductance at $H\neq 0$}

The behaviour of Fano resonances in magnetic fields was also studied
in Ref. \onlinecite{gores}.  It was found that the resonances exhibited a marked
response to extremely small magnetic fields ($g\mu H/ \Gamma \sim 10^{-2}$).
In particular they demonstrated that upon application of $H$, a small bipolar
Fano resonance was transformed into a much larger unipolar structure
(see inset to Figure 25).  The authors of Ref. \onlinecite{gores} suggested the origin of this behaviour
lay in the large effective area covered by an electron traversing the dot via the resonant channel.
This large area, $A$, needs a relatively small magnetic field, $H$, to produce a flux $A\cdot H$
significant enough to induce the breakdown of coherent backscattering.

We are able to offer a different explanation found in the Kondo-like physics present in our treatment of the problem.
From Section III we know that there exist wide ranges of $\vp$ where a doubling of the low-lying spectral weight
at positive energies occurs relative to $\vp = 0$.  This low-lying spectral weight can
be accessed by magnetic fields on the order of the Kondo temperature, $T_k$, as computed
in the $\vp=0$ Anderson model.  In such ranges of $\vp$, and as discussed in detail in Section III,
we expect a marked sensitivity to small magnetic fields.

It is then not surprising we are able to reproduce the magnetoconductance measurements of Ref. \onlinecite{gores}.
In Figure 25 we find that the conductance for $H=0$, $U/\Gamma = 30$, and $\vp = -3.6$ (with this choice of $\vp$
the low-lying spectral weight has been enhanced in comparison with $\vp=0$) is characterized
by a Fano resonance with a small bipolar structure.
Upon introducing a small field, $g\mu H=0.006\Gamma = 1.12T_k$, a unipolar peak is superimposed
over the bipolar structure.  As the field is increased further (although still kept to be the same
order as the Kondo temperature) this unipolar peak narrows in width.
Plotted in the inset to Figure 25 are the experimental measurements of the conductance in the presence of
a small magnetic field.  At $H=0$, Ref. \onlinecite{gores} sees a small bipolar peak which as a small
magnetic field is turned on is obliterated by a much larger unipolar structure.  That the bipolar
structure does not entirely disappear in our calculation is perhaps a reflection 
that our computation is done at $T=0$.  With finite temperature blurring, the bipolar peak
might completely merge into the larger unipolar structure.

In Ref. \onlinecite{clerk} this sensitivity to field is explained in a statistical sense.  The introduction
of a magnetic field (of any size) serves to break time reversal invariance.  
The breaking of this discrete symmetry changes the statistical S-matrix (i.e. it places it in a different
universality class) used by Ref. \onlinecite{clerk} to describe scattering through the dot.  At least in an
average sense this change tends to favour larger unipolar over smaller bipolar structures in the conductance.

\begin{acknowledgments}
The author acknowledges support 
from the NSF (DMR-0104799) and the DOE (DE-FG02-97ER41027) at the University
of Virginia and the DOE (DE-AC02-98CH10886) at Brookhaven National Laboratory.
He also acknowledges helpful discussions 
with A. Tsvelik, D. Haldane, C. Hooley, M. J. Bhaseen, A. Ludwig, A. Clerk, D. Goldhaber-Gordon, and in particular
F. Essler.
\end{acknowledgments}

\bibliographystyle{apsrev}

\end{document}